\documentclass[acmsmall]{acmart}

\AtBeginDocument{%
  \providecommand\BibTeX{{%
    \normalfont B\kern-0.5em{\scshape i\kern-0.25em b}\kern-0.8em\TeX}}}

\setcopyright{cc}
\setcctype{by-nd}
\acmJournal{PACMHCI}
\acmYear{2025} \acmVolume{9} \acmNumber{2} \acmArticle{CSCW095} \acmMonth{4} \acmDOI{10.1145/3710993}

\usepackage{ifthen}
 \newboolean{showcomments}
 \setboolean{showcomments}{False} % Set to false to hide all comments

\newcommand{\raj}[1]{\ifthenelse{\boolean{showcomments}}{\textcolor{purple}{[#1 —raj]}}{}}

\usepackage{multirow}
\usepackage{caption}
\usepackage{subcaption}
\usepackage{svg}
\usepackage{csquotes}

\begin{document}

%%
%% The "title" command has an optional parameter,
%% allowing the author to define a "short title" to be used in page headers.
\title[Helping the Helper]{\textit{Helping the Helper}: Supporting Peer Counselors via AI-Empowered Practice and Feedback}

%%
%% The "author" command and its associated commands are used to define
%% the authors and their affiliations.
%% Of note is the shared affiliation of the first two authors, and the
%% "authornote" and "authornotemark" commands
%% used to denote shared contribution to the research.
\author{Shang-Ling Hsu}
\email{hsushang@usc.edu}
\orcid{0009-0005-1096-0405}
\affiliation{
  \institution{University of Southern California}
  \country{USA}
}
\author{Raj Sanjay Shah}
\email{rajsanjayshah@gatech.edu}
\orcid{0000-0002-0847-8426}
\affiliation{
  \institution{Georgia Institute of Technology}
  \country{USA}
}
\author{Prathik Senthil}
\email{psenthil3@gatech.edu}
\orcid{0009-0007-7102-5773}
\affiliation{
  \institution{Georgia Institute of Technology}
  \country{USA}
}
\author{Zahra Ashktorab}
\email{Zahra.Ashktorab1@ibm.com}
\orcid{0000-0002-0686-7911}
\affiliation{
  \institution{IBM Research}
  \country{USA}
}
\author{Casey Dugan}
\email{cadugan@us.ibm.com}
\orcid{0000-0002-1508-2091}
\affiliation{
  \institution{IBM Research}
  \country{USA}
}
\author{Werner Geyer}
\email{werner.geyer@us.ibm.com}
\orcid{0000-0003-4699-5026}
\affiliation{
  \institution{IBM Research}
  \country{USA}
}
\author{Diyi Yang}
\email{diyiy@stanford.edu}
\orcid{0000-0003-1220-3983}
\affiliation{
  \institution{Stanford University}
  \country{USA}
}

%%
%% By default, the full list of authors will be used in the page
%% headers. Often, this list is too long, and will overlap
%% other information printed in the page headers. This command allows
%% the author to define a more concise list
%% of authors' names for this purpose.
% \renewcommand{\shortauthors}{Hsu et al.}
\renewcommand{\shortauthors}{Shang-Ling Hsu et al.} 

%%
%% The abstract is a short summary of the work to be presented in the
%% article.
\begin{abstract}
Millions of users come to online peer counseling platforms to seek support. 
However, studies show that online peer support groups are not always as effective as expected, largely due to users' negative experiences with unhelpful counselors.
Peer counselors are key to the success of online peer counseling platforms, but most often do not receive appropriate training. 
Hence, we introduce \texttt{CARE}: an AI-based tool to empower and train peer counselors through practice and feedback. 
Concretely, \texttt{CARE} helps diagnose which counseling strategies are needed in a given situation and suggests example responses to counselors during their practice sessions. %Counselors can select, modify, or ignore any suggestion before replying. 
Building upon the Motivational Interviewing framework, \texttt{CARE} utilizes large-scale counseling conversation data with text generation techniques to enable these functionalities. We demonstrate the efficacy of \texttt{CARE} by performing quantitative evaluations and qualitative user studies through simulated chats and semi-structured interviews, finding that \texttt{CARE} especially helps novice counselors in challenging situations.
The code is available at \url{https://github.com/SALT-NLP/CARE}.
\end{abstract}

%%
%% The code below is generated by the tool at http://dl.acm.org/ccs.cfm.
%% Please copy and paste the code instead of the example below.
%%


%%
%% Keywords. The author(s) should pick words that accurately describe
%% the work being presented. Separate the keywords with commas.

%% A "teaser" image appears between the author and affiliation
%% information and the body of the document, and typically spans the
%% page.
% \begin{teaserfigure}
%   \includegraphics[width=\textwidth]{sampleteaser}
%   \caption{Seattle Mariners at Spring Training, 2010.}
%   \Description{Enjoying the baseball game from the third-base
%   seats. Ichiro Suzuki preparing to bat.}
%   \label{fig:teaser}
% \end{teaserfigure}

% \received{20 February 2007}
% \received[revised]{12 March 2009}
% \received[accepted]{5 June 2009}

%%
%% This command processes the author and affiliation and title
%% information and builds the first part of the formatted document.
\maketitle

\section{Introduction} 
Millions of people (nearly 1 in 5 American adults) are touched by mental illness \citep{dieleman2016us}.  
The majority of people affected never access care \cite{center2015behavioral}, due to structural barriers including the high cost of treatment and the lack of trained professionals to meet the demand \cite{kazdin2011rebooting,mojtabai2011barriers,kantor2017perceived}. 
One promising remedy for under-treatment and, to meet the demand for mental healthcare, is the use of online peer counseling platforms, % \footnote{In this work, we use \emph{online peer counseling platform} and \emph{online peer support groups} interchangeably.}, 
which provide the benefits of anonymity, empowerment, and ease of access \cite{houston2002internet,melling2011online,powell2007investigating}. Online peer counseling platforms have been proven effective in reducing people's self-reported anxiety and depression and improving their quality of life \citep{clarke2005overcoming,eysenbach2004health}.
Nevertheless, studies show that online support groups are not always as effective as expected \citep{hogan2002social},
largely due to support seekers' negative experiences with unhelpful supporters \footnote{In this work, we use \textit{peer counselors} (or sometimes supporters, or support providers) to refer to the individuals providing help and use \textit{support seekers} to refer to people seeking or receiving support. }\citep{smithson2011problem,burke2010membership,burke2008mind}.
Supporters are the key to the success of online peer counseling platforms \cite{ali2015online}, and the quality of the support that seekers receive within a community heavily depends on those providing support.  
Thus, finding ways to help peer counselors be more helpful and supportive is important.

Evidence from volunteer-based support services indicates that even brief training may improve a supporter's ability to help seekers \citep{rodgers2010review} and may ultimately lead to improved mental health outcomes in seekers \citep{armstrong2010effective}.  
Thus, there is a growing trend in scaling effective peer support training online.  For instance, \textit{7 Cups} \cite{7cups}, an online peer-to-peer counseling platform, provides users with training on active listening techniques \cite{baumel2016adjusting}, and \textit{SAHAR}, an Israeli-based suicide-prevention initiative, trains its online counselors extensively during weeks of in-person training to handle supportive chats \cite{barak2007emotional}. 
However, supporters typically do not have the rigorous professional training of their offline counterparts \cite{pruksachatkun2019moments}, with very abbreviated training or no training at all for most online peer support groups \cite{wang2015eliciting,gould2013impact}. 
Thus, even if supporters want to help others in need, they might lack the necessary set of skills or the knowledge to use evidence-based counseling strategies \cite{theriault2009feelings}.
Moreover, unlike traditional counseling settings, online supporters often do not have systematic ways to receive supervision or guidance, especially during their conversations with seekers and when there are high levels of uncertainty about how to respond. Without appropriate guidance, supporters might develop biased or even inappropriate helping skills without being aware of it. This can lead to ineffective, or at worst, harmful, support, which might prevent seekers from receiving effective support or other types of support options sooner \citep{reynolds2004mismatches,sovold2021prioritizing}
and may also expose supporters to self-doubts \cite{theriault2009feelings}, increased stress and decreased self-efficacy \citep{curran2019does,rozental2016negative}. 
\begin{figure*}[t!]
  \centering
  \includegraphics[trim = {1cm 4cm 3cm 3cm}, clip, width=0.9\textwidth]{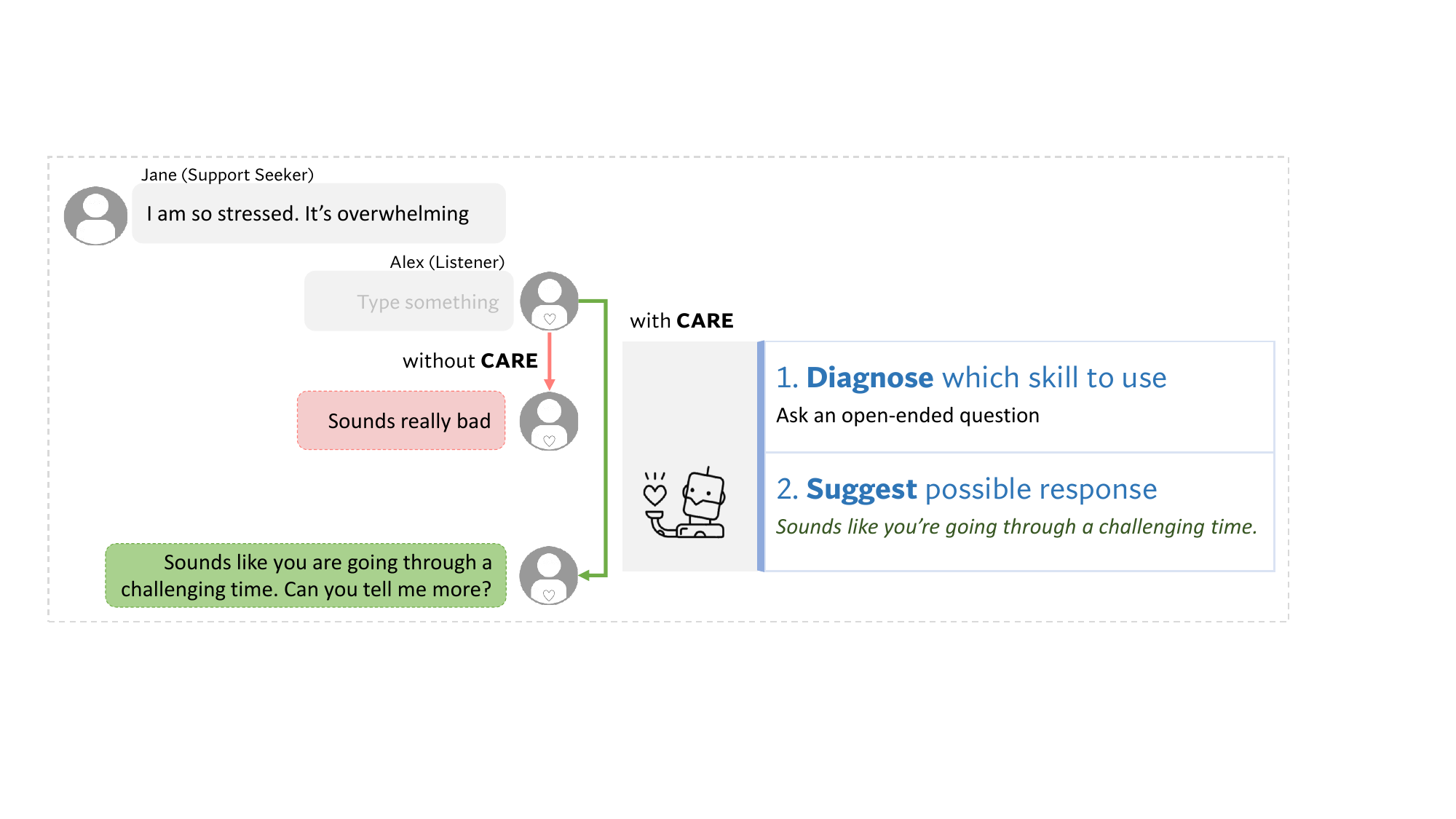}
  \caption{Interface of how \texttt{CARE} empowers supporters by diagnosing which skill to use and suggesting responses. 
}
\Description[Example interface of CARE]{Example interface of \texttt{CARE} to empower supporters by diagnosing which skill to use and suggesting responses.}
  \label{fig:teaser_2}
\end{figure*}

Existing mechanisms of training or scaffolding largely rely on human supervision, which requires an extensive amount of resources such as cost, time, labor, and expertise, 
making it hard to scale up to help the large number of supporters \cite{atkins2014scaling} who 
support millions of people in need of care
in text-based online peer counseling platforms. 
Social computing research on online peer support groups mainly looks at how supporters' language use or counseling strategies are associated with better outcomes for seekers \cite{althoff2016large,perez2019makes,perez2017predicting,chikersal2020understanding,pruksachatkun2019moments,o2017design}. 
In contrast, there have been relatively fewer studies on building scalable systems and tools to empower supporters to offer the most effective support in online peer counseling platforms, with a few exceptions \cite{peng2020exploring,sharma2023human}. More importantly, the majority of research has specifically been about designing virtual counselors that emulate a human counselor in a chatbot-like setting \cite{han2013counseling,han2015exploiting}. 
In contrast to prior research on chatbots focused on virtual counselors or client-clinician interaction \cite{coyle2012interaction}, our approach involves the development of an automated \textit{assistant} designed to enhance and support individuals by providing exposure to a diverse array of counseling strategies during training, with the intention of augmentation rather than replacement \cite{raisch2021artificial}.

Concretely, we propose to design an interactive training agent \texttt{CARE} to empower peer counselors in text-based, online peer-to-peer counseling platforms. Here, we use 7 Cups as a research site and build contextualized language generation approaches to provide tailored assistance for supporters by highlighting which counseling strategies are needed in a given situation and suggesting example responses.  Building upon psychotherapy theories and empirical studies on Motivational Interviewing (MI) \cite{MITI_4,shah2022modeling}, we select a set of counseling strategies from MI to help volunteer counselors. We build machine learning classifiers on top of the pre-trained language model BERT \cite{BERT} to predict the most suitable counseling strategies in a given context and then develop language generation approaches via a pre-trained generation model DialoGPT \cite{DialoGPT} for generating example responses that counselors can further use and edit. As shown in Figure \ref{fig:teaser_2}, we integrate different components into an interactive interface for \texttt{CARE} to enable supporters at scale. 

We work closely with different stakeholders from 7 Cups in the development process of \texttt{CARE}: platform developers and volunteer counselors. To evaluate whether the proposed system \texttt{CARE} works for peer counselors, we perform quantitative and qualitative evaluations, including system log analyses, questionnaires, and semi-structured interviews, and demonstrate the impacts of \texttt{CARE} on counselors. We find that \texttt{CARE} especially helps beginner and novice counselors to better deal with these challenging situations. To sum up, we make the following contributions:
\begin{itemize}\setlength\itemsep{0em}
    \item Develop contextualized language generation techniques that provide tailored assistance for peer counselors by highlighting which counseling strategies may be needed in a given situation and suggesting example responses. 
    \item Create an interactive tool \texttt{CARE} to empower peer counselors by offering on-the-fly suggestions during practice sessions on an online peer counseling platform. %  \slh{Is there a better way to frame this...?}
    \item Evaluate \texttt{CARE} via both quantitative system log analyses and qualitative user studies to demonstrate the efficacy of \texttt{CARE} on a set of representative counseling scenarios.    % \item Provide concrete design recommendations for online peer counseling platforms in terms of how to better support counselors using training and scaffolding. 
    % \item Demonstrate the operationalization of an effective human-AI collaboration paradigm for a socially important context around online peer-to-peer counseling.
\end{itemize}
% \vspace{-0.15in}
\section{Related Work}
\subsection{Peer Support and Online Peer Counseling Platforms}
Peer support has been shown to be effective in a wide range of mental health services \cite{repper2011review}, and is a major motivation for people who go for online mental health support \cite{powell2007investigating}.  
Prior work has provided some evidence for the effectiveness of online peer support groups. For instance, a recent survey on 7 Cups reported higher user satisfaction with the support provided by 7 Cups counselors, and users who indicated receiving psychotherapy in the past marked the peer counselors' support to be as helpful as psychotherapy \cite{baumel2015online}.
Most studies on online text-based peer counseling context have focused on the automatic modeling of behaviors from peer counselors and seekers and the correlations between the behaviors and successful outcomes \citep{klonek2015coding,xiao2014modeling,lord2015more,althoff2016large,yang2024makes}, user engagement \citep{sharma2023human,andalibi2018social}, or support provision \citep{linguistic_accommodation,kim2023supporters}. 
% Howes et al. \cite{howes-etal-2014-linguistic} investigate depression and anxiety symptom severity on online therapy platforms through linguistic analysis. 
% Similar to these studies, we model the behaviors of successful peer counselors to provide suggestions on counseling strategies and example responses.

An increasing volume of automated text-based analytical work has been carried out to identify behavioral codes derived from the Motivational Interviewing (MI) Skill Codes \citep{miller1996motivational, can2012case, tanana2016comparison, perez2017predicting, perez2019makes, schwalbe2014sustaining, huang2018modeling, fang2023makes, min-etal-2022-pair}. Some works specifically concentrate on automating the scoring of \emph{reflections}, a key listening skill, to offer numeric feedback on counselors' responses \cite{can2012case, perez2017predicting, shen2020counseling, min-etal-2022-pair}. Others explore the relation between the use of multiple MI strategies and the satisfaction of support seekers in counseling sessions, showcasing the effectiveness of MI strategies through extensive data analysis \cite{shah2022modeling, perez2019makes, fang2023makes}. In a recent study, Chen et al. \cite{scaffolding} delve into the challenges faced and MI strategies employed by novice peer counselors through qualitative user studies, discussing design implications to better prepare for the skill development of novice therapists. Building upon these prior works, we focus on a set of widely used MI strategies and further leverage these to guide the design of an interactive tool aimed at empowering peer counselors, especially inexperienced ones, with contextual examples for emulation during training phases.

% \vspace{-0.1in}
\subsection{Systems and Interventions for Peer Counseling Platforms}
Interventions for support are generally valued by people with mental health concerns \cite{o2017design}. For example, people with schizophrenia showed a positive attitude toward using technology for care \cite{gay2016digital}. 
Supporters show a substantial interest in learning helping skills and using web-disseminated peer support interventions to help one another online \cite{bernecker2017web}. Supportive chats guided by prompts in a design study were found to be associated with reduced anxiety, perceived as ``deeper'' with solutions to problems and new perspectives \cite{o2018suddenly} compared to unguided ones, calling for natural language processing empowered systems to provide appropriate scaffolds at scale. This is in line with a recent study on the impact of technology in psychotherapy \cite{imel2017technology}, which has identified the development of machine learning technologies for counselors' training and feedback as important needs where technology has a significant impact.
To address this, our work leverages natural language processing methods to provide peer counselors with contextualized training and to help them acquire or improve their counseling skills. %, in order to further promote many of the benefits of peer support online.

Additionally, there has been increased attention towards building automatic tools for online health communities \citep{peng2020exploring, Greer2019UseOT}, however, the majority of research has specifically been about designing virtual counselors that emulate humans in a chatbot-like setting. For instance, Han et al. \cite{han2013counseling, han2015exploiting} introduce a counseling dialogue system that engages users by recognizing their input and generating responses incorporating basic counseling techniques through rule-based decisions and extraction-based response templates. \citet{shen2020counseling} and \citet{oneil-etal-2023-automatic} employ GPT-based models to generate reflections. These \textbf{automation} approaches are efficient and labor-free \cite{raisch2021artificial}. However, in high-stakes and high-touch scenarios like counseling, \emph{automation} poses risks for support seekers and raises ethical and legal concerns \cite{liu2022will, imel2017technology, miner2019key, zohny2024generative, woodnutt2024could}. In contrast to \emph{automation} approaches, Saha et al. \cite{saha2022towards} and Sharma et al. \cite{sharma2021towards} devise response rewriting methods to enhance empathy to \emph{augment} trainee's candidate response. Similarly, to further increase interactiveness, Sharma et al. \cite{sharma2023human} propose HAILEY, a tool that suggests modifications to peer supporters' responses to help them respond more empathically to support seekers. 
Given the potential risks associated with \emph{automation}, \texttt{CARE} also focuses on \textbf{augmentation} of peer counselor support by proactively diagnosing counseling strategies to use and suggesting possible responses in a contextualized manner. Overall, we argue that an \emph{augmentation} approach offers a balanced solution between support at scale and risk management.

\subsection{Contextualized Language Generation}
To empower peer supporters, automatic assistants must generate situationally appropriate responses. Contextualized language generation is rooted in generating text that is closely tied to the surrounding context, which is particularly important in tasks that heavily rely on context, such as text summarization, machine translation, and chatbots.
Various strategies are employed for contextualized language generation. These strategies involve training recurrent models like RNNs \cite{rumelhart1986learning, shang2015neural}, LSTMs \cite{hochreiter1997long}, and GRUs \cite{cho2014learning} using sequence-to-sequence frameworks. Moreover, a prevalent approach involves utilizing pre-trained Transformer-based models \cite{vaswani2017attention} like BART \cite{BART}, T5 \cite{raffel2020exploring}, and the GPT series \cite{GPT2, brown2020language, ouyang2022training, openai2023gpt4}, where these models can be fine-tuned, trained from scratch, or queried for specific tasks.

The choice of context in contextualized language generation depends on the task and design considerations. Specifically, in generating responses for conversations, it's common to incorporate the conversation history, comprising prior utterances, to ensure coherent dialogues. In scenarios involving single-turn responses, researchers often employ RNNs \cite{shang2015neural} or LSTMs \cite{vinyals2015neural}, using the most recent utterance as input. For more extensive and ongoing conversations, approaches like XiaoIce \cite{zhou2020design} introduce encoded dialog states to the input of GRUs \cite{cho2014learning}. Addressing multi-turn conversations and aiming for diverse responses, DialoGPT \cite{DialoGPT} builds upon GPT-2 \cite{GPT2}, re-ranking generation options using a backward model.
BlenderBot \cite{roller2021recipes} enhances context by adding conversation topics and personas to the dialog history, incorporating a retrieval step to mitigate hallucinations. Its subsequent iteration \citep{xu2022beyond} continuously maintains a summary of conversation history for reference across chat sessions. InstructGPT \cite{ouyang2022training} adopts reinforcement learning to fine-tune GPT-3 \cite{brown2020language}, aligning responses better with human intent.
% LaMDA \cite{thoppilan2022lamda} enriches generated responses with knowledge from selected sources, bolstering the grounding of dialogues.
More recent models such as ChatGPT and GPT-4 \cite{openai2023gpt4} push the contextualized language generation to a new height by generating very human-like text based on context and past conversations. However, such models are only accessible via API calls, which raises concerns over data security and confidentiality and makes them difficult to tailor to mental health settings with model fine-tuning or few-shot learning.  
Thus to generate contextually relevant suggestions for mental health, we extend a controllable version of DialoGPT to fine-tune it to build \texttt{CARE}, which incorporates conversation history and integrates an MI strategy into the context for language generation.

Many studies have looked at language generation techniques to support real-time writing assistants, covering diverse applications such as keyboard prediction \cite{hard2018federated}, grammar and spelling error correction \cite{fitria2021grammarly, ji2021spellbert}, auto-completion \cite{chen2019gmail, chen2021evaluating, karapapa2015search, nazari2021application, wang2022clozesearch}, and response suggestion for messaging \cite{hohenstein2018ai} and emails \cite{kannan2016smart, robertson2021can}. Building on these studies, our interface design resembles that of response suggestion assistants. Similar to prior research on writing assistants, we quantitatively evaluate \texttt{CARE} with utilization-based metrics, such as click-through rate, visibility of suggestions, and response lengths. Hence, certain findings from previous studies, such as the trade-offs between suggestion savings and cognitive workload \cite{trnka2009user, quinn2016cost, kosch2023survey}, may be relevant to \texttt{CARE}.
Unlike other similar writing assistants, such as commercial software to generate suggestions in email \cite{robertson2021can} or controlling the diversity of suggestions through clustering \cite{kannan2016smart}, \texttt{CARE} aims to help novices gain skills by providing them with domain-aware generation conditioned on different Motivational Interviewing (MI) Skill Codes. 

\section{The Design of \texttt{CARE}}\label{sec:design}
Realistic practice and tailored feedback are key processes for training individuals with therapy skills \citep{reagans2005individual,kluger1996effects,bennett2006therapist}. As discussed earlier, existing mechanisms of providing feedback largely depend on human supervision. It often takes well-trained human raters up to ten times as long as the duration of the session itself to finish labeling that session on the involved counseling strategies \citep{atkins2014scaling}, let alone providing detailed feedback, making it difficult to scale up to help a large number of volunteer counselors who use online communities. Our work aims to use natural language processing tools to automatically provide suggestions to assist volunteer counselors during their practice sessions, especially for novices at scale \cite{hill2007training}.

\paragraph{Research Site} 
Our design of \texttt{CARE} will use the online peer-counseling platform - \emph{7 Cups} as a case study. 7 Cups is an online peer support service, where support seekers with a variety of mental health problems participate in text-based chats with peer counselors who have completed active listening and other therapeutic training. As of July 2024, 7 Cups had over \textbf{429,000 trained peer counselors}, supporting over a million people a month. 
\texttt{CARE} uses the conversational data obtained from 7 Cups to train suggestive models described in sections \ref{sec:data} and \ref{sec:response-generation}. The data used in this study to train our models was provided via a partnership with 7 Cups, following the Health Insurance Portability and Accountability Act (HIPAA)  and confidential data use agreements. 
% In addition, we have taken multiple steps to ensure data confidentiality and user privacy: 
% (1)  The data used has been de-identified and is available under strict confidentiality agreements with 7 Cups.
% (2) \texttt{CARE} is only used in supervised scenarios where support providers interact with support seekers (role-played by researchers). Data generated in this process will never be shared with anyone outside of the context in which it was generated. 
%     % The functionalities that our \texttt{CARE} prototype system provides are always used in supervised scenarios.  % , and are never exposed to the participants of the study without any researcher supervision,  to avoid accidental leakage of the training information through the agents.
% (3) Our user studies are approved by the institutional review board at the researcher's organization. All user studies follow the safety protocol described in Section \ref{sec:safety-protocol}.

\paragraph{Scope of \texttt{CARE}} We design \texttt{CARE} as a prototype system that can be used in \textbf{a training or simulated environment} where counselors can practice their skills in a safe environment with no risk to real people, and receive feedback or assistance. While prior research on chatbots for mental health support focuses on building virtual counselors (replacement) \cite{islam2021mobile, haque2023overview, abd2021perceptions}, we develop a \textbf{scaffolding tool} to aid in training (augmentation) for the following reasons. Firstly, the replacement of humans with AI carries serious ethical concerns \cite{imel2017technology, miner2019key, zohny2024generative, liu2022will, zohny2024generative} and massive risks \cite{blease2023chatgpt, woodnutt2024could} to the support seekers despite potential for scalability. Secondly, \texttt{CARE} allows for human oversight and control. We require counselors to engage with and review the suggested content if they want to use it during the training stage. We envision systems like \texttt{CARE} to be mainly used to train counselors who have already received very abbreviated default training on 7 Cups, \textbf{not} seekers or the vulnerable members who suffer from mental health issues; thus, there is no direct communication between support seekers or patients and \texttt{CARE}. 
% In other words,  counselors have full control and can decide whether and to what extent they would use the suggested content; counselors can easily turn on or turn off this feature. 
% \texttt{CARE} is designed to \emph{assist} counselors in \emph{training environments} by learning to improve their responses.
% We design \texttt{CARE} in a way to empower counselors in practice chats, rather than replacing counselors in realcounseling sessions or posing any interferences to counselor-seeker relations
In its current prototype stage, \texttt{CARE} aims to serve as an invaluable tool in a training or simulated environment, providing counselors with a risk-free interaction environment. Our controlled setting ensures safety and allows for essential real-time human oversight. There are many complexities involved in the transition from a prototype to a real-world application beyond \texttt{CARE}'s current scope, which we will discuss in depth in Section~\ref{sec:limit-future}.  

\paragraph{Overview of \texttt{CARE}}  \texttt{CARE} is an interactive tool that works in a private text-based session consisting of two speakers: a support seeker and a peer counselor.
\texttt{CARE} aids users by recommending up to three possible responses to counselors in a given situation. These suggestions are built upon psychotherapy literature on Motivational Interviewing \cite{MITI_4, shah2022modeling}, which is a client-centered counseling style for eliciting positive behavior change by helping support seekers to explore and resolve ambivalence. 
We design the tool interface to be the same as the UI of 7 Cups \cite{7cups} (figure \ref{fig:front_end}) to simulate real-life user studies and obtain genuine user feedback. \texttt{CARE} comprises a backend system with suggestion-generation modules and a front-end user interface. They together form a full stack system. Note that our backend system is designed to be independent of the frontend client for generalizable deployment. The same backend system can be connected to other front-end clients in a plug-and-play fashion.

\subsection{Design Considerations}\label{sec:design-process}
We start the design of \texttt{CARE} by engaging with peer counselors on 7 Cups. Our main considerations for the design of \texttt{CARE} are summarized as follows, and we discuss the limitation of our design process in Section~\ref{sec:limit-future}. 
% \raj{@shang-ling can we get rid of the understanding the setting paragraph? I am unsure about what we are trying to say here that we do not say in other places of the paper. Are we saying we do a field study? I only think we did (2). (1) is already expected in papers.}

% \raj{I did a pass on the three paragraphs below - please review}
\paragraph{Design and Evaluation Choices}
We concurrently consider the design and evaluation of the tool while building \texttt{CARE}. We evaluate \texttt{CARE} by recruiting peer counselors from 7 Cups as participants. To achieve this, we designed the trainees' experience with \texttt{CARE} to closely mirror the look and feel of 7 Cups sessions, minimizing the need for them to learn a new user interface. This ensures that the study focuses solely on changes in the user behavior due to the addition of AI suggestions, rather than the impact of a new user interface. To achieve these objectives, we aim to integrate \texttt{CARE} into the existing 7 Cups interface rather than designing the user interface from scratch. This approach % results in limited design choices, leading us to 
leads us to 
design \texttt{CARE} as an add-on panel that extends the 7 Cups chat interface. We discuss the interface design in Section~\ref{simulating_7cups} and suggestions for future work in Section~\ref{sec:limit-future}.

% \raj{Alternate design and evaluation choices below}
% \paragraph{Design Choices} We design \texttt{CARE} to simulate 

\paragraph{Challenges in existing training methods}
% \raj{What does authenticity mean here?}

In the typical educational process of a peer counselor on 7 Cups, a supporter seeking advice on a 1-1 chat shares an anonymized summary of a challenging situation in a chatroom or with a more experienced counselor, keeping the session confidential. Other counselors then provide feedback on handling the situation. While this promotes peer learning, it has some drawbacks: (1) It might be time-consuming for experienced counselors to give feedback, and sometimes there might be a mismatch between the expertise of a specific experienced counselor and the seeker's . (2) Trainees must wait for an experienced counselor to be available, which often delays the process of receiving feedback. Such issues around the efficiency of training and the quality of feedback leads us to design \texttt{CARE} in its current way. 
% we trade off the authentic suggestion-seeking user flow for a built-in collapsible panel that appears in the mock chat interface.
% \raj{@shang ling can you add a citation for authentic summarization for suggestion user flow?, I do not really understand it.}

% \paragraph{Amount of Information and Readability}
% To optimize learning and usability for trainees, information should be managed carefully. Following prior research in response suggestion applications \cite{kannan2016smart,hohenstein2018ai}, we (1) place the suggestion panel between prior messages and the input text box in the user interface and (2) limit the number of suggestions to three, considering the interface length and flexibility for counselors. Additionally, it is crucial to incorporate strategies that not only assist in training but also offer potential control over the learning process. Future work should explore the integration of these strategies, leveraging their demonstrated effectiveness in various contexts \cite{yang2024makes}. Combining effective information presentation with tailored strategies can significantly enhance the training process, providing a structured yet flexible learning environment for peer counselors. \raj{I suggest commenting out this paragraph for the following reasons - 1. it does not describe design limitations like the other two sections}

\subsection{Motivational Interviewing Strategies}\label{MI}
\renewcommand{\arraystretch}{1.3}
\begin{table*}
\small
  \caption{\textbf{Selected Motivational Interviewing strategies, their number of annotations from \cite{shah2022modeling}, and results of the most suitable counseling strategy prediction.
  } We use the human-annotated 7 Cups dataset from \citet{shah2022modeling}, to follow the distribution of counseling strategies on 7 Cups, and select all 8 \textit{MI-consistent} strategies from the 10 most frequent strategies on 7 Cups. The Motivational Interviewing Strategy, Description, and \# Instances are distilled from Table 1 and 2 of \cite{shah2022modeling}; \underline{Acc}uracy and \underline{F1} Score are the evaluation results of the Strategy Suggestion models trained on 7C-HQ, reported on 7C-MI in Section \ref{sec:strategy-suggestion}.
  }
  \label{tab:auto-label}
  \resizebox{\textwidth}{!}{
  \begin{tabular}{p{29mm} p{76mm} r c c}
    \toprule
Strategy & Description & \multicolumn{1}{l}{\begin{tabular}[l]{@{}l@{}}\# Instances \end{tabular}} &Acc. &F1\\
    \midrule
Open Questions & Open-ended questions that leave room for a response. & 2,507 (15.29\%) &0.632 & 0.686\\[5pt]%10.05\% (2,055,176)\\

Closed Questions & Questions with short specific answers.	& 1,954 (11.91\%) &0.612 &0.672\\[5pt]%10.68\% (2,183,629)\\

Persuade with Permission & Counselor explicitly tries to change members' opinions, attitudes, or behavior based on logical arguments and facts. The counselor asks for permission first or emphasizes collaboration.	& 1,918 (11.69\%) &0.692 &0.719\\[20pt]%9.32\% (1,905,013) \\

Reflection 	& Counselor captures the implicit meaning and feelings of client statements and returns it to the client through rephrases. &1,697 (10.35\%) &0.648 &0.695\\[10pt]%9.08\% (1,857,540)\\

Support & Sympathetic, compassionate, or understanding comments encouraging client behavior. & 1,493 (9.10\%) &0.672 &0.705\\[10pt]%8.51\% (1,739,270)\\

Introduction or Greeting & Counselor and seeker greet each other, exchange names etc. & 1,260 (7.68\%) &0.867 &0.874\\[10pt]%4.18\% (854,448)\\

Grounding & Counselor facilitates conversation through acknowledgments. & 1,027 (6.26\%) &0.721 &0.745\\[5pt]%7.06\% ( 1,442,678)\\

Affirm & Counselor compliments the seeker. & 346 (2.11\%) &0.722 &0.747\\[5pt]%1.64\% (334,715)\\

  \bottomrule
\end{tabular}
}
\end{table*}

Motivational Interviewing (MI) Strategies are a set of client-centric techniques used by therapists for client betterment \cite{MITI_4}. The underlying principle of MI techniques focuses on a therapeutic alliance between a patient and a therapist with an emphasis on patient autonomy. These techniques are suitable for online peer counseling platforms given the established effectiveness in reducing drug and alcohol abuse \cite{mi_abuse_1, mi_abuse_2, mi_abuse_3, effective_MI_1}, decrease in smoking \cite{mi_smoking_1}, and reduction of sexual risk behaviors \cite{mi_beh_se_1, mi_beh_se_2}. 
We used the corpus released by prior work on 7 Cups \cite{shah2022modeling}, which identifies and annotates 17 categories of counseling techniques on 14,797 utterances from 7 Cups chats (referred to as \emph{7C-MI} hereafter).
In this work, we selected a set of MI strategies mainly based on whether there are enough annotations from previous work to build reasonably accurate machine-learning models. % ;
% and (2) these MI strategies are shown to promote positive behavior change in clients \cite{magill_MI_causal_model}.
This led us to 8 strategies, as shown in Table \ref{tab:auto-label}. 
Future work can easily annotate other types of counseling strategies or employ other psychotherapy frameworks to expand this set. 

\subsubsection*{Data Preparation}\label{sec:data}

\begin{comment}
\begin{table}[t]\centering
\small
    \caption{\textbf{Datasets used in this study.} We obtained permission and release from \cite{shah2022modeling} for \textbf{7C-MI}.
    With the MI strategy classifiers from \cite{shah2022modeling}, we labeled \textbf{7C-HQ}.
    Finally, we ran the model selection experiments on \textbf{7C-HQ-small}, a 1-month subset. 
    To evaluate \texttt{CARE}, we trained and fine-tuned models on \textbf{7C-HQ} and reported results by evaluating them on \textbf{7C-MI}.}\label{tab:datasets}
    \begin{tabular}{lllrl}\toprule
    Abbreviation &Description &MI Strategy Annotations &\# Utterances &Usage \\\midrule
    7C-MI &From \cite{shah2022modeling} &Annotated by humans &14,797 &Evaluate models \\
    7C-HQ &Highly rated &Labeled by classifiers from \cite{shah2022modeling} &20,445,517 &Train models \\
    % 7C-HQ-small &Highly rated; 1 month &No &344,335 &Select models \\
    \bottomrule
    \end{tabular}
\end{table}
\end{comment}

To find out which MI strategies are suitable in a context or which example response might be appropriate, we can not ask annotators on crowdsourcing platforms to simply perform the annotation, given the sensitive nature of mental health data and data use agreement with 7 Cups. Furthermore, labeling which MI strategies are suitable or producing example responses in a given context by a third party requires a significant amount of effort and expertise. 
Thus, we make a strong assumption that if a conversation is of high quality as indicated by their satisfaction ratings from support seekers, then these MI strategies used by support providers in their utterances are appropriate. To this end, we collected all conversations in January 2021 that received a rating of 5 on a 5-point Likert scale from support seekers by working with 7 Cups, resulting in 20,445,517 utterances.
With written permission from \cite{shah2022modeling}, we use their released strategy classifiers trained on utterances between January 2020 and August 2020 to label these utterances in terms of their MI strategies, to construct our corpus (\emph{7C-HQ}) for predicting the suitable response a support provider should use in response to a message from a support seeker. 
% The information about the datasets is presented in Table \ref{tab:datasets}. 
We randomly split the dataset into training set (\%80), development set (\%10), and test set (\%10). Given the sensitive nature of the domain and the confidentiality of data, 
we will only release the datasets used in the study with researchers who have obtained data use agreements from 7 Cups.
%   train/dev/test = 8:1:1 throughout this study for model development.
% \newline
% \noindent\textit{Data release:} Due to the sensitive nature of the data and the data use agreement, we cannot release the datasets (7C-MI, 7C-HQ) used in this study.
% We use these high-quality conversations to serve as the candidate groundtruth for these most suitable responses. 
% we assign MI strategy labels to our data with their released strategy classifiers.
% The purpose of assigning labels to our data instead of using their annotations directly is to obtain a greater number of MI strategy annotations on highly rated conversations to build our tool. 
% For each strategy, we use the binary classifier of that strategy to assign labels to 20,445,517 utterances from 7 Cups conversations that receive a rating of 5 on a 5-point Likert scale from support seekers (\emph{7C-HQ} hereafter). 
% Lastly, we train and develop on 7C-HQ, setting aside 7C-MI as the ground-truth test set. We also created a 7C-HQ-small, a 1-month subset of high-quality conversations for response generation model selection (Section \ref{sec:response-generation}), which are explorative, computationally intensive experiments. \diyi{i am confused about this 7C-HQ-Small. need to revisit it after i finish the below sections}

\subsection{Generating Suggestions}
\label{Generating_Responses}
\texttt{CARE} suggests counseling strategies and potential responses for a conversation in real-time by using a multi-step process described as follows (Figure \ref{fig:architecture}):
\begin{itemize}
    \item [\textbf{Step 1.}] For each of the 8 counseling strategies described in Table \ref{tab:auto-label}, \texttt{CARE} predicts the probability of which strategy should be used in the next peer counselor' response using the 5 latest utterances in the chat and a strategy-specific fine-tuned machine learning classifier. %  Large Language Model (BERT).
    \item [\textbf{Step 2.}] \texttt{CARE} generates a potential counselor response for each of the top three most probable counseling strategies. This generation is conditioned with the 5 latest utterances in the chat and uses a DialoGPT model fine-tuned with one of the 8 counseling strategies.
    \item [\textbf{Step 3.}] \texttt{CARE} ensures another layer of support seeker safety by filtering all potentially inappropriate generations using a \textit{finetuned} HateBERT \cite{caselli2021hatebert}.
    \item [\textbf{Step 4.}] The de-duplicated responses are rendered on the front end in descending order of the probabilities from Step 1.
\end{itemize}

We now describe in detail the machine learning models involved in Step 1, Step 2, and Step 3. 
\subsubsection{Step 1: Predicting the Most Suitable Counseling Strategy}\label{sec:strategy-suggestion}
% Prior research shows that large pretrained language models ~ \cite{BERT,roberta} demonstrate state-of-the-art performances on a large set of language-based downstream classification tasks ~\cite{glue}.  Thus, 
We use one of the most popular pre-trained language models---Bidirectional Encoder Representations from Transformers (BERT)~\cite{BERT}, to predict the appropriate counseling strategy that a counselor should use in response to the utterance from the support seeker. 
We do not use closed-source API-based models like ChatGPT, given that such uses might raise concerns over data security and confidentiality and violate our data use agreement. Moreover, supervised fine-tuned models (like BERT) show state-of-the-art performance compared to in-context learning methods for most classification tasks in social domains \citep{ziems2023can}.
Mathematically, we frame it as a multi-label classification task: given the current conversation between a support seeker and a peer counselor, and a set of possible counseling strategies, identify a set of strategies that are most suitable for the given context.   Practically, we first use our automated models of identifying MI strategies mentioned in Section \ref{MI}, to identify (i.e., label) the strategies used in counselors' responses across all conversations. Then, we convert this newly labeled data into a classification task of predicting the strategies in a counselor's response given a seeker's message (context).
% Then, we treat pairs of seeker's messages and counselors' responses identified as utilizing a certain set of strategies as positive instances of those strategies and that are \emph{not} labeled with them as negative instances.

During the implementation stage, we transform this multi-label classification task into eight binary classification tasks, following the binary relevance paradigm for multi-label classification. We used the most recent five messages as the context for the current conversation, as suggested by prior work \cite{shah2022modeling}. 
To deal with any class imbalance issue, we used downsampling by sampling the negative class size to be the same as the positive automatic annotations. 
The final classifiers are built by fine-tuning a pre-trained \texttt{BERT-base-uncased} checkpoint from Hugging Face \cite{HuggingFace} on 7C-HQ. 
The results are reported on 7C-MI, the human-annotated instances from \cite{shah2022modeling} as the test set. We show the number of instances, model accuracy, and the F1 scores in Table \ref{tab:auto-label}. 
We found that our machine learning classifiers performed reasonably well, with an overall F1 score larger than 0.705. 
Given these reasonable predictions of the most appropriate counseling strategies, we then can move forward with the detailed feedback generation.

\subsubsection{Step 2: Generating Example Responses}\label{sec:response-generation}
\begin{table*}
  \caption{% Comparison of different models conditioned on predicted strategies on 7C-HQ-small, and 
  Results of strategy-conditioned next-utterance generation. \emph{Adherence} indicates agreement of predictors by Shah et al.~\cite{shah2022modeling} on the strategy of the generated response.  These BLEU scores are competitive to that of the related work \cite{EmpatheticDialogues, sharma2021towards, DialoGPT}. % , and the differences in numbers are statistically non-significant (t-test, p > 0.1). 
  }
  \label{tab:generator}
  \resizebox{\textwidth}{!}{%
  \begin{tabular}{lcccccccc}
    \toprule
   % \multicolumn{1}{l}{Models} &  Total \# Instances &Avg \# Words&ROUGE-1&ROUGE-2&ROUGE-L&BERT Score&BLEU&Positive\\ \midrule
   % \multicolumn{1}{l}{BART} & 14,797 & 7.710 & 0.128 & \textbf{0.017} & \textbf{0.117} & 0.845 & 0.074 & NA \\ 
   %     \multicolumn{1}{l}{GPT-2} & 14,797 & 9.728 & 0.111 & 0.013 & 0.100 & \textbf{0.879} & \textbf{0.084} & NA\\
    %    \multicolumn{1}{l}{DialoGPT}  & 14,797& 8.871 & \textbf{0.132}	&0.012	&0.114	&0.878	&\textbf{0.084} & NA\\ 
   % \bottomrule
    Strategy& \# Instances&Avg \# Words&ROUGE-1&ROUGE-2&ROUGE-L&BERT Score&BLEU&Adherence\\
    \midrule
Open Questions&	1,853&	7.991&	0.164	&0.034 &	0.155&	0.876&	0.188&	0.930\\

Closed Questions&	1,785&	10.754 &	0.138 &	0.020 &	0.124 &	0.870&	0.185&	0.880\\

Persuade&	1,883&	13.809&	0.111&	0.010&	0.093&	0.861&	0.177&	0.680\\

Reflection &	1,638&	13.417 &	0.107 &	0.010 &	0.090 &	0.863&	0.167&	0.642\\

Support &	1,364&	11.042 &	0.172 &	0.037 &	0.156 &	0.872&	0.167&	0.809\\

Introduction &	214 & 	4.734 &	0.248	& 0.092& 0.244 & 0.879&	0.191&	0.925\\

Grounding &	960&	1.984 &	0.143	& 0.035 &	0.143	& 0.883&	0.098&	0.948\\

Affirm & 324 &	12.320 &	0.172 &	0.040&	0.155 &	0.872 &	0.183 &	0.766\\

    % \midrule
% Overall & 10,021& 10.442 & 0.141 & 0.025 & 0.128 & 0.870& 0.180 & 0.781\\

  \bottomrule
\end{tabular}
}
\end{table*}
Supporters, especially beginners, might not be able to identify how to respond to a given situation \cite{naslund2019digital}, or they may not feel confident about providing an ideal response.
To this end,  we introduce the task of contextualized suggestion generation with humans in the loop. 
The goal here is % not to replace counselors but 
to \textbf{augment} them when they experience high uncertainty in conversation, as well as to provide help to these supporters in formulating their responses.
We formulate this task as a conditional text generation problem: given an input (a seekers' post and the context) and a suitable strategy, we generate a statement that provides a sample response so that volunteer counselors can review and use---or edit and then use---in a given session.
Similar to the setup in Section \ref{sec:data}, we identified a set of pairs (post, response) from high-quality conversations filtered by satisfaction ratings as our corpus. 

Auto-regressive models like GPT-2~\cite{GPT2} and BART \cite{BART} have shown great performance on generation tasks. Here, we choose a variant of GPT-2, DialoGPT~\cite{DialoGPT}, trained on multi-turn dialog conversations, as its domain shares a considerable quantity of similarities with the counseling setting and is open-sourced thus supporting the fine-tuning on our dataset.  
Specifically, our model generates responses conditioned on (1) the counseling strategy, and (2) the most recent five utterances in the conversation. 
We do this by appending the predicted MI strategy token to the dialog history at both training and inference time. 
By experimenting with different models, we found that 
% As shown in Tables \ref{tab:generator}, overall, 
DialoGPT\footnote{Table \ref{tab:best_generator} in Appendix provides a more detailed comparison in different settings for different models.} outperforms the GPT-2 and BART in terms of our automatic evaluation metrics~\cite{ROUGE, BERTScore, BLEU}, and thus use it to 
% We then use this best-performing model DialoGPT to 
generate the example response when conditioning on a specific counseling strategy. 
% The lower half of 
Table \ref{tab:generator} presents the performances of our generation models, with an overall semantic similarity score between a generated text and the ground truth text larger than 0.87 (indicated by BERTScore).
The column of \emph{Adherence}, which refers to whether the generated text exhibits the type of counseling strategy that it is supposed to have, demonstrates relatively good agreement with an overall score of 78\%. 
% Table \ref{tab:generator} also suggests that 
We also find that 
the quality of generation differs in strategies, with \emph{Grounding}, \emph{Open Questions}, and \emph{Introduction} being the strategies with the highest Adherence rates, and \emph{Reflection}, \emph{Persuade}, and \emph{Affirm} given the lowest scores.

\subsubsection{Step 3: Filtering Out Inappropriate Responses}
As an extra layer of safety protection, \texttt{CARE} filters out inappropriate and undesired responses before presenting them to peer counselors. A response is defined as \emph{inappropriate} if it consists of abusive behavior, the inquiry of unnecessary support seeker personal information/ personal identifiers, or contains unnecessary and excessive swear words. 
To avoid outputting inappropriate example responses, \texttt{CARE} employs the \textit{inappropriate} classifier \cite{shah2022modeling}, which is a HateBERT model \cite{caselli2021hatebert} \textit{finetuned} on human annotations of inappropriate messages on 7 Cups.
The classifier is recall-heavy to ensure that \texttt{CARE} refrains from suggesting possibly inappropriate
responses. It gives an F1 score of 84.21\% on the human-annotated test set of 7C-MI.

\begin{figure}
  \centering
  \includegraphics[width=\linewidth]{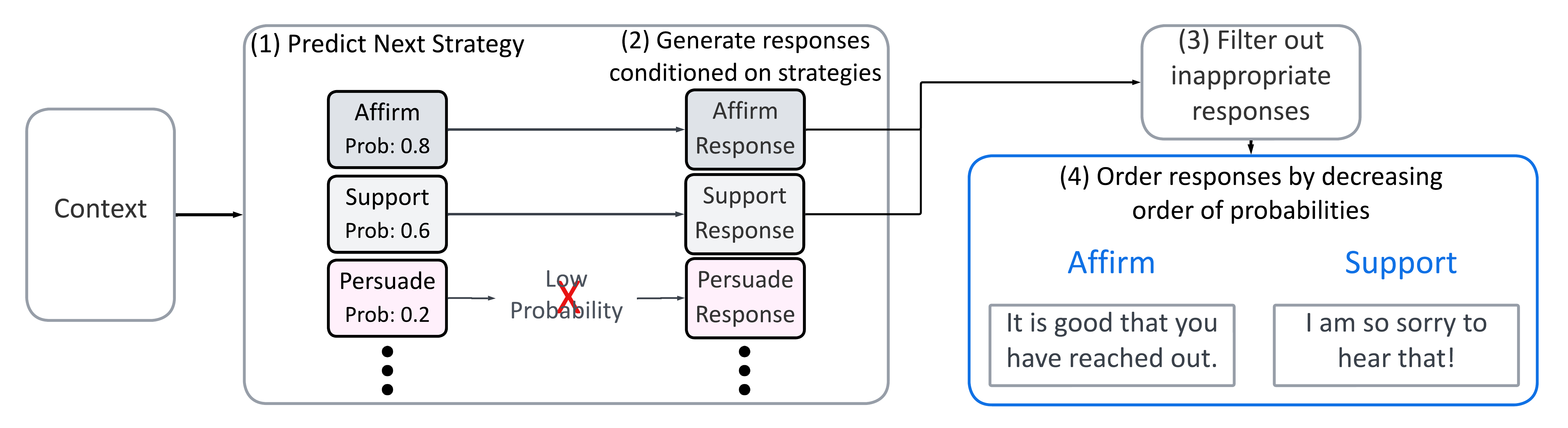}
  \caption{ \texttt{CARE} system architecture. \texttt{CARE} consists of 8 next-strategy predictors, one next-response generator, and one inappropriate response classifier. It predicts the probability distribution of the next counseling strategies and generates suggested responses for each confident strategy independently. In the end, the list of strategies and responses is filtered and ordered decreasingly by the confidence of the predictors.}

  \Description[\texttt{CARE} system architecture.]{\texttt{CARE} consists of 8 next-strategy predictors, one next-response generator, and one inappropriate response classifier. It predicts the probability distribution of the next counseling strategies and generates suggested responses for each confident strategy independently. In the end, the list of strategies and responses is filtered and ordered decreasingly by the confidence of the predictors.}
  \label{fig:architecture}
\end{figure}

\subsection{System Development}
Building upon the aforementioned steps, we design an interactive system \texttt{CARE} to empower counselors during their chats with seekers. 
\texttt{CARE} integrates front-end and back-end components to optimize the user experience and runtime efficiency\footnote{Details are provided in Appendix \ref{sec:sys-details}.}. The Server-side event handlers are flexible to connect with any type of front-end client. This allows \texttt{CARE} to be used with any 7 Cups-like online mental health platform and provides training to peer counselors in a platform-agnostic manner.
A client-side API reference is used to facilitate the processing of function calls to the back-end API. This enables functionalities such as the initialization of new chat sessions, the addition of new messages to the chat log, and the clearing of sessions. With this architecture, the chat system can return suggestions and generations in real-time. We ensure that the client-side API is lightweight for quick loading of the system on the client end. 

\subsubsection{Design of the System Front-End}
\label{simulating_7cups}
\begin{figure}
  \centering
  \includegraphics[width=\linewidth]{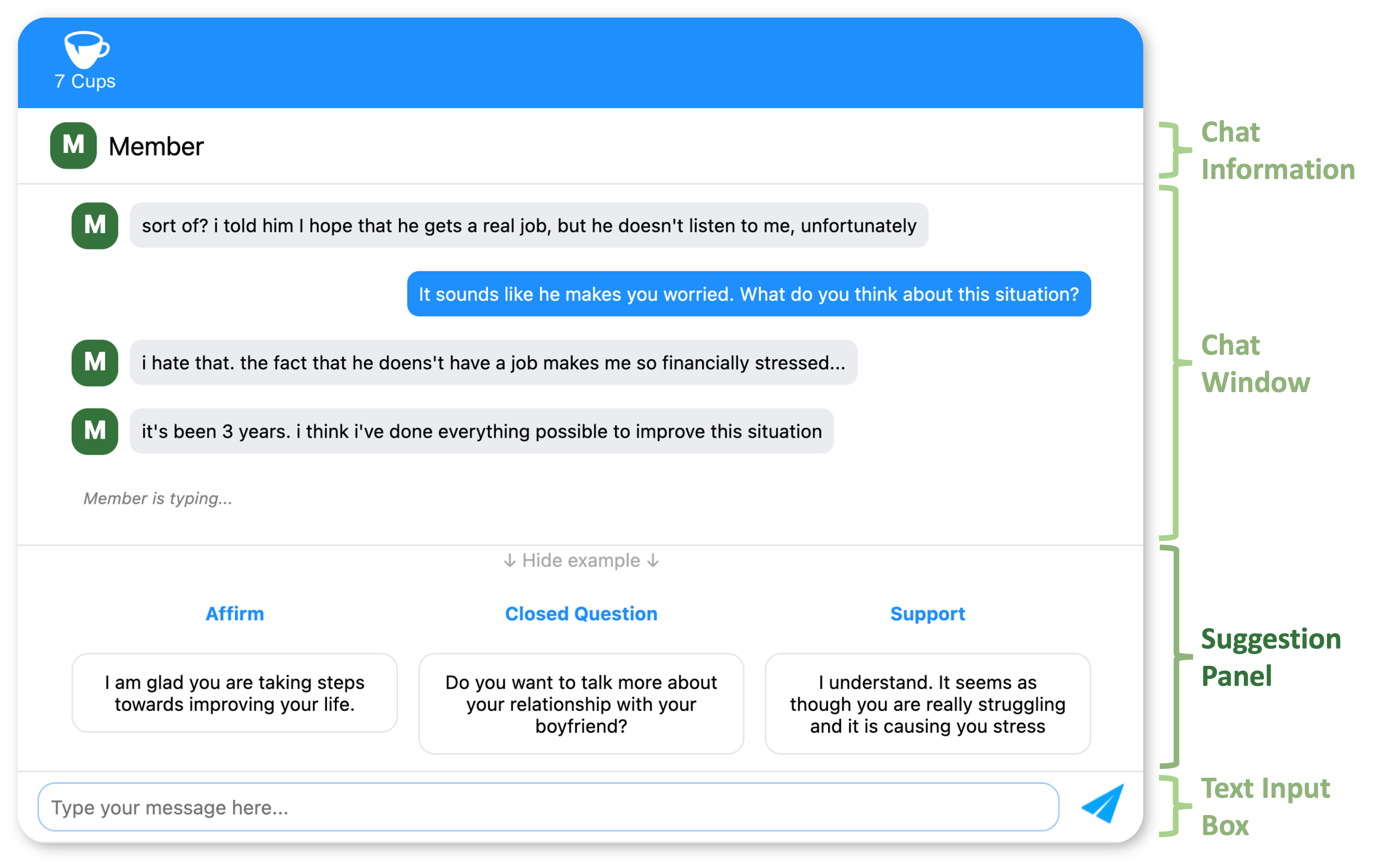}
  \caption{The peer counselors' frontend view of the tool. We design \texttt{CARE} as a \textbf{Suggestion Panel}. The other components (Chat Information, Chat Window, and Text Input Box) mimic the frontend view of 7 Cups.
  \texttt{CARE} is comprised of three elements: 1) \textbf{The blue words}: Suggested MI counseling strategy to be used in the next peer counselor response. Users can hover over the strategy to see its 1-sentence description. 2) \textbf{Black sentences in boxes}: A generated next peer counselor response that utilized the MI strategy above it. Users can click the response to auto-fill the input text box for modification before sending. 3) \textbf{Hide/Show example button}: Toggle for collapsing and expanding Suggestion Panel.
  }

   \Description[The peer counselors' frontend view of the tool.]{CARE's functionality comes from a suggestion panel. The other components include Chat Information, Chat Window, and Text Input Box and mimic the 7 Cups front-end. \texttt{CARE} is comprised of three elements: 1) The blue words: Suggested MI counseling strategy to be used in the next peer counselor response. Users can hover over the strategy to see its 1-sentence description. 2) Black sentences in boxes: A generated next peer counselor response that utilized the MI strategy above it. Users can click the response to auto-fill the input text box for modification before sending. 3) Hide/Show example button: Toggle for collapsing and expanding Suggestion Panel.}

  \label{fig:front_end}
\end{figure}
The front end has two views, one for the support seeker and one for the peer counselor. The view for the peer counselor is given in Figure \ref{fig:front_end}. The UI contains the four major components to replicate a chat environment like 7 Cups. These components are the chat information toolbar on the top, the chat window, the collapsible panel for suggestions and responses generated by \texttt{CARE}, and the text input box. The support seeker view of the platform is the same as the peer counselor view except for the absence of the collapsible panel.
Verbal feedback on the interface was obtained by working with counselors from 7 Cups to simulate the look and feel of the platform. For instance, based on insights from domain users, we have introduced several features: % Using the feedback, we added the following features, 
(1) an indicator of ``\emph{a user is typing}'', and (2) the option for users to hide the tool window. When a peer counselor clicks on one of the potential responses in the tool, the text input box gets populated with the response. The peer counselor can then choose to modify the response before sending it.
\subsubsection{Logging of Peer Counselor Actions}
% With the prior written consent of the participants, 
We log\footnote{We log all data with written user consent.} peer counselor actions on the suggested responses to understand the impact of our tool in chats. For each utterance, we log the \texttt{CARE} suggestions, the peer counselor's click (choice) in the suggested responses, the peer counselor's click to show and hide the collapsible panel, and the final peer counselor response. We use these logs for our subsequent quantitative evaluation.

\section{Evaluation of \texttt{CARE} via User Studies}
Evaluating the proposed system \texttt{CARE} requires rigorous examination and ethical consideration based on its sensitive and high-stakes cases. We conduct user studies to ensure that our tool can be used widely in practice by working closely with peer counselors from 7 Cups---the end-users of\texttt{CARE}. 
% Given the potential harm of AI exposure to vulnerable individuals, we do not engage with support seekers in this study.

% Note that although \texttt{CARE} is designed to be used for training in mock chats only, we also collect participants' opinions on the possibility of using \texttt{CARE}-like systems in real-time chats to understand its potential and risks more comprehensively, as an additional analysis.
\begin{table}[t]\centering
\caption{\textbf{Participant Information}. \textbf{ID} is the ID of the participant in this study. \textbf{Tenure} denotes the duration a participant has been a peer counselor on 7 Cups by the date they complete their user study session. These quantities are rounded to the closest integer in the largest non-zero unit. \textbf{\texttt{CARE} Chat} marks whether the participant sees \texttt{CARE}'s AI assistance in the first (1) or the second (2) chat in our user study session, which is randomly assigned. \textbf{Category} is the mock chat situation chosen by a participant from \textit{\textbf{A}nxiety} and \textit{\textbf{R}elationship Stress}. Information about the \textbf{Device(s) Used for 7 Cups} is collected because 7 Cups provides UIs for computers, tablets, and cellphones, but for consistency, we conduct our user studies using computers only. Half-and-half for this column indicates participants use computers and mobile devices in a Half-and-Half duration fashion. \textbf{Most Recent Background} denotes the participant's most recent, external, relevant academic or professional background. % Given the variety of responses and the sample size, we include all essential participant information here but do not analyze the relations between these variables and the participants' perceptions of \texttt{CARE}.
}
\label{tab:participants}
% \small
\resizebox{0.88\textwidth}{!}{
\begin{tabular}{p{5mm} r c cll}\toprule
ID &Tenure &\texttt{CARE} Chat &Category &Device(s) Used for 7 Cups &Most Recent Background \\\midrule
P1 &3 months &2 &R &Mostly computers &Degree in Sociology \\
P2 &5 years &1 &A &Always computers &No \\
P3 &3 years &2 &A &Mostly computers &Degree in Psychology \\
P4 &7 years &2 &R &Always computers &Degree in Counseling \\
P5 &8 years &2 &R &Half-and-half &Unknown \\
P6 &4 years &1 &A &Always cellphones/tablets &Student in Clinical Social Work \\
P7 &2 years &1 &R &Mostly computers &No \\
P8 &4 years &1 &R &Mostly computers &Student in Psychology \\
P9 &8 years &2 &A &Half-and-half &Crisis Counselor \\
P10 &3 years &2 &A &Half-and-half &No \\
P11 &1 year &2 &A &Mostly computers &Student in Psychology \\
P12 &5 days &1 &A &Always cellphones/tablets &No \\
P13 &8 years &1 &R &Always computers &Unknown \\
P14 &1 month &2 &R &Always cellphones/tablets &No \\
P15 &3 months &1 &R &Half-and-half &No \\
\bottomrule
\end{tabular}
}
\end{table}

\subsection{Recruitment of Participants}
% This study has been approved by the Institutional Review Board (IRB) at the authors' institution. %(Protocol H22041). 
We recruit peer counselors who have extensive experience with the platform and its expectations, as well as first-hand knowledge of the learning process. Our recruitment criteria required participants to have completed at least 25 conversations on 7 Cups to ensure familiarity with the platform, and to have a peer counselor rating of 4 or higher (out of 5) to ensure they have a moderate understanding of the expectations for effective peer counseling and can assess appropriate responses in various situations.
Additionally, we excluded any peer counselor who had been blocked by a support seeker in the past year. This set of inclusion and exclusion criteria positions them in a good position to effectively evaluate our training tool. Besides these criteria, we did not introduce any other exclusion criteria in order to obtain a natural sample of peer counselors. By collaborating with community moderators and platform owners on 7 Cups, we recruited 15 peer counselors from the United States on a first-come, first-served basis (11 females, 3 males, 1 unknown, ages 18-69).\footnote{We discuss the imbalance of this participants pool in Section~\ref{sec:representives} and \ref{sec:limit-future}.}
All of the participants were recruited via notifications on the 7 Cups mobile/web app and were compensated \$20 for their time (the average time duration of the study was a little under 1 hour). 
% Our recruitment strategies resulted in 15 virtually conducted participant studies. 
The experience level of these participants ranged from 5 days to 8 years on the platform at the time of their interview. We present their non-identifiable information in Table~\ref{tab:participants}.

\subsection{User Study Details}
We test how \texttt{CARE} empowers volunteer counselors by conducting user studies with two conditions: (1) no support, and (2) support provision with \texttt{CARE}. We now describe in detail the different components involved in this user study. % Note that since \texttt{CARE} is designed to assist \emph{peer counselors} in training, the user study and analysis do not involve support seekers.

\subsubsection{Procedure and Method}
For each user study session, based on their familiarity and comfort, a participant (peer counselor) first selects a chat category between \emph{anxiety} and \emph{relationship stress} (the two most frequent topics on 7Cups \cite{wang2023metrics}). The participant then tests the system by interacting with a researcher who acts as a support seeker. Both of the systems mimic the chat interface of 7 Cups as described in Section \ref{simulating_7cups}. The order of the chats is randomly assigned using a computer randomizer: one with \texttt{CARE}, and the other without \texttt{CARE}. 

\subsubsection{Onboarding Participants with \texttt{CARE}}
To ensure that each participant is familiar with all the features of \texttt{CARE}, we briefly introduce the tool before they experiment with the system. The participants are also asked to watch a tutorial video emphasizing the voluntary use of \texttt{CARE} and explicitly teaching how to hide and show \texttt{CARE}'s AI assistance (video transcript in Appendix Section~\ref{transcript}). The video also describes the different (8) counseling strategies along with their examples. The researchers also clarify any doubts and confusions a participant may have about the study. 
During the chat, when \texttt{CARE} shows AI assistance for the first time, the researchers verbally instruct the participants again that they may show/hide the tool, ignore/modify generated responses, or use it as they prefer.

\subsubsection{Conducting Simulated Chats}\label{sec:chats}
We select a larger set of representative seeker-supporter conversations from 7 Cups, remove any personal identifiers in these conversations, and manually paraphrase and fuse them, to obtain a smaller set of representative chats. 
% All researchers involved in this process have an active responsible code of conduct in research and have confidentiality data use agreements to preserve the privacy of 7 Cups users. 
In total, we devise four representative scenarios, two for each category and their corresponding scripts. In each simulated chat, the researchers act as support seekers by sending messages modified from a pre-defined script. 
In detail, for our focused two categories: relationship stress and anxiety, the research team read roughly 30+ conversations from actual chats in 7 Cups. Researchers involved in the user study in our team have signed up as volunteer counselors and support seekers and conducted observations with more than 100 cumulative hours of counseling time on the platform. This enables the researchers to be well-equipped to interact with participants while conducting the studies. 
While the prepared scripts help maintain a coherent theme, we also allow for variability in the conversation due to the variability in the peer counselor's responses. The scenarios are summarized in Table \ref{tab:mock-scenarios}, and the scripts are provided in Section~\ref{sec:script-mock-chats}.

\subsection{Measure and Evaluation}\label{ques-tool}
We perform quantitative analyses on counselors' use of suggested counseling strategies when they interact with seekers, and qualitative analyses on the participants' feedback on what they like and dislike about each session. 
We understand the participants' perceptions of \texttt{CARE} by administering a post-task, anonymous questionnaire (Table \ref{tab:questionnaire} in Appendix), and also conducting a semi-structured interview. 
The questionnaire does not ask for any personal user identifiers to allow participants to freely disagree/ agree to the use of \texttt{CARE} and leave their unbiased opinions and comments. The questionnaire contains a mixture of Likert scales, checkboxes, and an open-ended response textbox, depending on the nature of the question. The participants complete the questionnaire during the session and may opt to unmute and ask questions whenever the questionnaire is unclear to them. 
In the semi-structured interview, we prepare a list of questions\footnote{Provided in Section~\ref{sec:script-interview}.} and ask follow-up questions upon detecting a potential theme. 
% R2: How was the data handled and processed? How was the data analysed and knowledge created?
Afterward, the research team inductively develops a codebook and codes all interview transcriptions and questionnaire responses to the open-ended questions.

\subsection{Ethical Considerations and Safety Protocols}\label{sec:safety-protocol}

The Institutional Review Board has been approved for this study at the researchers' institution. The data for the study was collected in partnership with 7 Cups, following HIPAA and confidentiality data use agreements. To protect the privacy of the participants, the data has been anonymized and additional steps have been taken to ensure that it cannot be linked to any specific user. At all times, actual user data was present in secure servers, which could be only accessed by the researchers of this study. All the researchers involved in this study have completed CITI Program certifications on responsible code of conduct in research. All the participants in the user study are older than 18 and have signed a consent form stating their explicit consent. 
% The authors acknowledge and address the potential safety issues for support seekers and peer counselors when using such a system for training in the Safety Protocol in Section \ref{sec:safety-protocol} and the Limitations in Section \ref{sec:limit-future}. 

\paragraph{Safety Protocols:} We take careful steps to ensure the safety of our participants. First, we note that \texttt{CARE} is designed for \emph{training} and not for real-time use; thus, supporters are involved in a simulated environment with little risk. We collect feedback and refine \texttt{CARE} in an interactive process based on user insights to make sure it works as expected.
Second, at all times, the participants were free to leave the study (with the guarantee of full payment) if they felt uncomfortable with the topics discussed in the chats.  
Third, our research team was involved in every support provision conversation transcript; for any signs of negative impacts (e.g., negative mood, frustration), an established protocol was to be implemented - the study would be stopped immediately and the participants would then be referred to the consulting clinician and information about available national resources and help-lines was to be shared. None of the participants expressed discontent and seemed uncomfortable in the study.

% \section{Ethical Considerations}
% The Institutional Review Board has been approved for this study at the researchers' institution. The data for the study was collected in partnership with 7 Cups, following HIPAA and confidentiality data use agreements. To protect the privacy of the participants, the data has been anonymized and additional steps have been taken to ensure that it cannot be linked to any specific user. At all times, actual user data was present in secure servers, which could be only accessed by the researchers of this study. All the researchers involved in this study have completed CITI Program certifications on responsible code of conduct in research. All the participants in the user study are older than 18 and have signed a consent form stating their explicit consent. The authors acknowledge and address the potential safety issues for support seekers and peer counselors when using such a system for training in the Safety Protocol in Section \ref{sec:safety-protocol} and the Limitations in Section \ref{sec:limit-future}. 
% In addition to the consent form that disclose risks of this tool to users, the authors will advocate for practitioners to centrally host and log the content generated by our system so that it can be audited to determine whether there are any problematic behaviors in the tool use. 

\begin{table*}
  \caption{Summary of the mock chat scenarios.}
  \label{tab:mock-scenarios}
  % \small
  \resizebox{\textwidth}{!}{
  \begin{tabular}{lcl}
    \toprule
    Category & Order & Scenario Summary \\
    \midrule
    \multirow{2}{8em}{Anxiety} & 1 & A college student is worried about an upcoming admission exam and school-life balance. \\
     & 2 & A person feels lonely and seeks to take their friendship to a deeper level. \\
    \midrule
    \multirow{2}{8em}{Relationship Stress} & 1 & A college freshman struggles with a newly long-distance relationship. \\
     & 2 & A person feels financially stressed after their lover becomes unemployed for 3 years. \\
    \bottomrule
  \end{tabular}
  }
\end{table*}

\section{Results}\label{results}
As described in Section \ref{ques-tool}, our participants fill out an anonymous questionnaire after they complete the first 30 utterances of each of the two system-testing chats. 
There are 30 system testing chats from the 15 user studies, which result in 926 utterances sent in those mock chats, where 470 utterances (50.76\%) are sent by peer counselors, and the remaining 456 utterances (49.24\%) are sent by us, the simulated help seekers. 
% The median span of the total chat time is 15.78 minutes. 
On top of the obtained numerical data from log analysis (Section \ref{sec:log-analysis}) and Survey Responses (Section \ref{sec:survey-responses}), we also conduct semi-structured interviews to understand the impact of \texttt{CARE} (Section \ref{sec:interview-insights}). Detailed results are presented below. % The quantitative and qualitative results of our study are given below.

\subsection{Log Analysis}\label{sec:log-analysis}
We analyze the system logs to understand whether \texttt{CARE} has effects on peer counselors and show that peer counselors \textbf{make use of} \texttt{CARE} when it is provided. We report the aggregated numbers at both, the utterance level and the chat level, to give a sense of the overall statistics and account for the difference between individual preferences.

\subsubsection{\texttt{CARE} provides assistance 84\% of the time.}
\texttt{CARE}'s AI assistance is frequently provided to peer counselors during the experiment chats. \texttt{CARE} starts suggesting strategies and responses when a chat reaches at least 5 utterances. After that, \texttt{CARE} presents potential responses if the confidence score of the next strategy prediction is greater than a threshold of 0.5. Our analysis shows that across all utterances,  \texttt{CARE} gives suggestions for 374 out of 467 utterances (80.09\%) that were sent in chats, which is 84.38\% of the time if we compute the median at the chat level.

\subsubsection{Peer counselors choose to see \texttt{CARE} 93\% of the time.}\label{sec:log-show}
Despite being told that they have the choice to see the tool or not see the tool, most peer counselors almost always keep the \texttt{CARE} panel open during the mock chats. The peer counselors choose to turn on \texttt{CARE} for 340 utterances out of 374 utterances (90.91\%) when \texttt{CARE} is available during the experimental chats, which takes up 93.20\% of the time span on average across chats (std=23.376\%, median=100\%). Notably, 13/15 peer counselors keep the AI assistance panel shown throughout their experimental chat.

\subsubsection{Peer counselors check \texttt{CARE}'s suggestions before sending in around 47\% of responses.}\label{sec:log-click}
We join system logs on clicks and messages together to analyze their usage of \texttt{CARE}. When it is the peer counselors' turn, and they see \texttt{CARE}'s suggestion, they click them before sending their response 75 out of 199 times, making the total click-through rate 37.69\%. The median across the 15 experimental chats with \texttt{CARE} yields a 46.67\% click-through rate.
Appendix Section \ref{sec:pos-exceprts} and \ref{sec:neg-exceprts} summarize the situations where \texttt{CARE} is clicked and not clicked, respectively.

\subsubsection{Peer counselors use a \texttt{CARE}'s response directly or edit it.}
We compare the text similarity between what \texttt{CARE} suggests and what peer counselors finally send out. 
From the utterances peer counselors sent after clicking responses suggested by \texttt{CARE} (click-through rate\footnote{We discuss click-through rate in learning context in Appendix Section~\ref{sec:click-through}.}=37.69\%), we found that \textbf{most (45/75; 60\%) of the responses were sent without modification from \texttt{CARE}'s suggestion},\footnote{We use \textbf{bold font} to highlight text segments that we find particularly interesting.} while the others were altered before being sent.
For the 30/75 (40\%) edited responses, we calculate the length of the Longest Common Subsequence (LCS) between the clicked suggested response and the actual utterance sent by the peer counselor. This calculation yields a median difference of 41.5 characters between the two strings.
On one hand, if we divide the length of each LCS by the length of its corresponding generated response, we get the median 
99.3\% (mean=87.7\%, std=19.6\%). % 99.296\% (mean=87.704\%, std=19.584\%
On the other hand, if each LCS length is instead divided by the length of the corresponding actual response peer counselors sent, the resulting median is 47.7\% (mean=50.8\%, std=20.6\%). % 47.681\% (mean=50.882\%, std=20.639\%). 
These two ratios together advise that \textbf{when peer counselors modify a \texttt{CARE}'s suggestion, they usually build upon it by preserving almost the entire suggestion and adding more content to it.}

\subsubsection{Peer counselors send longer responses with \texttt{CARE}.}
The median length of peer counselors' responses increases with \texttt{CARE} compared with its counterpart without \texttt{CARE}.
Concretely, when peer counselors see \texttt{CARE}'s suggestions,  
the median length of responses (77 characters) is significantly longer than the response length without \texttt{CARE} (61 characters) (N = (199, 271), p < 0.01; Mann-Whitney U rank test).

\subsection{Survey Responses}\label{sec:survey-responses}
% This section summarizes the responses to the questionnaires in terms of Likert chart questions (Section \ref{sec:survey-likert}) and check box questions (Section \ref{sec:survey-aspects} and \ref{sec:survey-helps-more}). % Responses to the open-ended questions are rather free of form and highly relevant to the interview contents, so they are coded with interview transcripts in Section \ref{sec:interview-insights}.

\subsubsection{Overall Perceptions}\label{sec:survey-likert}
\begin{figure}
  \centering
  \begin{subfigure}[t]{0.6\textwidth}
      \includesvg[width=\textwidth]{images/tool-quality.svg}
      \caption{Perceptions of Models and Future Usage}
      \label{fig:model-perception}
  \end{subfigure}
  \hfill
  \begin{subfigure}[t]{0.34\textwidth}
    \includesvg[width=\textwidth]{images/was-the-tool.svg}
    \caption{Perceptions of \texttt{CARE}}
    \label{fig:overall-perception}
  \end{subfigure}
  \caption{The responses to perception questions. a) A bar chart of the participants' answers to Likert scale questions regarding their perception of the suggested counseling strategies (\textit{strategies}), example responses (\textit{examples}), and future usage of \texttt{CARE}. b) A bar chart of participants' overall perception of \texttt{CARE}. The dashed lines mark the medians of responses.}

  \Description[The responses to perception questions.]{Responses to the questions on the user perception of\texttt{CARE}. Most users feel that examples frequently look natural, the strategies occasionally suit the situations and help, and the examples occasionally fit topics and help. Most users agree that they find \texttt{CARE} helpful and will use it.}
  \label{fig:quality}
\end{figure}
As shown in Figure \ref{fig:model-perception}, the responses to the model perception questions show that for most participants, the strategy prediction and suggestion generation of \texttt{CARE} are helpful in terms of whether the counseling strategies suit the situation, and whether the generated responses fit the conversation topic, and help peer counselors. % The figure also shows that most participants feel that the model suggestions look natural for a 7 Cups-like peer counseling platform.
As shown in Figure \ref{fig:overall-perception}, all participants agree that \texttt{CARE} is straightforward to use, while more than half of them think it is overall helpful.
The median response to ``Will you use \texttt{CARE}'' (\textit{If the tool is provided to you, how often do you think you will make use of it?} in the questionnaire) tilts towards positive.
Lastly, responses to the question ``Do the examples contain *harmful* message'' are: Never (4/15), Very Rarely (6/15), Rarely (3/15), Occasionally (1/15), Frequently (1/15), and Very Frequently (0/15). This indicates that in general the example responses \texttt{CARE} suggested are unlikely to cause harm, but more advanced filters are still needed.

\subsubsection{Like \& Dislike}\label{sec:survey-aspects}
\begin{figure}
  \centering
  \begin{subfigure}[t]{0.35\textwidth}
      \includesvg[width=\textwidth]{images/like.svg}  
      \caption{Like Aspects}
  \end{subfigure}
  \hfill
  \begin{subfigure}[t]{0.33\textwidth}
      \includesvg[width=\textwidth]{images/dislike.svg}
      \caption{Dislike Aspects}
  \end{subfigure}
  \hfill
  \begin{subfigure}[t]{0.30\textwidth}
      \includesvg[width=\textwidth]{images/helps-more.svg}
      \caption{Helpful Occasions}\label{fig:helpful}
  \end{subfigure}
  \caption{The responses to multi-select checkbox questions regarding preferences. The dashed lines mark the medians of responses.
  }
  \Description[The responses to multi-select checkbox questions regarding preferences.]{The responses to multi-select checkbox questions regarding preferences. Users like \texttt{CARE} because it reminds them of strategies, inspires responses, and increases confidence. Users dislike \texttt{CARE} because it interferes and disrupts thinking. Users agree that \texttt{CARE} helps more when users are newer, stressed, and underconfident.}
  \label{fig:like-dislike}
\end{figure}

To deeply understand what features or aspects of \texttt{CARE} are more appreciated by peer counselors, we visualized their preference responses 
in Figure \ref{fig:like-dislike}.
We find that more than half of the participants agree that \texttt{CARE} reminds them of counseling strategies (12/15), inspires them to better responses (10/15), and increases their confidence (8/15).
While about half of the participants think \texttt{CARE} can disrupt the thought process during conversations, as discussed in Section \ref{sec:interview-dislikes}, its seamlessness may increase as users become more familiar with it in the training environment.  

\subsubsection{When Is \texttt{CARE} Helpful?}\label{sec:survey-helps-more}
% \slh{Describ
As shown in Figure \ref{fig:helpful}, for the question of when \texttt{CARE} is helpful, we saw from the top three selections to our question ``\texttt{CARE} helps more when...?'' that, 13/15 said YES to \textit{the peer counselor is new to listening chats}, 10/15 mentioned \textit{the peer counselor feels more stressed}, and 10/15 emphasized when \textit{the peer counselor is less confident in the chat's category}. This result suggests the effectiveness of \texttt{CARE} in helping, especially novice peer counselors.
\subsection{In-Depth Interview Results}\label{sec:interview-insights}
% \diyi{did we do the interview, or is this mainly open-ended responses on their questionnaire? }
To provide an in-depth understanding of our findings from the log analyses and questionnaires, this section takes a deep dive into what participants shared with us during their semi-structured interviews. 
\subsubsection{How Peer Counselors Use \texttt{CARE}}
\paragraph{Preference for and against showing \texttt{CARE}}
Participant P7 attributed why they keep \texttt{CARE} open throughout the simulated chat to potential inspirations, which also explains %  This brings a potential explanation for 
the result about the major user preference of showing \texttt{CARE} more often than hiding it in Section \ref{sec:log-show}.
\begin{quote}
    \textit{``I think I would probably \textbf{have it on more often than not}. I'm, just in case it suggested something that I hadn't, that hadn't popped into my mind.
    '' -- P7}
\end{quote}
Alternatively, P6 and P14 prefer to use \texttt{CARE} as a safety net and refer to \texttt{CARE} when there are uncertainties.
\begin{quote}
    \textit{``Only if I'm really on, like, lost on what to say, I don't have any ideas or strategies to go, I would just go there [\texttt{CARE}'s suggestions]. Other than that, I prefer to go with my train of thought without looking at it.'' -- P14}.
\end{quote}

\paragraph{Importance of Optionality}
Although most (13/15) participants chose to show \texttt{CARE} instead of hiding it, giving peer counselors an option to hide the tool is still essential in the training process. Some peer counselors (P3, P4, P6, P10) pointed out the need for this option to switch the tool on/off.
\begin{quote}
    \textit{``You can switch it on and off as needed. If you pick up a chat with a subject you're not as familiar with, if you decide \textbf{you want to put it on, you have that option}. If you don't want to, you don't have to.'' -- P4}
\end{quote}

\paragraph{Adopting a \texttt{CARE} Response}
Section \ref{sec:log-click} shows that peer counselors click \texttt{CARE}'s suggestion before sending in around 47\% of responses.
P7 demystifies this behavior by illustrating their usage of \texttt{CARE}:
\begin{quote}
    \textit{``I might take a quick glance at it [CARE] and see if it's the same thing I was going to type anyway, and then \textbf{click on it} [the suggestion]. That's definitely a \textbf{time saver}.'' -- P7}
\end{quote}

In terms of intentions behind the modification, 
P8 mentioned their rationale for doing so is 
because \textit{``There were some open-ended questions that the tool suggested that I didn't think of, that I thought is really good, and \textbf{sometimes I use exactly what was suggested}.''}.
At other times, P8 modifies the suggestion for \textbf{personalization}: \textit{``The \textbf{personalization} part, I still think that the listener would have to be able to write the words out themselves.''} 
Thus, when P8 wants to make the response conform to their style more, they \textit{``fix up the wording to make it a bit warmer and less cold''} once in a while. One such modification P8 makes is shown in Fig. \ref{fig:chat-5} in Appendix \ref{sec:pos-exceprts}.

\subsubsection{Peer counselors think \texttt{CARE} is straightforward and helpful.}\label{sec:interview-perceptions}
% Peer counselors discuss their perception of \texttt{CARE} from different aspects. 
Participants' interview comments on the straightforwardness and helpfulness of \texttt{CARE} are consistent with the survey response results in Section \ref{sec:survey-likert}.

\paragraph{Straightforwardness of \texttt{CARE}}
When being asked to comment on the UI/UX of \texttt{CARE}, P10 compliments on the UI design as they \textit{``like how the suggestions are right above the chat bar.''}. P12 describes the intuitive nature of \texttt{CARE} as \textit{``Even if you didn't know about it [how to use the tool] before, you can [directly] use this technology here.''}. 

\paragraph{Helpfulness of \texttt{CARE}}
Many participants (P1, P6, P7, P8, P9, P15) also agree that \texttt{CARE} is helpful and also mention the utility of the strategies. Others show satisfaction in the quality of \texttt{CARE}'s suggested responses.
\begin{quote}
    \textit{``I think it could be a \textbf{great tool}. I really do. I think it's something that we need in 7 Cups because we don't really train listeners very much. We, it's very minimal training, so we kind of grab listeners once problems are occurring. So if we could put this at the front end, that would be phenomenal.'' -- P1}
% \end{itemize}
% A few participants mentioned the utility of the strategies \texttt{CARE} suggests:
% \begin{itemize}
%    \item[] \emph{``I think the \textbf{strategies are good}, like, a way to ask a question or to like, like, reassure someone with, you know, the, just like confirming like their feelings and things like that?'' -- P10}
% \end{itemize}
% Others show satisfaction in the quality of \texttt{CARE}'s suggested responses:
% \begin{itemize}
  %   \item[] \emph{``Um, I think the \textbf{language was very natural}. It was like, I was speaking to another human. It flowed.'' -- P12}
\end{quote}

\paragraph{Risks of \texttt{CARE}}\label{sec:risks}
While \texttt{CARE} is perceived rarely harmful (Section \ref{sec:survey-likert}), peer counselors who see \texttt{CARE}'s mistakes think \texttt{CARE} may not actually cause negative consequences \emph{during training}:
\begin{quote}
    \emph{``I did notice just a couple of suggested responses that didn't quite fit, but um, I mean, I think the average listener would know not to click on them. \textbf{I don't think that they would just randomly start clicking.}'' -- P7}
\end{quote}
% Overall, the peer counselors' impression of \texttt{CARE} is positive. In the following Section, we present the specific dimensions where the participants find \texttt{CARE} helpful.

\subsubsection{Interpretation of Likes}\label{sec:interview-likes}
This part extends Section \ref{sec:survey-aspects} from what to why to gauge the reason behind peer counselors' preference for \texttt{CARE}.

\paragraph{Reminds peer counselors of counseling strategies.}
The top questionnaire answer for what peer counselors like about \texttt{CARE} finds that it \textit{reminds peer counselors of counseling strategies} (Section \ref{sec:survey-aspects}). $80\%$ of peer counselors like the fact that \texttt{CARE} reminds them of MI strategies, indicating that the tool's ability to suggest relevant MI strategies is a helpful feature that supplements the short training these counselors receive. 
% Additionally, \texttt{CARE}'s ability to support counselors by reminding them of MI strategies could also improve how peer counselors respond on their own over time. This also gives high-level feedback to peer counselors and trains them in real-time on MI strategies. 
Peer counselors P6 and P12 mentioned in their interview \textbf{how \texttt{CARE} helped them with learning MI strategies}.
\begin{quote}
    \emph{"The tool is using what the member [support seeker] is saying to come up with these prompts, and it's giving me open question options, it's giving me the different active listening skills that you're supposed to use when you're listening. And people often forget that as listeners. They get stuck on asking questions all the time, or just validating all the time, or just reflecting all the time. And I think this gives me the confidence \textbf{to explore the different skills} that you should have as a listener." -- P6}
\end{quote}

\begin{quote}
    \emph{"I think there were a \textbf{lot of good [strategy] suggestions,} and it helped remind me of different coping mechanisms that were appropriate to like, mention in the use with the member [support seeker]. For example, the whole challenging negative thought thing. The suggestions, you know, reminded me of that, because with affirmation, and as, being one tool, for example, it really helped me." -- P12}
\end{quote}

\paragraph{Inspire better responses}
In Section \ref{sec:survey-aspects}, we bring out that 10/15 of the surveyed participants reported that they liked how \texttt{CARE} \textit{inspires better responses}. \texttt{CARE}'s dual-functioning of providing the MI strategy and potential response of what to say is a helpful feature. Even if peer counselors know which MI strategy they want to use, their training and experience may not be extensive enough to adequately apply the strategy to the situation. This is where \texttt{CARE} is able to support counselors by providing examples of using MI in the specific scenarios at hand. Even if counselors do not directly use the suggested utterance, it gives insight into what an alternative/better response might be. Peer counselor P6 mentioned multiple times how \texttt{CARE} \textbf{helped write better responses, even in situations where the generated responses are not a perfect fit}, they inspire the choice of response and the listener can then modify the suggestions, as P6 mentioned: \emph{"[it] inspires me to use different strategies and \textbf{gives me ideas} and then I take that and modify it."}

\paragraph{Increases peer counselors' confidence}
More than half of the participants mentioned that using \texttt{CARE} increases their confidence (8/15). During the interviews, many peer counselors (P1, P3, P6, P7, P8, P9, P10, P11) supported that the tool would be greatly beneficial to novice counselors by exposing them to new situations and would help build confidence. Peer counselor P3, who often coaches peer counselors to be better, also mentioned how it helps in new situations. Peer counselor P10 stated that \texttt{CARE} would help provide better service to the support seekers.
\begin{quote}
    \emph{"I liked it honestly. I do listener coaching and mentoring. And I liked that this really prompted, I think, a nice variety of responses depending on how comfortable the person is, and diving more deeply into whatever the issue that the member [support seeker] has. So I thought that as I was going through, I thought, this is a \textbf{great tool for listeners who are newer or are not feeling comfortable with the topic.}" -- P3}
\end{quote}
\begin{quote}
    \emph{"I feel like it [deploying \texttt{CARE} on 7 Cups as part of the training program] would make them [new peer counselors] more knowledgeable. It would make it easier for them to \textbf{provide better service} in a way for the members [support seekers]." -- P10}    
\end{quote}

\subsubsection{Interpretation of Dislikes and Potential Failure Modes of \texttt{CARE}}\label{sec:interview-dislikes}
Similar to Section \ref{sec:interview-dislikes}, we also highlight why some peer counselors are inclined to disfavor \texttt{CARE}.
We will discuss mitigation strategies in Section \ref{sec:limit-future}. % , provide corresponding chat examples in Appendix Section \ref{sec:neg-exceprts}, and summarize recommended future directions in Appendix Section~\ref{tab:future-summary}.

\paragraph{Disrupts thinking}\label{sec:disrupt}
Some participants (8/15) pointed out that the addition of an interactive tool may distract a peer counselor from the actual conversation. %  They reported that the tool may potentially disrupt the peer counselor's thought process. 
The same peer counselor, P6, who said that the tool gave ideas, also thinks that it leads to an unconscious comparison between the peer counselor's response and the suggested response, which can lead to distractions---\emph{"I am always \textbf{comparing what it was saying to what I was going to say}. And just like having another thing to see." }
However, a few participants (P3, P8) disagree. This is supported by prior research that shows similar user feedback on the introduction of any new AI assistance tools: users feel that the tool disrupts thinking initially; over time, they feel that the tool is indispensable as they get used to it \cite{recommender_system, trust_on_ai}. P3 attributes this to friction during the process of getting used to AI assistance. 
\begin{quote}
    \emph{``It's interesting that you say that [whether \texttt{CARE} disrupts thinking] because at first, I was thinking, `oh, this is going to be distracting.' But then when I didn't have it, I was like, well, \textbf{I have to work a little harder because I have to come up with everything on my own.}'' -- P3}
\end{quote}

\paragraph{Does not capture nuances.}
Several participants (P1, P2, P3, P6, P8, P10, P11) mention responses not apprehending the message underlying the conversation. 
While P1 thinks \emph{``it just that-, it just seemed that there were certain situations or certain, like, nuances of the conversation that the AI just didn't really pick up on.''}, P6 sees the difficulties as an inherent property of AI -- \emph{``As a listener, we're able to detect what this person might mean based on what they've already said, based on our whole conversation, but it's more difficult for AI to do that. \textbf{They [AI] just take the words interpreted differently.}''} Despite this, P10 still thinks \emph{``having that general basis of knowledge \textbf{is useful for new listeners} when they are building their confidence''}, and P3 also believes that for new peer counselors \texttt{CARE} is already more humanlike than the existing assistance provided by 7 Cups, which are a set of pre-defined, generic tips randomly prompted to peer counselors during chats:
\begin{quote}
    \emph{``\textbf{In 7 Cups, we have something sort of like this [\texttt{CARE}], but it's always the same, and people complain that it sounds like a bot because it doesn't sound very authentic.} It [the random tip suggestion in 7 Cups] sounds pretty, you know, like a scripted type of thing. And people will always say like don't send those bot messages, and, so it's hard when you're new, for someone [new] to be comfortable enough to do this. And I liked that these messages [from \texttt{CARE}] sounded very genuine and related to the issue and not so abstract almost.''} -- P3
\end{quote}

\paragraph{Contains made-up information.}
A few participants (P8, P9, P11) noticed some false information in \texttt{CARE}'s suggestions. For example, the suggested responses may be sometimes based on inaccurate assumptions.
% the following is more like discussion?
This may be a consequence of the insufficient dialogue context given to \texttt{CARE}. For example, if the gender of the support seeker is not mentioned in the latest message context, \texttt{CARE} may infer one. However, peer counselors can refer to the earlier context of the chat and correct this.
Future work may incorporate a dialog summarization model \cite{chen2022human} to integrate the conversation context.
\begin{comment}
 \begin{quote}
    \emph{``At one point, I remember, during the chat, it said the person was saying how they were stressed and the prompt, the suggestion was, `your anxiety and depression seem really bad,' which-, and if a listener selects that, that is kinda assuming that the person who's talking to, who has anxiety and depression, which is probably not good.''} -- P9
\end{quote}
\end{comment}
% \paragraph{Quality varies between strategies.}
% Participants' perceptions of \texttt{CARE}'s quality can vary from strategy to strategy. For example, P7 says, \emph{``I think it [\texttt{CARE}] initiates the right kind of responses, especially the support ones. I liked those the best I think, I think the one I didn't like was the reflective one.''}.

\subsubsection{\texttt{CARE} helps both new and experienced peer counselors.}\label{sec:interivew-occasions}
% We analyze the occasions where peer counselors think \texttt{CARE} helps more in this section.
\paragraph{\texttt{CARE} helps inexperienced peer counselors \textbf{learn and grow}.}
As most peer counselors (13/15), experienced and inexperienced, believe that \texttt{CARE} helps more \textit{the peer counselor is newer to listening chats} as it \textit{``gives them suggestions about what to say or how to respond''} (P1). \emph{Aiding new peer counselors} is also a common theme mentioned in interviews and open-ended questionnaire responses. 
\begin{quote}
    \textit{``New people haven't really dealt with the situation before. And they're not listening and they don't really know exactly what to say yet. [\texttt{CARE}] can be quite a \textbf{guide}.'' -- P2}
\end{quote}
A couple of peer counselors, especially those who are involved in the new peer counselor training and mentoring efforts, think \texttt{CARE} will be \textbf{useful in training new peer counselors}.
Some compare \texttt{CARE} with the current training modules and reach-out resources on 7 Cups, saying that \textbf{learning through real-time experience with \texttt{CARE} is more natural and practical than current static training modules}:
\begin{quote}
    \textit{``With this tool, I think it'll be really helpful because this helps, like, when they are taking the chat instead of, like, help before or after that. Because this is \textbf{real-time help}, so it will be a lot more helpful for the new listeners to get familiar with the process.'' -- P8}
\end{quote}

\begin{quote}
    \textit{``As most listeners, you know, they don't devote the time [on training modules]. Maybe not most, but some of them at least don't. So this way, you'll just, it's much \textbf{better to pick up through experience}.'' -- P12}
\end{quote}

Others even make recommendations on how to incorporate \texttt{CARE} into the peer counselor training process. 
P3 mentions how suggested responses give new peer counselors a sense of the expected language and the degree of formality on 7 Cups, making it a helpful tool for new peer counselors to refer to during the practice chats before they are verified.
%\begin{itemize}
 %   \item[] \textit{``It [\texttt{CARE}] gives them a sense of like, what the expectation is, like what you should be saying, because the other complaint we get is some listeners are way too casual, you know with like texts speak or just like "hey," one-word answers and this is giving you really, I think a \textbf{nice foundation for what the expectations look like}.'' -- P3}
% \end{itemize}
Still, others, especially those who join 7 Cups newly, think other than helping them to learn, hypothetically, the fact that \texttt{CARE} is available on 7 Cups will itself increase their willingness to contribute more to chats. This can potentially help mitigate the high dropout rate or low engagement of peer counselors.
\begin{quote}
    \textit{``It makes the whole process likely easier. It could only be helpful. So I guess it would \textbf{make me feel more open to doing more chats than I usually do}.'' -- P15}
\end{quote}
Peer counselors also foresee that new peer counselors' dependence on \texttt{CARE} can decrease over time since they gradually pick up what and how to use different strategies and respond with \texttt{CARE}, which meet our goal of helping new peer counselors to grow. One listener depicts their personal experience with \texttt{CARE} during the mock chats.
\begin{quote}
    \emph{``When I started listening, it was helpful because when, \textbf{when you first get into the listening, it's kind of hard to figure out immediately what you're going to say. So I did read the suggestion that was there}. But once I'm more comfortable, I was more comfortable listening, and I kind of got the idea of how to be empathetic, and how to ask open or closed questions, or just sympathize.'' -- P14}
\end{quote}

\paragraph{\texttt{CARE} helps experienced peer counselors \textbf{in difficult situations}.} %\raj{Why is this in the failure modes of CARE}
In terms of the \texttt{CARE}'s effect on experienced peer counselors, other than the aforementioned responses in the questionnaire, peer counselors think of other occasions where \texttt{CARE} can help, such as dealing with culture/ language barriers, unfamiliar topics, or context switches when being in multiple chats.
By suggesting suitable responses, \texttt{CARE} can help mitigate the language barrier or age gap between a peer counselor and a help seeker. % . P1 and P4 elucidate each of these aspects respectively:
\begin{quote}
    \textit{``It [CARE] could be really helpful. So I know for some people on 7 Cups, they don't always have the same native language. Sometimes there are \textbf{language differences} or, like slang words that aren't always understood.'' -- P1}
\end{quote}

\begin{quote}
    \textit{``Sometimes you have trouble \textbf{connecting with people who are not your age} or even closer. I want to feel like a colleague and not a mother. You know, so maybe the tool would help in that in that instance.'' -- P4}
\end{quote}

Peer counselors often specialize in a set of topics, yet they often have to converse on topics beyond that. A few participants describe difficulties in new topics which \texttt{CARE} helps mitigate: \textit{``[listeners often] feel stuck in a chat, which happens quite often''} (P9). Specifically, P4 encountered this situation during the simulated chat and reflects that: 
\begin{quote}
    \textit{``My specialty is grief. Relationship stress, [the category of the simulated chat], is not something that I'm as familiar with. I mean, I am, but I'm better on certain topics. So I think \textbf{it would help me, you know, with the category being as not as familiar}.'' -- P4}
\end{quote}

Many peer counselors manage multiple chats concurrently, P3 talks about how \texttt{CARE} can help in such context switches of peer counselors by hinting according to their experience from peer counselor coaching.
\begin{quote}
    \textit{``I have some people that I coach who take multiple chats and have a hard time managing them. \textbf{This [CARE] would allow them, I think, to manage them more properly} because they would have some kind of preset options to choose from and not have to come up with everything originally.'' -- P3}
\end{quote}

To sum up, via our system log analyses, questionnaires, and semi-structured interviews, we found that:
\begin{itemize}\setlength\itemsep{0em}
    \item \textbf{General Perception}: Overall, peer counselors think \texttt{CARE} is  helpful and often use the tool when provided. The tool also leads to longer peer counselor responses on average. Peer counselors also agree that the tool is easy to use and these suggestions are natural. Analysis of system logs and surveys both showed that users choose to use \texttt{CARE} more often than not when given an option.
    \item \textbf{Situations where \texttt{CARE} helps}: 
    Peer counselors think that \texttt{CARE} inspires better responses and reminds them of relevant counseling strategies. This helps novice counselors gain experience and confidence.  Additionally, experienced peer counselors agree that \texttt{CARE} will help them with difficult and new topics. % Semi-structured interviews also revealed that \texttt{CARE} holds the possibility to help bridge language, age, and cultural barriers. 
    % \item \textbf{Vision of Future \texttt{CARE}}: 
    % Peer counselors recommend measures that mitigate concerns about \texttt{CARE} and hope for suggestions on nuanced situations, such as crisis referral.
\end{itemize}

\section{Conclusion and Discussion}

This work develops an interactive training agent \texttt{CARE} for empowering volunteer counselors on online peer-to-peer counseling platforms
For any given chat between a support seeker and a peer counselor, 
\texttt{CARE} automatically suggests counseling strategies and provides tailored responses based on the conversation context to assist counselors.
We built upon the motivational interviewing framework and focused on a set of representative counseling strategies. We then develop contextualized classifiers to highlight which counseling strategies are needed in a given situation and scalable language generation approaches to generate example responses on the fly. 
We integrate different components into the interactive counseling support tool \texttt{CARE} to enable a realistic user experience. 
Through both quantitative analysis and qualitative user studies, we demonstrate the efficacy of \texttt{CARE} on a set of representative counseling scenarios.

\subsection{Implications and Design Recommendation} 
This work combines psychotherapy theories and advances in natural language processing to build an interactive assistant \texttt{CARE}  that empowers peer counselors on online peer counseling platforms. Our contributions consist of operationalizing theory-driven motivational interviewing strategies, developing contextualized natural language models, designing an interactive HCI system, analyzing observational data from system logs and questionnaire results, as well as performing semi-structured interviews---all of these demonstrate an example paradigm on how to build both theory-driven and model-driven systems to augment human capabilities via AI.

One straightforward design recommendation is to use \texttt{CARE} to complement the current training programs on 7 Cups and many other similar online peer counseling platforms. As we discussed earlier, existing mechanisms of training or scaffolding largely rely on human supervision, which is costly and difficult to scale. Volunteer counselors often do not have the rigorous professional training as their offline counterparts and most online counseling platforms only provide very abbreviated training. By evaluating it with 15 stakeholders both quantitatively and qualitatively, we demonstrate the possibility of \texttt{CARE} as an initial effective effort to improve these training mechanisms. For instance, integrating \texttt{CARE}-assisted practice chats into the training curriculum for new peer counselors could facilitate simulation-based learning \cite{lee2020translating, lateef2010simulation}. This approach can be further enhanced by incorporating support seeker bots \cite{demasi2020multi, tanana2019development}.

% Building an end-to-end working system sheds light on how \texttt{CARE} may be integrated into platforms such as 7 Cups to better support peer counselors. 
With appropriate ethical consideration, \texttt{CARE} also holds great potential to be deployed for \textit{real-time scaffolding} which supporters can turn to during their interactions with seekers. This provides supporters with actionable suggestions on which skill best fits in a given scenario, and how to formulate what they are going to say, with the long-term goal of increasing the number of people they can help and the effectiveness of their interactions in addressing the mental health needs of support seekers.

% The human-AI collaboration paradigms and design recommendations presented in this paper can be extended to other scaffolding setups where users would benefit from AI mentors \cite{yang2024social}. These setups include areas such as active listening skills, cultural sensitivity, negotiation, conflict resolution, empathy, persuasion, public relations, feedback delivery, and crisis communication. System characteristics like real-time feedback, tailored strategy recommendations, and addressing user over-reliance on AI can be effectively used in these contexts. 
We design \texttt{CARE} to be primarily used as a training or simulated roleplay environment for support providers so that they can practice their skills in a safe environment, without harming real people \cite{ronning2019use}. 
While the user studies and findings are grounded in the context of 7 Cups as the primary platform, findings on the impact of such an AI tool, design recommendations for the deployment of the tool, and the general perception of the tool are broadly applicable to other scaffolding and online communities with similar setups of private text-based communication channels. These channels include active listening skills, cultural sensitivity, negotiation, conflict resolution, empathy, persuasion, public relations, feedback delivery, and crisis communication. In other words, this system is generic and agnostic to any online peer counseling platform and can be used in many similar environments\cite{yang2024social}.

From a technical perspective, \texttt{CARE} consists of two main functions: highlighting which counseling strategies are most suitable and suggesting example responses in a given context. Each function can be used as an individual sub-tool to help specific use cases in online support groups. For instance, the resulting machine learning classifiers in detecting the most suitable strategies can also be used to understand the relationship between different strategies and their influence on conversation success or the well-being of support seekers. 

% This work also provides insights and raises questions for the field of natural language processing and machine learning on how to generate responsible and contextualized responses in high-stakes scenarios. 

\subsection{Ethical and Societal Considerations for Tools like \texttt{CARE}}\label{sec:improve}
While this work shows promise in using AI to provide feedback to peer counselors, societal and ethical considerations must also be considered as we deploy systems like \texttt{CARE}: 

% We summarize the features participants desired in the application of \texttt{CARE} as a peer counseling training assistance tool, addressing some concerns raised in previous sections.

\paragraph{Distributional Shift}
We design \texttt{CARE} as a scaffolding practice tool that peer counselors can use during training, with the hope that they will be able to transfer these skills to the real world. As with other training environments, we acknowledge this shift between training and the real world. We have communicated to our users that our training system is only intended for practices and that instead of completely relying on this feedback, they should use it to learn how to use these strategies effectively.   

\paragraph{Deployment Considerations} This work demonstrates how \texttt{CARE} can empower peer counselors as proof of concept. The leap from a prototype to a real-world application involves complexities that transcend the current scope of\texttt{CARE}.
For one, the dynamics of counselor-seeker interactions in real-world settings are far more nuanced and unpredictable than those used in our simulated environment. 
Furthermore, real-world applications' ethical considerations, data privacy, and security aspects demand rigorous scrutiny and adaptations beyond what we have introduced to \texttt{CARE}. We urge future work to investigate these dimensions seriously when considering the deployment of tools like \texttt{CARE}, such as how to maintain privacy and prevent data leakage when using AI models, how to continuously update the model to correspond to real-world events like the pandemic, as well as how to moderate and audit the processes of how these models are used so that norms and policies are updated accordingly. 

\paragraph{Over-Reliance} In addition, users may be heavily reliant on our tools, which may make them uncomfortable upon lack of access to the tool in real-world situations \cite{passi2022overreliance}. Furthermore, we recommend tools like \texttt{CARE} as a complement to existing training tools, as AI blind reliance can lead to cognitive atrophy \cite{ahmad2023impact}. To encourage peer counselors to learn with \texttt{CARE} rather than to rely on it, developers can incorporate  various measures like warnings and cognitive forcing functions to ensure user engagement with \texttt{CARE} with an emphasis on the unavailability of \texttt{CARE} in actual chats\cite{buccinca2021trust}.

% \paragraph{Minimizing Disruptions and Reliance}
% \raj{@diyi, reviewer R1 asks for this in the design recommendations. I will add another version of this in the recommendations, but with a different interpretation--Semistructured interviews with counselors show that such interactions often disrupt thought processes for counselors and we suggest looking at decreasing the cognitive load of AI suggestions. However, I am not sure if this is the best place for the text.}
% To mitigate concerns on disruptions (Section~\ref{sec:disrupt}) and over-reliance (Section~\ref{sec:risks}), P7 suggests that \texttt{CARE} \textit{``provide prompts [suggestions] when there are long pauses [in a conversation].''} While prior studies have investigated the trade-offs between the benefits and visual disruptions brought by an assistant tool \cite{trnka2009user, quinn2016cost}, we conjecture that deliberately introducing such a temporal barrier may encourage counselors to reflect and formulate their own thoughts before seeking guidance, thereby reducing the concerns about the indiscriminate use of \texttt{CARE}-provided suggestions during training. %Additionally, to encourage peer counselors to learn with \texttt{CARE} rather than rely on it, developers can incorporate warnings that emphasize that \texttt{CARE} will not be available in actual chats.
% An alternative approach is to have the suggestion panel toggle closed by default at each turn.

\paragraph{Personalized Feedback}
% \paragraph{Two-Way Feedback for Strategies}
To improve the effectiveness and user satisfaction of 
% \texttt{CARE} as a personalized training tool, % P7 recommended incorporating mechanisms that allow two-way feedback for strategies. For the system-to-user direction, one approach is to report the frequency of strategies chosen by the trainees during the practice chat, which can help users understand their own decision-making patterns on the choice of strategies.
% In the other direction, the system
feedback generation tools like \texttt{CARE}, they can be further improved to learn user preferences over time, similar to recommender systems, which adapt based on user behavior patterns.  %Ultimately, providing such feedback and adaptability would encourage a more engaged and reflective training process, enhancing the overall utility of \texttt{CARE} for peer counselors.
Just as some participants suggested---``\textit{every listener has a different way of responding and showing empathy}'', adaptive feedback generation based on individual counseling styles could make \texttt{CARE} more responsive and personalized. 
% \paragraph{Tailoring Strategy Recommendations}
% To enhance the relevance and effectiveness of \texttt{CARE}'s training suggestions, participants are advised to tailor the suggestions to specific situations and individual counselor styles. P12 expressed a desire for a more detailed set of strategies that suited the situation. P6 highlighted the ability to personalize suggestions  as a crucial feature because By integrating these features, a system can foster better user satisfaction. 

% \paragraph{Incorporating Crisis Referral}
% Addressing the critical need for effective crisis referral mechanisms, P6 and P9 recommended that \texttt{CARE} incorporate responses for handling crisis situations to ensure that trainees understand the guidelines and limitations of their role. By incorporating these considerations, \texttt{CARE} can guide peer counselors to provide crisis referrals that respect the sensitive nature of such interactions.

\paragraph{Biases and Stereotypes with AI Suggestion} Given the sensitive nature of this study on online peer counseling platforms, any use of AI models in providing assistance might introduce concerns about user safety and suffer from the inherent limitation of unwanted biases. We have followed responsible AI and safety best practices in this work by implementing appropriate guardrails and filtering unwanted output before presenting any content to supporters, as shown in Figure \ref{fig:architecture} and by maintaining safety protocols as described in Section \ref{sec:safety-protocol} to disclose risks to users. We recommend the development and use of better algorithms for debiasing and content filtering \cite{inan2023llama}, as well as introducing more real-time human oversight and control when deploying similar systems. While the exact safety requirements vary by task configuration, we recommend (1) model validation with experienced peer counselors and platform developers, (2) AI and human content moderation and filtering, and (3) humans at the center of control.

\paragraph{Augmentation and Automation}
Prior studies have explored the use of chatbots or large language models as peer counselor surrogates \cite{islam2021mobile, haque2023overview, abd2021perceptions}, to make peer counseling labor-free and scalable. Nevertheless, in high-stake and high-touch scenarios like counseling, such \textbf{automation} can pose significant risks for support seekers who suffer from mental health issues and raise ethical and legal concerns \cite{liu2022will, imel2017technology, miner2019key, zohny2024generative, woodnutt2024could}. Despite extensive discussions and reported uses\footnote{\url{https://www.newyorker.com/magazine/2023/03/06/can-ai-treat-mental-illness}} for using LLMs as therapists, recent research has shown that LLM simulated therapists suffer from hallucination, caricature \citep{cheng2023compost}, stereotypes, and low-quality output \citep{chiu2024computational}.  
Given all of these risks associated with \emph{automation}, our work takes an \textbf{augmentation} perspective to use AI to help the helper. 
\citet{stade2023artificial} used an analogy of building an autonomous vehicle to discuss the role of AI automation for counseling and categorized the space into \emph{assistive, collaborative}, and \emph{fully autonomous}; our augmentation perspective is similar to the \emph{assistive} dimension. Overall, we argue that such an augmentation route in the current form of AI tools can provide a balance between scale and risk management, and automation would require more advanced AI capabilities, stronger interdisciplinary collaboration, and attention to issues such as evaluation, risk, bias, and transparency, which remain challenges for LLMs today.

\subsection{Limitations and Future Work}\label{sec:limit-future}
There are a few limitations to this work, which we discuss in detail here with recommendations for future work. % We discuss potential future work in this section and curate a complete list of future directions mentioned throughout this article in Appendix Table~\ref{tab:future-summary}.

% While we have taken very careful steps for this work, there might still be potentially undesirable AI outputs.
% We require supporters to review the suggested content before using it and warn against directly using them without any modification. \texttt{CARE} is mainly used by supporters who have already received very abbreviated default training on 7 Cups, \textbf{not} seekers or the vulnerable members who suffer from mental health issues; thus, there is no direct communication between people who have mental health concerns and \texttt{CARE}. 
% Future work can help mitigate this concern by developing better algorithms for debiasing and content filtering. %  as well as introducing more real-time human oversight and control. 

\paragraph{Multi-party and Long Conversations} We design \texttt{CARE} as a training tool primarily for one-to-one conversations and evaluate it with relatively short conversations. However, some online support provisions might happen in multi-user chat forums like Reddit or public chat groups. 
Furthermore, conversations between support seekers and peer counselors can be very long in the real world; such long-term dependencies may not be captured well in the current setup when providing assistance. Generalization of \texttt{CARE} to multi-party and long conversations requires further investigation.  

\paragraph{Support Seekers' Perspectives} The primary goal of \texttt{CARE} is the training of novice peer counselors and not support seekers. Therefore, various experiments focus on supporter perception and preferences for the use of \texttt{CARE}, including Wizard-of-oz operationalization of the end-to-end system. While future studies could include support seekers - especially their perceptions about peer counselors' AI-supported training, we believe that is beyond the scope of this study.

\paragraph{Design Choices} Although we utilized the current state-of-the-art models for recommending counseling strategies and example responses, some generation results are still not very satisfactory as pointed out by participants. 
Such limited quality of generations may negatively affect the training or suggestion provision to peer counselors. Future work can further improve these models by performing extensive parameter searches and utilizing more recent language models. % with most recent large language models. 
In terms of multiple design choices such as counseling strategies, we mainly focus on a subset of motivational interviewing strategies, which provide a solid foundation for \texttt{CARE}. However, there are other types of MI strategies or other prominent counseling techniques such as cognitive behavioral therapy which could be strong alternatives for empowering peer counselors. While the chosen subset of strategies covers most messages in the 7 Cups dataset, its potential influence on the trainees' variety of strategies remains unknown, such as whether they may be biased against the strategies exclusive of our chosen subset. Similarly, for multiple threshold choices such as the number of responses to show on the \texttt{CARE} interface,  we chose them based on the researchers' expertise in design and feedback from stakeholders. % we chose the number of prior contexts as 5 when predicting the most suitable counseling strategies by following prior studies. 
We plan to iteratively refine them and perform more robustness checks for the next version of \texttt{CARE}. 

\paragraph{Long-Term Impact} As an initial effort to empower peer counselors, our work mainly focuses on the development and design of \texttt{CARE} as a prototype system, and the evaluation of it with domain users via interviews and system logs in a simulated one-time chat environment. 
Our work has not evaluated any \emph{long-term} effects of \texttt{CARE}, such as its influence on certain skill acquisition of peer counselors, on the mental health outcome changes in support seekers, or the change in listener engagement to a conversation with the availability of \texttt{CARE}. 
We plan to continue improving  \texttt{CARE} and evaluate its long-term impact by longitudinally working with more domain users. 

% \paragraph{Explicit Feedback} In contrast to the expert knowledge - and labor-intensive training under human supervision, the \texttt{CARE}-based training system depends on peer counselors learning from contextual examples through practice rather than receiving explicit feedback on their skills and performance. Future research can investigate the methods and consequences of generating context- and strategy-specific feedback for peer counselor training.

% \paragraph{Impartiality} The user studies are conducted online to better align with the nature of online peer counseling, with the majority of the authors having a background in computer science. Despite our efforts to maintain impartiality, the experimental results and analysis may lean towards a pro-technology perspective and might not be directly applicable to in-person peer counseling. We encourage future researchers to explore similar tools in a physical setting and invite a more diverse group of researchers to contribute to the field.

\paragraph{Generalization to Other Platforms} % As a pilot study, our design process of \texttt{CARE} does not involve peer counselors other than the researchers ourselves.
We designed and evaluated \texttt{CARE} by working with domain users---peer counselors from 7 Cups, as a result, our system design principles can be applied to similar online peer counseling platforms, especially those that share similar training and chatting feature, such as Supportiv\footnote{\url{https://www.supportiv.com/}} and TalkLife\footnote{\url{https://www.talklife.com/}}. 
While we expect many findings to be platform-agnostic, it is difficult to generalize without rigorous testing with specific domain users. We leave this to future work to examine the generalizability of similar tools across different peer counseling platforms.
% While we believe our contributions are platform-agnostic, the chat interface used in our user studies is similar to that of 7 Cups, and we recruit peer counselors from 7 Cups to be our participants. We encourage future work to  % (1) adopt a user-centered, iterative design process, such as Research through Design (RtD), (2) include more stakeholders in the design process for scaffolding--end users, educators, platform developers, etc. to provide seamless user experience, and (3) 
% examine the generalizability of similar tools across different peer counseling platforms, such as BetterHelp\footnote{\url{https://www.betterhelp.com/}} and Talkspace.\footnote{\url{https://www.talkspace.com/}} %  Feedback loops, user studies, and prototype testing across platforms can ensure the tool meets user educational needs effectively and is generalizable.

\paragraph{Representativeness} During our user studies, we asked peer counselors to focus on two of the most popular topics: relationship stress and anxiety. However, online peer counseling platforms cover a wide variety of issues, including depression, health, LGBTQ+ matters, dissociative identity, home life, self-improvement, and more. Future research could explore the efficacy of our system across a broader range of topics to improve generalizability.
Additionally, our user studies included only 15 peer counselors, located in the United States, and using English in their sessions. We do not have data on the actual distribution of gender, location, and language of peer counselors on 7 Cups. Significant differences between our participants and the actual counselor population could introduce biases into our observations.
Furthermore, the participating peer counselors varied in their length of service, with most lacking formal counseling training. Expert opinions on the quality of \texttt{CARE} may more accurately reflect its potential impacts. Expanding this work to include more participants with diverse demographics, counseling backgrounds, languages, locations, and expertise could provide more nuanced and generalizable findings.

\begin{acks}
The authors would like to thank the reviewers and members of SALT Lab for their feedback. This work was supported by an NSF grant IIS-2247357.
\end{acks}

\bibliographystyle{ACM-Reference-Format}
\bibliography{ref,diyi}

%%% -*-BibTeX-*-
%%% Do NOT edit. File created by BibTeX with style
%%% ACM-Reference-Format-Journals [18-Jan-2012].

\begin{thebibliography}{153}

%%% ====================================================================
%%% NOTE TO THE USER: you can override these defaults by providing
%%% customized versions of any of these macros before the \bibliography
%%% command.  Each of them MUST provide its own final punctuation,
%%% except for \shownote{}, \showDOI{}, and \showURL{}.  The latter two
%%% do not use final punctuation, in order to avoid confusing it with
%%% the Web address.
%%%
%%% To suppress output of a particular field, define its macro to expand
%%% to an empty string, or better, \unskip, like this:
%%%
%%% \newcommand{\showDOI}[1]{\unskip}   % LaTeX syntax
%%%
%%% \def \showDOI #1{\unskip}           % plain TeX syntax
%%%
%%% ====================================================================

\ifx \showCODEN    \undefined \def \showCODEN     #1{\unskip}     \fi
\ifx \showDOI      \undefined \def \showDOI       #1{#1}\fi
\ifx \showISBNx    \undefined \def \showISBNx     #1{\unskip}     \fi
\ifx \showISBNxiii \undefined \def \showISBNxiii  #1{\unskip}     \fi
\ifx \showISSN     \undefined \def \showISSN      #1{\unskip}     \fi
\ifx \showLCCN     \undefined \def \showLCCN      #1{\unskip}     \fi
\ifx \shownote     \undefined \def \shownote      #1{#1}          \fi
\ifx \showarticletitle \undefined \def \showarticletitle #1{#1}   \fi
\ifx \showURL      \undefined \def \showURL       {\relax}        \fi
% The following commands are used for tagged output and should be
% invisible to TeX
\providecommand\bibfield[2]{#2}
\providecommand\bibinfo[2]{#2}
\providecommand\natexlab[1]{#1}
\providecommand\showeprint[2][]{arXiv:#2}

\bibitem[Abd-Alrazaq et~al\mbox{.}(2021)]%
        {abd2021perceptions}
\bibfield{author}{\bibinfo{person}{Alaa~A Abd-Alrazaq}, \bibinfo{person}{Mohannad Alajlani}, \bibinfo{person}{Nashva Ali}, \bibinfo{person}{Kerstin Denecke}, \bibinfo{person}{Bridgette~M Bewick}, {and} \bibinfo{person}{Mowafa Househ}.} \bibinfo{year}{2021}\natexlab{}.
\newblock \showarticletitle{Perceptions and opinions of patients about mental health chatbots: scoping review}.
\newblock \bibinfo{journal}{\emph{Journal of medical Internet research}} \bibinfo{volume}{23}, \bibinfo{number}{1} (\bibinfo{year}{2021}), \bibinfo{pages}{e17828}.
\newblock


\bibitem[Ahmad et~al\mbox{.}(2023)]%
        {ahmad2023impact}
\bibfield{author}{\bibinfo{person}{Sayed~Fayaz Ahmad}, \bibinfo{person}{Heesup Han}, \bibinfo{person}{Muhammad~Mansoor Alam}, \bibinfo{person}{Mohd Rehmat}, \bibinfo{person}{Muhammad Irshad}, \bibinfo{person}{Marcelo Arra{\~n}o-Mu{\~n}oz}, \bibinfo{person}{Antonio Ariza-Montes}, {et~al\mbox{.}}} \bibinfo{year}{2023}\natexlab{}.
\newblock \showarticletitle{Impact of artificial intelligence on human loss in decision making, laziness and safety in education}.
\newblock \bibinfo{journal}{\emph{Humanities and Social Sciences Communications}} \bibinfo{volume}{10}, \bibinfo{number}{1} (\bibinfo{year}{2023}), \bibinfo{pages}{1--14}.
\newblock


\bibitem[Ali et~al\mbox{.}(2015)]%
        {ali2015online}
\bibfield{author}{\bibinfo{person}{Kathina Ali}, \bibinfo{person}{Louise Farrer}, \bibinfo{person}{Amelia Gulliver}, {and} \bibinfo{person}{Kathleen~M Griffiths}.} \bibinfo{year}{2015}\natexlab{}.
\newblock \showarticletitle{Online peer-to-peer support for young people with mental health problems: a systematic review}.
\newblock \bibinfo{journal}{\emph{JMIR mental health}} \bibinfo{volume}{2}, \bibinfo{number}{2} (\bibinfo{year}{2015}), \bibinfo{pages}{e4418}.
\newblock


\bibitem[Althoff et~al\mbox{.}(2016)]%
        {althoff2016large}
\bibfield{author}{\bibinfo{person}{Tim Althoff}, \bibinfo{person}{Kevin Clark}, {and} \bibinfo{person}{Jure Leskovec}.} \bibinfo{year}{2016}\natexlab{}.
\newblock \showarticletitle{Large-scale analysis of counseling conversations: An application of natural language processing to mental health}.
\newblock \bibinfo{journal}{\emph{Transactions of the Association for Computational Linguistics}}  \bibinfo{volume}{4} (\bibinfo{year}{2016}), \bibinfo{pages}{463--476}.
\newblock


\bibitem[Andalibi et~al\mbox{.}(2018)]%
        {andalibi2018social}
\bibfield{author}{\bibinfo{person}{Nazanin Andalibi}, \bibinfo{person}{Oliver~L Haimson}, \bibinfo{person}{Munmun~De Choudhury}, {and} \bibinfo{person}{Andrea Forte}.} \bibinfo{year}{2018}\natexlab{}.
\newblock \showarticletitle{Social support, reciprocity, and anonymity in responses to sexual abuse disclosures on social media}.
\newblock \bibinfo{journal}{\emph{ACM Transactions on Computer-Human Interaction (TOCHI)}} \bibinfo{volume}{25}, \bibinfo{number}{5} (\bibinfo{year}{2018}), \bibinfo{pages}{1--35}.
\newblock


\bibitem[Armstrong(2010)]%
        {armstrong2010effective}
\bibfield{author}{\bibinfo{person}{Joe Armstrong}.} \bibinfo{year}{2010}\natexlab{}.
\newblock \showarticletitle{How effective are minimally trained/experienced volunteer mental health counsellors? Evaluation of CORE outcome data}.
\newblock \bibinfo{journal}{\emph{Counselling and Psychotherapy Research}} \bibinfo{volume}{10}, \bibinfo{number}{1} (\bibinfo{year}{2010}), \bibinfo{pages}{22--31}.
\newblock


\bibitem[Atkins et~al\mbox{.}(2014)]%
        {atkins2014scaling}
\bibfield{author}{\bibinfo{person}{David~C Atkins}, \bibinfo{person}{Mark Steyvers}, \bibinfo{person}{Zac~E Imel}, {and} \bibinfo{person}{Padhraic Smyth}.} \bibinfo{year}{2014}\natexlab{}.
\newblock \showarticletitle{Scaling up the evaluation of psychotherapy: evaluating motivational interviewing fidelity via statistical text classification}.
\newblock \bibinfo{journal}{\emph{Implementation Science}} \bibinfo{volume}{9}, \bibinfo{number}{1} (\bibinfo{year}{2014}), \bibinfo{pages}{1--11}.
\newblock


\bibitem[Barak(2007)]%
        {barak2007emotional}
\bibfield{author}{\bibinfo{person}{Azy Barak}.} \bibinfo{year}{2007}\natexlab{}.
\newblock \showarticletitle{Emotional support and suicide prevention through the Internet: A field project report}.
\newblock \bibinfo{journal}{\emph{Computers in Human Behavior}} \bibinfo{volume}{23}, \bibinfo{number}{2} (\bibinfo{year}{2007}), \bibinfo{pages}{971--984}.
\newblock


\bibitem[Baumel(2015)]%
        {baumel2015online}
\bibfield{author}{\bibinfo{person}{Amit Baumel}.} \bibinfo{year}{2015}\natexlab{}.
\newblock \showarticletitle{Online emotional support delivered by trained volunteers: users’ satisfaction and their perception of the service compared to psychotherapy}.
\newblock \bibinfo{journal}{\emph{Journal of mental health}} \bibinfo{volume}{24}, \bibinfo{number}{5} (\bibinfo{year}{2015}), \bibinfo{pages}{313--320}.
\newblock


\bibitem[Baumel and Schueller(2016)]%
        {baumel2016adjusting}
\bibfield{author}{\bibinfo{person}{Amit Baumel} {and} \bibinfo{person}{Stephen~M Schueller}.} \bibinfo{year}{2016}\natexlab{}.
\newblock \showarticletitle{Adjusting an available online peer support platform in a program to supplement the treatment of perinatal depression and anxiety}.
\newblock \bibinfo{journal}{\emph{JMIR mental health}} \bibinfo{volume}{3}, \bibinfo{number}{1} (\bibinfo{year}{2016}), \bibinfo{pages}{e5335}.
\newblock


\bibitem[Bennett-Levy(2006)]%
        {bennett2006therapist}
\bibfield{author}{\bibinfo{person}{James Bennett-Levy}.} \bibinfo{year}{2006}\natexlab{}.
\newblock \showarticletitle{Therapist skills: A cognitive model of their acquisition and refinement}.
\newblock \bibinfo{journal}{\emph{Behavioural and Cognitive Psychotherapy}} \bibinfo{volume}{34}, \bibinfo{number}{1} (\bibinfo{year}{2006}), \bibinfo{pages}{57--78}.
\newblock


\bibitem[Bernecker et~al\mbox{.}(2017)]%
        {bernecker2017web}
\bibfield{author}{\bibinfo{person}{Samantha~L Bernecker}, \bibinfo{person}{Kaitlin Banschback}, \bibinfo{person}{Gennarina~D Santorelli}, {and} \bibinfo{person}{Michael~J Constantino}.} \bibinfo{year}{2017}\natexlab{}.
\newblock \showarticletitle{A web-disseminated self-help and peer support program could fill gaps in mental health care: Lessons from a consumer survey}.
\newblock \bibinfo{journal}{\emph{JMIR mental health}} \bibinfo{volume}{4}, \bibinfo{number}{1} (\bibinfo{year}{2017}), \bibinfo{pages}{e4751}.
\newblock


\bibitem[Blease and Torous(2023)]%
        {blease2023chatgpt}
\bibfield{author}{\bibinfo{person}{Charlotte Blease} {and} \bibinfo{person}{John Torous}.} \bibinfo{year}{2023}\natexlab{}.
\newblock \showarticletitle{ChatGPT and mental healthcare: balancing benefits with risks of harms}.
\newblock \bibinfo{journal}{\emph{BMJ Ment Health}} \bibinfo{volume}{26}, \bibinfo{number}{1} (\bibinfo{year}{2023}).
\newblock


\bibitem[Brown et~al\mbox{.}(2020)]%
        {brown2020language}
\bibfield{author}{\bibinfo{person}{Tom Brown}, \bibinfo{person}{Benjamin Mann}, \bibinfo{person}{Nick Ryder}, \bibinfo{person}{Melanie Subbiah}, \bibinfo{person}{Jared~D Kaplan}, \bibinfo{person}{Prafulla Dhariwal}, \bibinfo{person}{Arvind Neelakantan}, \bibinfo{person}{Pranav Shyam}, \bibinfo{person}{Girish Sastry}, \bibinfo{person}{Amanda Askell}, {et~al\mbox{.}}} \bibinfo{year}{2020}\natexlab{}.
\newblock \showarticletitle{Language models are few-shot learners}.
\newblock \bibinfo{journal}{\emph{Advances in neural information processing systems}}  \bibinfo{volume}{33} (\bibinfo{year}{2020}), \bibinfo{pages}{1877--1901}.
\newblock


\bibitem[Bruning et~al\mbox{.}(2013)]%
        {bruning2013examining}
\bibfield{author}{\bibinfo{person}{Roger Bruning}, \bibinfo{person}{Michael Dempsey}, \bibinfo{person}{Douglas~F Kauffman}, \bibinfo{person}{Courtney McKim}, {and} \bibinfo{person}{Sharon Zumbrunn}.} \bibinfo{year}{2013}\natexlab{}.
\newblock \showarticletitle{Examining dimensions of self-efficacy for writing.}
\newblock \bibinfo{journal}{\emph{Journal of educational psychology}} \bibinfo{volume}{105}, \bibinfo{number}{1} (\bibinfo{year}{2013}), \bibinfo{pages}{25}.
\newblock


\bibitem[Bu{\c{c}}inca et~al\mbox{.}(2021)]%
        {buccinca2021trust}
\bibfield{author}{\bibinfo{person}{Zana Bu{\c{c}}inca}, \bibinfo{person}{Maja~Barbara Malaya}, {and} \bibinfo{person}{Krzysztof~Z Gajos}.} \bibinfo{year}{2021}\natexlab{}.
\newblock \showarticletitle{To trust or to think: cognitive forcing functions can reduce overreliance on AI in AI-assisted decision-making}.
\newblock \bibinfo{journal}{\emph{Proceedings of the ACM on Human-computer Interaction}} \bibinfo{volume}{5}, \bibinfo{number}{CSCW1} (\bibinfo{year}{2021}), \bibinfo{pages}{1--21}.
\newblock


\bibitem[Burke et~al\mbox{.}(2003)]%
        {mi_abuse_1}
\bibfield{author}{\bibinfo{person}{Brian~L Burke}, \bibinfo{person}{Hal Arkowitz}, {and} \bibinfo{person}{Marisa Menchola}.} \bibinfo{year}{2003}\natexlab{}.
\newblock \showarticletitle{The efficacy of motivational interviewing: a meta-analysis of controlled clinical trials}.
\newblock \bibinfo{journal}{\emph{J Consult Clin Psychol}} \bibinfo{volume}{71}, \bibinfo{number}{5} (\bibinfo{date}{Oct.} \bibinfo{year}{2003}), \bibinfo{pages}{843--861}.
\newblock


\bibitem[Burke and Kraut(2008)]%
        {burke2008mind}
\bibfield{author}{\bibinfo{person}{Moira Burke} {and} \bibinfo{person}{Robert Kraut}.} \bibinfo{year}{2008}\natexlab{}.
\newblock \showarticletitle{Mind your Ps and Qs: the impact of politeness and rudeness in online communities}. In \bibinfo{booktitle}{\emph{Proceedings of the 2008 ACM conference on Computer supported cooperative work}}. \bibinfo{pages}{281--284}.
\newblock


\bibitem[Burke et~al\mbox{.}(2010)]%
        {burke2010membership}
\bibfield{author}{\bibinfo{person}{Moira Burke}, \bibinfo{person}{Robert Kraut}, {and} \bibinfo{person}{Elisabeth Joyce}.} \bibinfo{year}{2010}\natexlab{}.
\newblock \showarticletitle{Membership claims and requests: Conversation-level newcomer socialization strategies in online groups}.
\newblock \bibinfo{journal}{\emph{Small group research}} \bibinfo{volume}{41}, \bibinfo{number}{1} (\bibinfo{year}{2010}), \bibinfo{pages}{4--40}.
\newblock


\bibitem[Can et~al\mbox{.}(2012)]%
        {can2012case}
\bibfield{author}{\bibinfo{person}{Do{\u{g}}an Can}, \bibinfo{person}{Panayiotis~G Georgiou}, \bibinfo{person}{David~C Atkins}, {and} \bibinfo{person}{Shrikanth~S Narayanan}.} \bibinfo{year}{2012}\natexlab{}.
\newblock \showarticletitle{A case study: Detecting counselor reflections in psychotherapy for addictions using linguistic features}. In \bibinfo{booktitle}{\emph{Thirteenth Annual Conference of the International Speech Communication Association}}.
\newblock


\bibitem[Carlini et~al\mbox{.}(2021)]%
        {carlini2021extracting}
\bibfield{author}{\bibinfo{person}{Nicholas Carlini}, \bibinfo{person}{Florian Tramer}, \bibinfo{person}{Eric Wallace}, \bibinfo{person}{Matthew Jagielski}, \bibinfo{person}{Ariel Herbert-Voss}, \bibinfo{person}{Katherine Lee}, \bibinfo{person}{Adam Roberts}, \bibinfo{person}{Tom Brown}, \bibinfo{person}{Dawn Song}, \bibinfo{person}{Ulfar Erlingsson}, {et~al\mbox{.}}} \bibinfo{year}{2021}\natexlab{}.
\newblock \showarticletitle{Extracting training data from large language models}. In \bibinfo{booktitle}{\emph{30th USENIX Security Symposium (USENIX Security 21)}}. \bibinfo{pages}{2633--2650}.
\newblock


\bibitem[Caselli et~al\mbox{.}(2021)]%
        {caselli2021hatebert}
\bibfield{author}{\bibinfo{person}{Tommaso Caselli}, \bibinfo{person}{Valerio Basile}, \bibinfo{person}{Jelena Mitrovi{\'c}}, {and} \bibinfo{person}{Michael Granitzer}.} \bibinfo{year}{2021}\natexlab{}.
\newblock \showarticletitle{HateBERT: Retraining BERT for Abusive Language Detection in English}. In \bibinfo{booktitle}{\emph{Proceedings of the 5th Workshop on Online Abuse and Harms (WOAH 2021)}}. \bibinfo{pages}{17--25}.
\newblock


\bibitem[Chen et~al\mbox{.}(2022)]%
        {chen2022human}
\bibfield{author}{\bibinfo{person}{Jiaao Chen}, \bibinfo{person}{Mohan Dodda}, {and} \bibinfo{person}{Diyi Yang}.} \bibinfo{year}{2022}\natexlab{}.
\newblock \showarticletitle{Human-in-the-loop Abstractive Dialogue Summarization}.
\newblock \bibinfo{journal}{\emph{arXiv e-prints}} (\bibinfo{year}{2022}), \bibinfo{pages}{arXiv--2212}.
\newblock


\bibitem[Chen et~al\mbox{.}(2021a)]%
        {chen2021evaluating}
\bibfield{author}{\bibinfo{person}{Mark Chen}, \bibinfo{person}{Jerry Tworek}, \bibinfo{person}{Heewoo Jun}, \bibinfo{person}{Qiming Yuan}, \bibinfo{person}{Henrique Ponde De~Oliveira Pinto}, \bibinfo{person}{Jared Kaplan}, \bibinfo{person}{Harri Edwards}, \bibinfo{person}{Yuri Burda}, \bibinfo{person}{Nicholas Joseph}, \bibinfo{person}{Greg Brockman}, {et~al\mbox{.}}} \bibinfo{year}{2021}\natexlab{a}.
\newblock \showarticletitle{Evaluating large language models trained on code}.
\newblock \bibinfo{journal}{\emph{arXiv preprint arXiv:2107.03374}} (\bibinfo{year}{2021}).
\newblock


\bibitem[Chen et~al\mbox{.}(2019)]%
        {chen2019gmail}
\bibfield{author}{\bibinfo{person}{Mia~Xu Chen}, \bibinfo{person}{Benjamin~N Lee}, \bibinfo{person}{Gagan Bansal}, \bibinfo{person}{Yuan Cao}, \bibinfo{person}{Shuyuan Zhang}, \bibinfo{person}{Justin Lu}, \bibinfo{person}{Jackie Tsay}, \bibinfo{person}{Yinan Wang}, \bibinfo{person}{Andrew~M Dai}, \bibinfo{person}{Zhifeng Chen}, {et~al\mbox{.}}} \bibinfo{year}{2019}\natexlab{}.
\newblock \showarticletitle{Gmail smart compose: Real-time assisted writing}. In \bibinfo{booktitle}{\emph{Proceedings of the 25th ACM SIGKDD International Conference on Knowledge Discovery \& Data Mining}}. \bibinfo{pages}{2287--2295}.
\newblock


\bibitem[Chen et~al\mbox{.}(2021b)]%
        {scaffolding}
\bibfield{author}{\bibinfo{person}{Tianying Chen}, \bibinfo{person}{Kristy Zhang}, \bibinfo{person}{Robert~E. Kraut}, {and} \bibinfo{person}{Laura Dabbish}.} \bibinfo{year}{2021}\natexlab{b}.
\newblock \showarticletitle{Scaffolding the Online Peer-Support Experience: Novice Supporters' Strategies and Challenges}.
\newblock \bibinfo{journal}{\emph{Proc. ACM Hum.-Comput. Interact.}} \bibinfo{volume}{5}, \bibinfo{number}{CSCW2}, Article \bibinfo{articleno}{366} (\bibinfo{date}{oct} \bibinfo{year}{2021}), \bibinfo{numpages}{30}~pages.
\newblock
\urldef\tempurl%
\url{https://doi.org/10.1145/3479510}
\showDOI{\tempurl}


\bibitem[Cheng et~al\mbox{.}(2023)]%
        {cheng2023compost}
\bibfield{author}{\bibinfo{person}{Myra Cheng}, \bibinfo{person}{Tiziano Piccardi}, {and} \bibinfo{person}{Diyi Yang}.} \bibinfo{year}{2023}\natexlab{}.
\newblock \showarticletitle{CoMPosT: Characterizing and evaluating caricature in LLM simulations}.
\newblock \bibinfo{journal}{\emph{arXiv preprint arXiv:2310.11501}} (\bibinfo{year}{2023}).
\newblock


\bibitem[Chikersal et~al\mbox{.}(2020)]%
        {chikersal2020understanding}
\bibfield{author}{\bibinfo{person}{Prerna Chikersal}, \bibinfo{person}{Danielle Belgrave}, \bibinfo{person}{Gavin Doherty}, \bibinfo{person}{Angel Enrique}, \bibinfo{person}{Jorge~E Palacios}, \bibinfo{person}{Derek Richards}, {and} \bibinfo{person}{Anja Thieme}.} \bibinfo{year}{2020}\natexlab{}.
\newblock \showarticletitle{Understanding client support strategies to improve clinical outcomes in an online mental health intervention}. In \bibinfo{booktitle}{\emph{Proceedings of the 2020 CHI conference on human factors in computing systems}}. \bibinfo{pages}{1--16}.
\newblock


\bibitem[Chiu et~al\mbox{.}(2024)]%
        {chiu2024computational}
\bibfield{author}{\bibinfo{person}{Yu~Ying Chiu}, \bibinfo{person}{Ashish Sharma}, \bibinfo{person}{Inna~Wanyin Lin}, {and} \bibinfo{person}{Tim Althoff}.} \bibinfo{year}{2024}\natexlab{}.
\newblock \showarticletitle{A Computational Framework for Behavioral Assessment of LLM Therapists}.
\newblock \bibinfo{journal}{\emph{arXiv preprint arXiv:2401.00820}} (\bibinfo{year}{2024}).
\newblock


\bibitem[Cho et~al\mbox{.}(2014)]%
        {cho2014learning}
\bibfield{author}{\bibinfo{person}{Kyunghyun Cho}, \bibinfo{person}{Bart van Merrienboer}, \bibinfo{person}{Caglar Gulcehre}, \bibinfo{person}{Dzmitry Bahdanau}, \bibinfo{person}{Fethi Bougares}, \bibinfo{person}{Holger Schwenk}, {and} \bibinfo{person}{Yoshua Bengio}.} \bibinfo{year}{2014}\natexlab{}.
\newblock \showarticletitle{Learning Phrase Representations using RNN Encoder--Decoder for Statistical Machine Translation}. In \bibinfo{booktitle}{\emph{Proceedings of the 2014 Conference on Empirical Methods in Natural Language Processing (EMNLP)}}. Association for Computational Linguistics, \bibinfo{pages}{1724}.
\newblock


\bibitem[Clarke et~al\mbox{.}(2005)]%
        {clarke2005overcoming}
\bibfield{author}{\bibinfo{person}{Greg Clarke}, \bibinfo{person}{Donna Eubanks}, \bibinfo{person}{Ed Reid}, \bibinfo{person}{Chris Kelleher}, \bibinfo{person}{Elizabeth O'connor}, \bibinfo{person}{Lynn~L DeBar}, \bibinfo{person}{Frances Lynch}, \bibinfo{person}{Sonia Nunley}, {and} \bibinfo{person}{Christina Gullion}.} \bibinfo{year}{2005}\natexlab{}.
\newblock \showarticletitle{Overcoming Depression on the Internet (ODIN)(2): a randomized trial of a self-help depression skills program with reminders}.
\newblock \bibinfo{journal}{\emph{Journal of medical Internet research}} \bibinfo{volume}{7}, \bibinfo{number}{2} (\bibinfo{year}{2005}), \bibinfo{pages}{e16}.
\newblock


\bibitem[Coyle et~al\mbox{.}(2012)]%
        {coyle2012interaction}
\bibfield{author}{\bibinfo{person}{David Coyle}, \bibinfo{person}{Conor Linehan}, \bibinfo{person}{Karen Tang}, {and} \bibinfo{person}{Sian Lindley}.} \bibinfo{year}{2012}\natexlab{}.
\newblock \showarticletitle{Interaction design and emotional wellbeing}.
\newblock In \bibinfo{booktitle}{\emph{CHI'12 Extended Abstracts on Human Factors in Computing Systems}}. \bibinfo{pages}{2775--2778}.
\newblock


\bibitem[Curran et~al\mbox{.}(2019)]%
        {curran2019does}
\bibfield{author}{\bibinfo{person}{Joe Curran}, \bibinfo{person}{Glenys~D Parry}, \bibinfo{person}{Gillian~E Hardy}, \bibinfo{person}{Jennifer Darling}, \bibinfo{person}{Ann-Marie Mason}, {and} \bibinfo{person}{Eleni Chambers}.} \bibinfo{year}{2019}\natexlab{}.
\newblock \showarticletitle{How does therapy harm? A model of adverse process using task analysis in the meta-synthesis of service users' experience}.
\newblock \bibinfo{journal}{\emph{Frontiers in psychology}}  \bibinfo{volume}{10} (\bibinfo{year}{2019}), \bibinfo{pages}{347}.
\newblock


\bibitem[Demasi et~al\mbox{.}(2020)]%
        {demasi2020multi}
\bibfield{author}{\bibinfo{person}{Orianna Demasi}, \bibinfo{person}{Yu Li}, {and} \bibinfo{person}{Zhou Yu}.} \bibinfo{year}{2020}\natexlab{}.
\newblock \showarticletitle{A multi-persona chatbot for hotline counselor training}. In \bibinfo{booktitle}{\emph{Findings of the Association for Computational Linguistics: EMNLP 2020}}. \bibinfo{pages}{3623--3636}.
\newblock


\bibitem[Devlin et~al\mbox{.}(2018)]%
        {BERT}
\bibfield{author}{\bibinfo{person}{Jacob Devlin}, \bibinfo{person}{Ming{-}Wei Chang}, \bibinfo{person}{Kenton Lee}, {and} \bibinfo{person}{Kristina Toutanova}.} \bibinfo{year}{2018}\natexlab{}.
\newblock \showarticletitle{{BERT:} Pre-training of Deep Bidirectional Transformers for Language Understanding}.
\newblock \bibinfo{journal}{\emph{CoRR}}  \bibinfo{volume}{abs/1810.04805} (\bibinfo{year}{2018}).
\newblock
\showeprint[arXiv]{1810.04805}
\urldef\tempurl%
\url{http://arxiv.org/abs/1810.04805}
\showURL{%
\tempurl}


\bibitem[Dewi et~al\mbox{.}(2020)]%
        {dewi2020implementation}
\bibfield{author}{\bibinfo{person}{Ratna~Sari Dewi}, \bibinfo{person}{Zaharil An'asy}, {et~al\mbox{.}}} \bibinfo{year}{2020}\natexlab{}.
\newblock \emph{\bibinfo{title}{The Implementation of Grammarly Tool to Boost Students’ Writing Skill of Exposition Text (A Classroom Action Research at the Eleventh Grade Students of SMA Dharma Karya UT in Academic Year 2019/2020)}}.
\newblock {B.S.} thesis. \bibinfo{school}{Jakarta: FITK UIN Syarif Hidayatullah Jakarta}.
\newblock


\bibitem[Dieleman et~al\mbox{.}(2016)]%
        {dieleman2016us}
\bibfield{author}{\bibinfo{person}{Joseph~L Dieleman}, \bibinfo{person}{Ranju Baral}, \bibinfo{person}{Maxwell Birger}, \bibinfo{person}{Anthony~L Bui}, \bibinfo{person}{Anne Bulchis}, \bibinfo{person}{Abigail Chapin}, \bibinfo{person}{Hannah Hamavid}, \bibinfo{person}{Cody Horst}, \bibinfo{person}{Elizabeth~K Johnson}, {and} \bibinfo{person}{Jonathan Joseph}.} \bibinfo{year}{2016}\natexlab{}.
\newblock \showarticletitle{US spending on personal health care and public health, 1996-2013}.
\newblock \bibinfo{journal}{\emph{Jama}} \bibinfo{volume}{316}, \bibinfo{number}{24} (\bibinfo{year}{2016}), \bibinfo{pages}{2627--2646}.
\newblock


\bibitem[Eysenbach et~al\mbox{.}(2004)]%
        {eysenbach2004health}
\bibfield{author}{\bibinfo{person}{Gunther Eysenbach}, \bibinfo{person}{John Powell}, \bibinfo{person}{Marina Englesakis}, \bibinfo{person}{Carlos Rizo}, {and} \bibinfo{person}{Anita Stern}.} \bibinfo{year}{2004}\natexlab{}.
\newblock \showarticletitle{Health related virtual communities and electronic support groups: systematic review of the effects of online peer to peer interactions}.
\newblock \bibinfo{journal}{\emph{Bmj}} \bibinfo{volume}{328}, \bibinfo{number}{7449} (\bibinfo{year}{2004}), \bibinfo{pages}{1166}.
\newblock


\bibitem[Fang et~al\mbox{.}(2023)]%
        {fang2023makes}
\bibfield{author}{\bibinfo{person}{Anna Fang}, \bibinfo{person}{Wenjie Yang}, \bibinfo{person}{Raj~Sanjay Shah}, \bibinfo{person}{Yash Mathur}, \bibinfo{person}{Diyi Yang}, \bibinfo{person}{Haiyi Zhu}, {and} \bibinfo{person}{Robert Kraut}.} \bibinfo{year}{2023}\natexlab{}.
\newblock \showarticletitle{What Makes Digital Support Effective? How Therapeutic Skills Affect Clinical Well-Being}.
\newblock \bibinfo{journal}{\emph{arXiv preprint arXiv:2312.10775}} (\bibinfo{year}{2023}).
\newblock


\bibitem[Fisher et~al\mbox{.}(2006)]%
        {mi_beh_se_1}
\bibfield{author}{\bibinfo{person}{Jeffrey~D Fisher}, \bibinfo{person}{William~A Fisher}, \bibinfo{person}{Deborah~H Cornman}, \bibinfo{person}{Rivet~K Amico}, \bibinfo{person}{Angela Bryan}, {and} \bibinfo{person}{Gerald~H Friedland}.} \bibinfo{year}{2006}\natexlab{}.
\newblock \showarticletitle{Clinician-delivered intervention during routine clinical care reduces unprotected sexual behavior among {HIV-infected} patients}.
\newblock \bibinfo{journal}{\emph{J Acquir Immune Defic Syndr}} \bibinfo{volume}{41}, \bibinfo{number}{1} (\bibinfo{date}{Jan.} \bibinfo{year}{2006}), \bibinfo{pages}{44--52}.
\newblock


\bibitem[Fitria(2021)]%
        {fitria2021grammarly}
\bibfield{author}{\bibinfo{person}{Tira~Nur Fitria}.} \bibinfo{year}{2021}\natexlab{}.
\newblock \showarticletitle{Grammarly as AI-powered English writing assistant: Students’ alternative for writing English}.
\newblock \bibinfo{journal}{\emph{Metathesis: Journal of English Language, Literature, and Teaching}} \bibinfo{volume}{5}, \bibinfo{number}{1} (\bibinfo{year}{2021}), \bibinfo{pages}{65--78}.
\newblock


\bibitem[for Behavioral Health~Statistics and Quality(2015)]%
        {center2015behavioral}
\bibfield{author}{\bibinfo{person}{Center for Behavioral Health~Statistics} {and} \bibinfo{person}{Quality}.} \bibinfo{year}{2015}\natexlab{}.
\newblock \showarticletitle{Behavioral health trends in the United States: results from the 2014 National Survey on Drug Use and Health}.
\newblock \bibinfo{journal}{\emph{HHS Publication No. SMA 15-4927, NSDUH Series H-50}} (\bibinfo{year}{2015}).
\newblock


\bibitem[Fredricks et~al\mbox{.}(2005)]%
        {fredricks2005school}
\bibfield{author}{\bibinfo{person}{Jennifer~A Fredricks}, \bibinfo{person}{Phyllis Blumenfeld}, \bibinfo{person}{Jeanne Friedel}, {and} \bibinfo{person}{Alison Paris}.} \bibinfo{year}{2005}\natexlab{}.
\newblock \showarticletitle{School engagement}.
\newblock \bibinfo{journal}{\emph{What do children need to flourish? Conceptualizing and measuring indicators of positive development}} (\bibinfo{year}{2005}), \bibinfo{pages}{305--321}.
\newblock


\bibitem[Gay et~al\mbox{.}(2016)]%
        {gay2016digital}
\bibfield{author}{\bibinfo{person}{Katrina Gay}, \bibinfo{person}{John Torous}, \bibinfo{person}{Adam Joseph}, \bibinfo{person}{Anand Pandya}, {and} \bibinfo{person}{Ken Duckworth}.} \bibinfo{year}{2016}\natexlab{}.
\newblock \showarticletitle{Digital technology use among individuals with schizophrenia: results of an online survey}.
\newblock \bibinfo{journal}{\emph{JMIR mental health}} \bibinfo{volume}{3}, \bibinfo{number}{2} (\bibinfo{year}{2016}), \bibinfo{pages}{e5379}.
\newblock


\bibitem[Gould et~al\mbox{.}(2013)]%
        {gould2013impact}
\bibfield{author}{\bibinfo{person}{Madelyn~S Gould}, \bibinfo{person}{Wendi Cross}, \bibinfo{person}{Anthony~R Pisani}, \bibinfo{person}{Jimmie~Lou Munfakh}, {and} \bibinfo{person}{Marjorie Kleinman}.} \bibinfo{year}{2013}\natexlab{}.
\newblock \showarticletitle{Impact of applied suicide intervention skills training on the national suicide prevention lifeline}.
\newblock \bibinfo{journal}{\emph{Suicide and Life-Threatening Behavior}} \bibinfo{volume}{43}, \bibinfo{number}{6} (\bibinfo{year}{2013}), \bibinfo{pages}{676--691}.
\newblock


\bibitem[Gray et~al\mbox{.}(2005)]%
        {mi_smoking_1}
\bibfield{author}{\bibinfo{person}{Emily Gray}, \bibinfo{person}{Jim McCambridge}, {and} \bibinfo{person}{John Strang}.} \bibinfo{year}{2005}\natexlab{}.
\newblock \showarticletitle{The effectiveness of motivational interviewing delivered by youth workers in reducing drinking, cigarette and cannabis smoking among young people: quasi-experimental pilot study}.
\newblock \bibinfo{journal}{\emph{Alcohol Alcohol}} \bibinfo{volume}{40}, \bibinfo{number}{6} (\bibinfo{date}{Aug.} \bibinfo{year}{2005}), \bibinfo{pages}{535--539}.
\newblock


\bibitem[Greer et~al\mbox{.}(2019)]%
        {Greer2019UseOT}
\bibfield{author}{\bibinfo{person}{Stephanie Greer}, \bibinfo{person}{Danielle~E. Ramo}, \bibinfo{person}{Yin-Juei Chang}, \bibinfo{person}{Michael Fu}, \bibinfo{person}{Judith~Tedlie Moskowitz}, {and} \bibinfo{person}{Jana Haritatos}.} \bibinfo{year}{2019}\natexlab{}.
\newblock \showarticletitle{Use of the Chatbot “Vivibot” to Deliver Positive Psychology Skills and Promote Well-Being Among Young People After Cancer Treatment: Randomized Controlled Feasibility Trial}.
\newblock \bibinfo{journal}{\emph{JMIR mHealth and uHealth}}  \bibinfo{volume}{7} (\bibinfo{year}{2019}).
\newblock


\bibitem[Han et~al\mbox{.}(2015)]%
        {han2015exploiting}
\bibfield{author}{\bibinfo{person}{Sangdo Han}, \bibinfo{person}{Jeesoo Bang}, \bibinfo{person}{Seonghan Ryu}, {and} \bibinfo{person}{Gary~Geunbae Lee}.} \bibinfo{year}{2015}\natexlab{}.
\newblock \showarticletitle{Exploiting knowledge base to generate responses for natural language dialog listening agents}. In \bibinfo{booktitle}{\emph{Proceedings of the 16th Annual Meeting of the Special Interest Group on Discourse and Dialogue}}. \bibinfo{pages}{129--133}.
\newblock


\bibitem[Han et~al\mbox{.}(2013)]%
        {han2013counseling}
\bibfield{author}{\bibinfo{person}{Sangdo Han}, \bibinfo{person}{Kyusong Lee}, \bibinfo{person}{Donghyeon Lee}, {and} \bibinfo{person}{Gary~Geunbae Lee}.} \bibinfo{year}{2013}\natexlab{}.
\newblock \showarticletitle{Counseling dialog system with 5w1h extraction}. In \bibinfo{booktitle}{\emph{Proceedings of the SIGDIAL 2013 Conference}}. \bibinfo{pages}{349--353}.
\newblock


\bibitem[Haque and Rubya(2023)]%
        {haque2023overview}
\bibfield{author}{\bibinfo{person}{MD~Romael Haque} {and} \bibinfo{person}{Sabirat Rubya}.} \bibinfo{year}{2023}\natexlab{}.
\newblock \showarticletitle{An overview of chatbot-based mobile mental health apps: insights from app description and user reviews}.
\newblock \bibinfo{journal}{\emph{JMIR mHealth and uHealth}} \bibinfo{volume}{11}, \bibinfo{number}{1} (\bibinfo{year}{2023}), \bibinfo{pages}{e44838}.
\newblock


\bibitem[Hard et~al\mbox{.}(2018)]%
        {hard2018federated}
\bibfield{author}{\bibinfo{person}{Andrew Hard}, \bibinfo{person}{Kanishka Rao}, \bibinfo{person}{Rajiv Mathews}, \bibinfo{person}{Swaroop Ramaswamy}, \bibinfo{person}{Fran{\c{c}}oise Beaufays}, \bibinfo{person}{Sean Augenstein}, \bibinfo{person}{Hubert Eichner}, \bibinfo{person}{Chlo{\'e} Kiddon}, {and} \bibinfo{person}{Daniel Ramage}.} \bibinfo{year}{2018}\natexlab{}.
\newblock \showarticletitle{Federated learning for mobile keyboard prediction}.
\newblock \bibinfo{journal}{\emph{arXiv preprint arXiv:1811.03604}} (\bibinfo{year}{2018}).
\newblock


\bibitem[Hettema et~al\mbox{.}(2005)]%
        {mi_abuse_2}
\bibfield{author}{\bibinfo{person}{Jennifer Hettema}, \bibinfo{person}{Julie Steele}, {and} \bibinfo{person}{William~R Miller}.} \bibinfo{year}{2005}\natexlab{}.
\newblock \showarticletitle{Motivational interviewing}.
\newblock \bibinfo{journal}{\emph{Annu Rev Clin Psychol}}  \bibinfo{volume}{1} (\bibinfo{year}{2005}), \bibinfo{pages}{91--111}.
\newblock


\bibitem[Hill et~al\mbox{.}(2007)]%
        {hill2007training}
\bibfield{author}{\bibinfo{person}{Clara~E Hill}, \bibinfo{person}{Jessica Stahl}, {and} \bibinfo{person}{Melissa Roffman}.} \bibinfo{year}{2007}\natexlab{}.
\newblock \showarticletitle{Training novice psychotherapists: Helping skills and beyond.}
\newblock \bibinfo{journal}{\emph{Psychotherapy: Theory, Research, Practice, Training}} \bibinfo{volume}{44}, \bibinfo{number}{4} (\bibinfo{year}{2007}), \bibinfo{pages}{364}.
\newblock


\bibitem[Hochreiter and Schmidhuber(1997)]%
        {hochreiter1997long}
\bibfield{author}{\bibinfo{person}{Sepp Hochreiter} {and} \bibinfo{person}{J{\"u}rgen Schmidhuber}.} \bibinfo{year}{1997}\natexlab{}.
\newblock \showarticletitle{Long short-term memory}.
\newblock \bibinfo{journal}{\emph{Neural computation}} \bibinfo{volume}{9}, \bibinfo{number}{8} (\bibinfo{year}{1997}), \bibinfo{pages}{1735--1780}.
\newblock


\bibitem[Hogan et~al\mbox{.}(2002)]%
        {hogan2002social}
\bibfield{author}{\bibinfo{person}{Brenda~E Hogan}, \bibinfo{person}{Wolfgang Linden}, {and} \bibinfo{person}{Bahman Najarian}.} \bibinfo{year}{2002}\natexlab{}.
\newblock \showarticletitle{Social support interventions: Do they work?}
\newblock \bibinfo{journal}{\emph{Clinical psychology review}} \bibinfo{volume}{22}, \bibinfo{number}{3} (\bibinfo{year}{2002}), \bibinfo{pages}{381--440}.
\newblock


\bibitem[Hohenstein and Jung(2018)]%
        {hohenstein2018ai}
\bibfield{author}{\bibinfo{person}{Jess Hohenstein} {and} \bibinfo{person}{Malte Jung}.} \bibinfo{year}{2018}\natexlab{}.
\newblock \showarticletitle{AI-supported messaging: An investigation of human-human text conversation with AI support}. In \bibinfo{booktitle}{\emph{Extended abstracts of the 2018 CHI conference on human factors in computing systems}}. \bibinfo{pages}{1--6}.
\newblock


\bibitem[Houston et~al\mbox{.}(2002)]%
        {houston2002internet}
\bibfield{author}{\bibinfo{person}{Thomas~K Houston}, \bibinfo{person}{Lisa~A Cooper}, {and} \bibinfo{person}{Daniel~E Ford}.} \bibinfo{year}{2002}\natexlab{}.
\newblock \showarticletitle{Internet support groups for depression: a 1-year prospective cohort study}.
\newblock \bibinfo{journal}{\emph{American Journal of Psychiatry}} \bibinfo{volume}{159}, \bibinfo{number}{12} (\bibinfo{year}{2002}), \bibinfo{pages}{2062--2068}.
\newblock


\bibitem[Huang et~al\mbox{.}(2020)]%
        {huang2020effectiveness}
\bibfield{author}{\bibinfo{person}{Hui-Wen Huang}, \bibinfo{person}{Zehui Li}, {and} \bibinfo{person}{Linda Taylor}.} \bibinfo{year}{2020}\natexlab{}.
\newblock \showarticletitle{The effectiveness of using grammarly to improve students' writing skills}. In \bibinfo{booktitle}{\emph{Proceedings of the 5th International Conference on Distance Education and Learning}}. \bibinfo{pages}{122--127}.
\newblock


\bibitem[Huang et~al\mbox{.}(2018)]%
        {huang2018modeling}
\bibfield{author}{\bibinfo{person}{Xiaolei Huang}, \bibinfo{person}{Lixing Liu}, \bibinfo{person}{Kate Carey}, \bibinfo{person}{Joshua Woolley}, \bibinfo{person}{Stefan Scherer}, {and} \bibinfo{person}{Brian Borsari}.} \bibinfo{year}{2018}\natexlab{}.
\newblock \showarticletitle{Modeling temporality of human intentions by domain adaptation}. In \bibinfo{booktitle}{\emph{Proceedings of the 2018 Conference on Empirical Methods in Natural Language Processing}}. \bibinfo{pages}{696--701}.
\newblock


\bibitem[Imel et~al\mbox{.}(2017)]%
        {imel2017technology}
\bibfield{author}{\bibinfo{person}{Zac~E Imel}, \bibinfo{person}{Derek~D Caperton}, \bibinfo{person}{Michael Tanana}, {and} \bibinfo{person}{David~C Atkins}.} \bibinfo{year}{2017}\natexlab{}.
\newblock \showarticletitle{Technology-enhanced human interaction in psychotherapy.}
\newblock \bibinfo{journal}{\emph{Journal of counseling psychology}} \bibinfo{volume}{64}, \bibinfo{number}{4} (\bibinfo{year}{2017}), \bibinfo{pages}{385}.
\newblock


\bibitem[Inan et~al\mbox{.}(2023)]%
        {inan2023llama}
\bibfield{author}{\bibinfo{person}{Hakan Inan}, \bibinfo{person}{Kartikeya Upasani}, \bibinfo{person}{Jianfeng Chi}, \bibinfo{person}{Rashi Rungta}, \bibinfo{person}{Krithika Iyer}, \bibinfo{person}{Yuning Mao}, \bibinfo{person}{Michael Tontchev}, \bibinfo{person}{Qing Hu}, \bibinfo{person}{Brian Fuller}, \bibinfo{person}{Davide Testuggine}, {et~al\mbox{.}}} \bibinfo{year}{2023}\natexlab{}.
\newblock \showarticletitle{Llama guard: Llm-based input-output safeguard for human-ai conversations}.
\newblock \bibinfo{journal}{\emph{arXiv preprint arXiv:2312.06674}} (\bibinfo{year}{2023}).
\newblock


\bibitem[Islam et~al\mbox{.}(2021)]%
        {islam2021mobile}
\bibfield{author}{\bibinfo{person}{Muhammad~Nazrul Islam}, \bibinfo{person}{Shahriar~Rahman Khan}, \bibinfo{person}{Noor~Nafiz Islam}, \bibinfo{person}{Md Rezwan-A-Rownok}, \bibinfo{person}{Syed~Rohit Zaman}, {and} \bibinfo{person}{Samiha~Raisa Zaman}.} \bibinfo{year}{2021}\natexlab{}.
\newblock \showarticletitle{A mobile application for mental health care during covid-19 pandemic: Development and usability evaluation with system usability scale}. In \bibinfo{booktitle}{\emph{Computational Intelligence in Information Systems: Proceedings of the Computational Intelligence in Information Systems Conference (CIIS 2020)}}. Springer, \bibinfo{pages}{33--42}.
\newblock


\bibitem[Jakesch et~al\mbox{.}(2019)]%
        {trust_on_ai}
\bibfield{author}{\bibinfo{person}{Maurice Jakesch}, \bibinfo{person}{Megan French}, \bibinfo{person}{Xiao Ma}, \bibinfo{person}{Jeffrey~T. Hancock}, {and} \bibinfo{person}{Mor Naaman}.} \bibinfo{year}{2019}\natexlab{}.
\newblock \showarticletitle{AI-Mediated Communication: How the Perception That Profile Text Was Written by AI Affects Trustworthiness}. In \bibinfo{booktitle}{\emph{Proceedings of the 2019 CHI Conference on Human Factors in Computing Systems}} (Glasgow, Scotland Uk) \emph{(\bibinfo{series}{CHI '19})}. \bibinfo{publisher}{Association for Computing Machinery}, \bibinfo{address}{New York, NY, USA}, \bibinfo{pages}{1–13}.
\newblock
\showISBNx{9781450359702}
\urldef\tempurl%
\url{https://doi.org/10.1145/3290605.3300469}
\showDOI{\tempurl}


\bibitem[Jensen et~al\mbox{.}(2011)]%
        {mi_abuse_3}
\bibfield{author}{\bibinfo{person}{Chad~D Jensen}, \bibinfo{person}{Christopher~C Cushing}, \bibinfo{person}{Brandon~S Aylward}, \bibinfo{person}{James~T Craig}, \bibinfo{person}{Danielle~M Sorell}, {and} \bibinfo{person}{Ric~G Steele}.} \bibinfo{year}{2011}\natexlab{}.
\newblock \showarticletitle{Effectiveness of motivational interviewing interventions for adolescent substance use behavior change: a meta-analytic review}.
\newblock \bibinfo{journal}{\emph{J Consult Clin Psychol}} \bibinfo{volume}{79}, \bibinfo{number}{4} (\bibinfo{date}{Aug.} \bibinfo{year}{2011}), \bibinfo{pages}{433--440}.
\newblock


\bibitem[Ji et~al\mbox{.}(2021)]%
        {ji2021spellbert}
\bibfield{author}{\bibinfo{person}{Tuo Ji}, \bibinfo{person}{Hang Yan}, {and} \bibinfo{person}{Xipeng Qiu}.} \bibinfo{year}{2021}\natexlab{}.
\newblock \showarticletitle{s}. In \bibinfo{booktitle}{\emph{Proceedings of the 2021 conference on empirical methods in natural language processing}}. \bibinfo{pages}{3544--3551}.
\newblock


\bibitem[Kannan et~al\mbox{.}(2016)]%
        {kannan2016smart}
\bibfield{author}{\bibinfo{person}{Anjuli Kannan}, \bibinfo{person}{Karol Kurach}, \bibinfo{person}{Sujith Ravi}, \bibinfo{person}{Tobias Kaufmann}, \bibinfo{person}{Andrew Tomkins}, \bibinfo{person}{Balint Miklos}, \bibinfo{person}{Greg Corrado}, \bibinfo{person}{Laszlo Lukacs}, \bibinfo{person}{Marina Ganea}, \bibinfo{person}{Peter Young}, {et~al\mbox{.}}} \bibinfo{year}{2016}\natexlab{}.
\newblock \showarticletitle{Smart reply: Automated response suggestion for email}. In \bibinfo{booktitle}{\emph{Proceedings of the 22nd ACM SIGKDD international conference on knowledge discovery and data mining}}. \bibinfo{pages}{955--964}.
\newblock


\bibitem[Kantor et~al\mbox{.}(2017)]%
        {kantor2017perceived}
\bibfield{author}{\bibinfo{person}{Viktoria Kantor}, \bibinfo{person}{Matthias Knefel}, {and} \bibinfo{person}{Brigitte Lueger-Schuster}.} \bibinfo{year}{2017}\natexlab{}.
\newblock \showarticletitle{Perceived barriers and facilitators of mental health service utilization in adult trauma survivors: A systematic review}.
\newblock \bibinfo{journal}{\emph{Clinical Psychology Review}}  \bibinfo{volume}{52} (\bibinfo{year}{2017}), \bibinfo{pages}{52--68}.
\newblock


\bibitem[Karapapa and Borghi(2015)]%
        {karapapa2015search}
\bibfield{author}{\bibinfo{person}{Stavroula Karapapa} {and} \bibinfo{person}{Maurizio Borghi}.} \bibinfo{year}{2015}\natexlab{}.
\newblock \showarticletitle{Search engine liability for autocomplete suggestions: personality, privacy and the power of the algorithm}.
\newblock \bibinfo{journal}{\emph{International Journal of Law and Information Technology}} \bibinfo{volume}{23}, \bibinfo{number}{3} (\bibinfo{year}{2015}), \bibinfo{pages}{261--289}.
\newblock


\bibitem[Kazdin and Blase(2011)]%
        {kazdin2011rebooting}
\bibfield{author}{\bibinfo{person}{Alan~E Kazdin} {and} \bibinfo{person}{Stacey~L Blase}.} \bibinfo{year}{2011}\natexlab{}.
\newblock \showarticletitle{Rebooting psychotherapy research and practice to reduce the burden of mental illness}.
\newblock \bibinfo{journal}{\emph{Perspectives on psychological science}} \bibinfo{volume}{6}, \bibinfo{number}{1} (\bibinfo{year}{2011}), \bibinfo{pages}{21--37}.
\newblock


\bibitem[Kim et~al\mbox{.}(2023)]%
        {kim2023supporters}
\bibfield{author}{\bibinfo{person}{Meeyun Kim}, \bibinfo{person}{Koustuv Saha}, \bibinfo{person}{Munmun De~Choudhury}, {and} \bibinfo{person}{Daejin Choi}.} \bibinfo{year}{2023}\natexlab{}.
\newblock \showarticletitle{Supporters First: Understanding Online Social Support on Mental Health from a Supporter Perspective}.
\newblock \bibinfo{journal}{\emph{Proceedings of the ACM on Human-Computer Interaction}} \bibinfo{volume}{7}, \bibinfo{number}{CSCW1} (\bibinfo{year}{2023}), \bibinfo{pages}{1--28}.
\newblock


\bibitem[Klonek et~al\mbox{.}(2015)]%
        {klonek2015coding}
\bibfield{author}{\bibinfo{person}{Florian~E Klonek}, \bibinfo{person}{Vicen{\c{c}} Quera}, {and} \bibinfo{person}{Simone Kauffeld}.} \bibinfo{year}{2015}\natexlab{}.
\newblock \showarticletitle{Coding interactions in Motivational Interviewing with computer-software: What are the advantages for process researchers?}
\newblock \bibinfo{journal}{\emph{Computers in Human Behavior}}  \bibinfo{volume}{44} (\bibinfo{year}{2015}), \bibinfo{pages}{284--292}.
\newblock


\bibitem[Kluger and DeNisi(1996)]%
        {kluger1996effects}
\bibfield{author}{\bibinfo{person}{Avraham~N Kluger} {and} \bibinfo{person}{Angelo DeNisi}.} \bibinfo{year}{1996}\natexlab{}.
\newblock \showarticletitle{The effects of feedback interventions on performance: A historical review, a meta-analysis, and a preliminary feedback intervention theory.}
\newblock \bibinfo{journal}{\emph{Psychological bulletin}} \bibinfo{volume}{119}, \bibinfo{number}{2} (\bibinfo{year}{1996}), \bibinfo{pages}{254}.
\newblock


\bibitem[Kosch et~al\mbox{.}(2023)]%
        {kosch2023survey}
\bibfield{author}{\bibinfo{person}{Thomas Kosch}, \bibinfo{person}{Jakob Karolus}, \bibinfo{person}{Johannes Zagermann}, \bibinfo{person}{Harald Reiterer}, \bibinfo{person}{Albrecht Schmidt}, {and} \bibinfo{person}{Pawe{\l}~W Wo{\'z}niak}.} \bibinfo{year}{2023}\natexlab{}.
\newblock \showarticletitle{A survey on measuring cognitive workload in human-computer interaction}.
\newblock \bibinfo{journal}{\emph{Comput. Surveys}} \bibinfo{volume}{55}, \bibinfo{number}{13s} (\bibinfo{year}{2023}), \bibinfo{pages}{1--39}.
\newblock


\bibitem[Kuyper et~al\mbox{.}(2009)]%
        {mi_beh_se_2}
\bibfield{author}{\bibinfo{person}{Lisette Kuyper}, \bibinfo{person}{John de Wit}, \bibinfo{person}{Titia Heijman}, \bibinfo{person}{Han Fennema}, \bibinfo{person}{Jan van Bergen}, {and} \bibinfo{person}{Ine Vanwesenbeeck}.} \bibinfo{year}{2009}\natexlab{}.
\newblock \showarticletitle{Influencing risk behavior of sexually transmitted infection clinic visitors: efficacy of a new methodology of motivational preventive counseling}.
\newblock \bibinfo{journal}{\emph{AIDS Patient Care STDS}} \bibinfo{volume}{23}, \bibinfo{number}{6} (\bibinfo{date}{June} \bibinfo{year}{2009}), \bibinfo{pages}{423--431}.
\newblock


\bibitem[Lateef(2010)]%
        {lateef2010simulation}
\bibfield{author}{\bibinfo{person}{Fatimah Lateef}.} \bibinfo{year}{2010}\natexlab{}.
\newblock \showarticletitle{Simulation-based learning: Just like the real thing}.
\newblock \bibinfo{journal}{\emph{Journal of emergencies, trauma, and shock}} \bibinfo{volume}{3}, \bibinfo{number}{4} (\bibinfo{year}{2010}), \bibinfo{pages}{348--352}.
\newblock


\bibitem[Lee et~al\mbox{.}(2020)]%
        {lee2020translating}
\bibfield{author}{\bibinfo{person}{Eunjung Lee}, \bibinfo{person}{Toula Kourgiantakis}, {and} \bibinfo{person}{Marion Bogo}.} \bibinfo{year}{2020}\natexlab{}.
\newblock \showarticletitle{Translating knowledge into practice: Using simulation to enhance mental health competence through social work education}.
\newblock \bibinfo{journal}{\emph{Social Work Education}} \bibinfo{volume}{39}, \bibinfo{number}{3} (\bibinfo{year}{2020}), \bibinfo{pages}{329--349}.
\newblock


\bibitem[Lewis et~al\mbox{.}(2020)]%
        {BART}
\bibfield{author}{\bibinfo{person}{Mike Lewis}, \bibinfo{person}{Yinhan Liu}, \bibinfo{person}{Naman Goyal}, \bibinfo{person}{Marjan Ghazvininejad}, \bibinfo{person}{Abdelrahman Mohamed}, \bibinfo{person}{Omer Levy}, \bibinfo{person}{Veselin Stoyanov}, {and} \bibinfo{person}{Luke Zettlemoyer}.} \bibinfo{year}{2020}\natexlab{}.
\newblock \showarticletitle{BART: Denoising Sequence-to-Sequence Pre-training for Natural Language Generation, Translation, and Comprehension}. In \bibinfo{booktitle}{\emph{Proceedings of the 58th Annual Meeting of the Association for Computational Linguistics}}. \bibinfo{pages}{7871--7880}.
\newblock


\bibitem[Lin and Hovy(2003)]%
        {ROUGE}
\bibfield{author}{\bibinfo{person}{Chin-Yew Lin} {and} \bibinfo{person}{Eduard Hovy}.} \bibinfo{year}{2003}\natexlab{}.
\newblock \showarticletitle{Automatic evaluation of summaries using n-gram co-occurrence statistics}. In \bibinfo{booktitle}{\emph{Proceedings of the 2003 Human Language Technology Conference of the North American Chapter of the Association for Computational Linguistics}}. \bibinfo{pages}{150--157}.
\newblock


\bibitem[Liu et~al\mbox{.}(2022)]%
        {liu2022will}
\bibfield{author}{\bibinfo{person}{Yihe Liu}, \bibinfo{person}{Anushk Mittal}, \bibinfo{person}{Diyi Yang}, {and} \bibinfo{person}{Amy Bruckman}.} \bibinfo{year}{2022}\natexlab{}.
\newblock \showarticletitle{Will AI console me when I lose my pet? Understanding perceptions of AI-mediated email writing}. In \bibinfo{booktitle}{\emph{Proceedings of the 2022 CHI conference on human factors in computing systems}}. \bibinfo{pages}{1--13}.
\newblock


\bibitem[Lord et~al\mbox{.}(2015)]%
        {lord2015more}
\bibfield{author}{\bibinfo{person}{Sarah~Peregrine Lord}, \bibinfo{person}{Elisa Sheng}, \bibinfo{person}{Zac~E Imel}, \bibinfo{person}{John Baer}, {and} \bibinfo{person}{David~C Atkins}.} \bibinfo{year}{2015}\natexlab{}.
\newblock \showarticletitle{More than reflections: empathy in motivational interviewing includes language style synchrony between therapist and client}.
\newblock \bibinfo{journal}{\emph{Behavior therapy}} \bibinfo{volume}{46}, \bibinfo{number}{3} (\bibinfo{year}{2015}), \bibinfo{pages}{296--303}.
\newblock


\bibitem[Lundahl et~al\mbox{.}(2010)]%
        {effective_MI_1}
\bibfield{author}{\bibinfo{person}{Brad Lundahl}, \bibinfo{person}{Chelsea Kunz}, \bibinfo{person}{Cynthia Brownell}, \bibinfo{person}{Derrik Tollefson}, {and} \bibinfo{person}{Brian Burke}.} \bibinfo{year}{2010}\natexlab{}.
\newblock \showarticletitle{A Meta-Analysis of Motivational Interviewing: Twenty-Five Years of Empirical Studies}.
\newblock \bibinfo{journal}{\emph{Research on Social Work Practice - RES SOCIAL WORK PRAC}}  \bibinfo{volume}{20} (\bibinfo{date}{03} \bibinfo{year}{2010}), \bibinfo{pages}{137--160}.
\newblock
\urldef\tempurl%
\url{https://doi.org/10.1177/1049731509347850}
\showDOI{\tempurl}


\bibitem[Melling and Houguet-Pincham(2011)]%
        {melling2011online}
\bibfield{author}{\bibinfo{person}{Belinda Melling} {and} \bibinfo{person}{Terry Houguet-Pincham}.} \bibinfo{year}{2011}\natexlab{}.
\newblock \showarticletitle{Online peer support for individuals with depression: A summary of current research and future considerations.}
\newblock \bibinfo{journal}{\emph{Psychiatric rehabilitation journal}} \bibinfo{volume}{34}, \bibinfo{number}{3} (\bibinfo{year}{2011}), \bibinfo{pages}{252}.
\newblock


\bibitem[Miller(1996)]%
        {miller1996motivational}
\bibfield{author}{\bibinfo{person}{William~R Miller}.} \bibinfo{year}{1996}\natexlab{}.
\newblock \showarticletitle{Motivational interviewing: research, practice, and puzzles}.
\newblock \bibinfo{journal}{\emph{Addictive behaviors}} \bibinfo{volume}{21}, \bibinfo{number}{6} (\bibinfo{year}{1996}), \bibinfo{pages}{835--842}.
\newblock


\bibitem[Min et~al\mbox{.}(2022)]%
        {min-etal-2022-pair}
\bibfield{author}{\bibinfo{person}{Do~June Min}, \bibinfo{person}{Ver{\'o}nica P{\'e}rez-Rosas}, \bibinfo{person}{Kenneth Resnicow}, {and} \bibinfo{person}{Rada Mihalcea}.} \bibinfo{year}{2022}\natexlab{}.
\newblock \showarticletitle{{PAIR}: Prompt-Aware marg{I}n Ranking for Counselor Reflection Scoring in Motivational Interviewing}. In \bibinfo{booktitle}{\emph{Proceedings of the 2022 Conference on Empirical Methods in Natural Language Processing}}, \bibfield{editor}{\bibinfo{person}{Yoav Goldberg}, \bibinfo{person}{Zornitsa Kozareva}, {and} \bibinfo{person}{Yue Zhang}} (Eds.). \bibinfo{publisher}{Association for Computational Linguistics}, \bibinfo{address}{Abu Dhabi, United Arab Emirates}, \bibinfo{pages}{148--158}.
\newblock
\urldef\tempurl%
\url{https://doi.org/10.18653/v1/2022.emnlp-main.11}
\showDOI{\tempurl}


\bibitem[Miner et~al\mbox{.}(2019)]%
        {miner2019key}
\bibfield{author}{\bibinfo{person}{Adam~S Miner}, \bibinfo{person}{Nigam Shah}, \bibinfo{person}{Kim~D Bullock}, \bibinfo{person}{Bruce~A Arnow}, \bibinfo{person}{Jeremy Bailenson}, {and} \bibinfo{person}{Jeff Hancock}.} \bibinfo{year}{2019}\natexlab{}.
\newblock \showarticletitle{Key considerations for incorporating conversational AI in psychotherapy}.
\newblock \bibinfo{journal}{\emph{Frontiers in psychiatry}}  \bibinfo{volume}{10} (\bibinfo{year}{2019}), \bibinfo{pages}{746}.
\newblock


\bibitem[Mojtabai et~al\mbox{.}(2011)]%
        {mojtabai2011barriers}
\bibfield{author}{\bibinfo{person}{Ramin Mojtabai}, \bibinfo{person}{Mark Olfson}, \bibinfo{person}{Nancy~A Sampson}, \bibinfo{person}{Robert Jin}, \bibinfo{person}{Benjamin Druss}, \bibinfo{person}{Philip~S Wang}, \bibinfo{person}{Kenneth~B Wells}, \bibinfo{person}{Harold~A Pincus}, {and} \bibinfo{person}{Ronald~C Kessler}.} \bibinfo{year}{2011}\natexlab{}.
\newblock \showarticletitle{Barriers to mental health treatment: results from the National Comorbidity Survey Replication}.
\newblock \bibinfo{journal}{\emph{Psychological medicine}} \bibinfo{volume}{41}, \bibinfo{number}{8} (\bibinfo{year}{2011}), \bibinfo{pages}{1751--1761}.
\newblock


\bibitem[Moyers et~al\mbox{.}(2014)]%
        {MITI_4}
\bibfield{author}{\bibinfo{person}{T.B. Moyers}, \bibinfo{person}{J.K. Manuel}, {and} \bibinfo{person}{D. Ernst}.} \bibinfo{year}{2014}\natexlab{}.
\newblock \bibinfo{booktitle}{\emph{Motivational Interviewing Treatment Integrity Coding Manual 4.1. Unpublished manual}}.
\newblock University of New Mexico Center on Alcoholism, Substance Abuse, and Addictions (CASAA).
\newblock


\bibitem[Naslund et~al\mbox{.}(2019)]%
        {naslund2019digital}
\bibfield{author}{\bibinfo{person}{John~A Naslund}, \bibinfo{person}{Rahul Shidhaye}, {and} \bibinfo{person}{Vikram Patel}.} \bibinfo{year}{2019}\natexlab{}.
\newblock \showarticletitle{Digital technology for building capacity of non-specialist health workers for task-sharing and scaling up mental health care globally}.
\newblock \bibinfo{journal}{\emph{Harvard review of psychiatry}} \bibinfo{volume}{27}, \bibinfo{number}{3} (\bibinfo{year}{2019}), \bibinfo{pages}{181}.
\newblock


\bibitem[Nazari et~al\mbox{.}(2021)]%
        {nazari2021application}
\bibfield{author}{\bibinfo{person}{Nabi Nazari}, \bibinfo{person}{Muhammad~Salman Shabbir}, {and} \bibinfo{person}{Roy Setiawan}.} \bibinfo{year}{2021}\natexlab{}.
\newblock \showarticletitle{Application of Artificial Intelligence powered digital writing assistant in higher education: randomized controlled trial}.
\newblock \bibinfo{journal}{\emph{Heliyon}} \bibinfo{volume}{7}, \bibinfo{number}{5} (\bibinfo{year}{2021}).
\newblock


\bibitem[of~Tea(2021)]%
        {7cups}
\bibfield{author}{\bibinfo{person}{7~Cups of Tea}.} \bibinfo{year}{2021}\natexlab{}.
\newblock \bibinfo{booktitle}{\emph{7cups}}.
\newblock
\urldef\tempurl%
\url{https://www.7cups.com/}
\showURL{%
\tempurl}


\bibitem[O'Leary et~al\mbox{.}(2017)]%
        {o2017design}
\bibfield{author}{\bibinfo{person}{Kathleen O'Leary}, \bibinfo{person}{Arpita Bhattacharya}, \bibinfo{person}{Sean~A Munson}, \bibinfo{person}{Jacob~O Wobbrock}, {and} \bibinfo{person}{Wanda Pratt}.} \bibinfo{year}{2017}\natexlab{}.
\newblock \showarticletitle{Design opportunities for mental health peer support technologies}. In \bibinfo{booktitle}{\emph{Proceedings of the 2017 ACM conference on computer supported cooperative work and social computing}}. \bibinfo{pages}{1470--1484}.
\newblock


\bibitem[O'Leary et~al\mbox{.}(2018)]%
        {o2018suddenly}
\bibfield{author}{\bibinfo{person}{Kathleen O'Leary}, \bibinfo{person}{Stephen~M Schueller}, \bibinfo{person}{Jacob~O Wobbrock}, {and} \bibinfo{person}{Wanda Pratt}.} \bibinfo{year}{2018}\natexlab{}.
\newblock \showarticletitle{“Suddenly, we got to become therapists for each other” Designing Peer Support Chats for Mental Health}. In \bibinfo{booktitle}{\emph{Proceedings of the 2018 CHI Conference on Human Factors in Computing Systems}}. \bibinfo{pages}{1--14}.
\newblock


\bibitem[O{'}neil et~al\mbox{.}(2023)]%
        {oneil-etal-2023-automatic}
\bibfield{author}{\bibinfo{person}{Emma O{'}neil}, \bibinfo{person}{Jo{\~a}o Sedoc}, \bibinfo{person}{Diyi Yang}, \bibinfo{person}{Haiyi Zhu}, {and} \bibinfo{person}{Lyle Ungar}.} \bibinfo{year}{2023}\natexlab{}.
\newblock \showarticletitle{Automatic Reflection Generation for Peer-to-Peer Counseling}. In \bibinfo{booktitle}{\emph{Proceedings of the Third Workshop on Natural Language Generation, Evaluation, and Metrics (GEM)}}, \bibfield{editor}{\bibinfo{person}{Sebastian Gehrmann}, \bibinfo{person}{Alex Wang}, \bibinfo{person}{Jo{\~a}o Sedoc}, \bibinfo{person}{Elizabeth Clark}, \bibinfo{person}{Kaustubh Dhole}, \bibinfo{person}{Khyathi~Raghavi Chandu}, \bibinfo{person}{Enrico Santus}, {and} \bibinfo{person}{Hooman Sedghamiz}} (Eds.). \bibinfo{publisher}{Association for Computational Linguistics}, \bibinfo{address}{Singapore}, \bibinfo{pages}{62--75}.
\newblock
\urldef\tempurl%
\url{https://aclanthology.org/2023.gem-1.6}
\showURL{%
\tempurl}


\bibitem[OpenAI(2023)]%
        {openai2023gpt4}
\bibfield{author}{\bibinfo{person}{OpenAI}.} \bibinfo{year}{2023}\natexlab{}.
\newblock \bibinfo{title}{GPT-4 Technical Report}.
\newblock
\newblock
\showeprint[arxiv]{2303.08774}~[cs.CL]


\bibitem[Ouyang et~al\mbox{.}(2022)]%
        {ouyang2022training}
\bibfield{author}{\bibinfo{person}{Long Ouyang}, \bibinfo{person}{Jeffrey Wu}, \bibinfo{person}{Xu Jiang}, \bibinfo{person}{Diogo Almeida}, \bibinfo{person}{Carroll Wainwright}, \bibinfo{person}{Pamela Mishkin}, \bibinfo{person}{Chong Zhang}, \bibinfo{person}{Sandhini Agarwal}, \bibinfo{person}{Katarina Slama}, \bibinfo{person}{Alex Ray}, {et~al\mbox{.}}} \bibinfo{year}{2022}\natexlab{}.
\newblock \showarticletitle{Training language models to follow instructions with human feedback}.
\newblock \bibinfo{journal}{\emph{Advances in Neural Information Processing Systems}}  \bibinfo{volume}{35} (\bibinfo{year}{2022}), \bibinfo{pages}{27730--27744}.
\newblock


\bibitem[Papineni et~al\mbox{.}(2002)]%
        {BLEU}
\bibfield{author}{\bibinfo{person}{Kishore Papineni}, \bibinfo{person}{Salim Roukos}, \bibinfo{person}{Todd Ward}, {and} \bibinfo{person}{Wei-Jing Zhu}.} \bibinfo{year}{2002}\natexlab{}.
\newblock \showarticletitle{Bleu: a method for automatic evaluation of machine translation}. In \bibinfo{booktitle}{\emph{Proceedings of the 40th annual meeting of the Association for Computational Linguistics}}. \bibinfo{pages}{311--318}.
\newblock


\bibitem[Passi and Vorvoreanu(2022)]%
        {passi2022overreliance}
\bibfield{author}{\bibinfo{person}{Samir Passi} {and} \bibinfo{person}{Mihaela Vorvoreanu}.} \bibinfo{year}{2022}\natexlab{}.
\newblock \showarticletitle{Overreliance on AI literature review}.
\newblock \bibinfo{journal}{\emph{Microsoft Research}} (\bibinfo{year}{2022}).
\newblock


\bibitem[Pekrun et~al\mbox{.}(2011)]%
        {pekrun2011measuring}
\bibfield{author}{\bibinfo{person}{Reinhard Pekrun}, \bibinfo{person}{Thomas Goetz}, \bibinfo{person}{Anne~C Frenzel}, \bibinfo{person}{Petra Barchfeld}, {and} \bibinfo{person}{Raymond~P Perry}.} \bibinfo{year}{2011}\natexlab{}.
\newblock \showarticletitle{Measuring emotions in students’ learning and performance: The Achievement Emotions Questionnaire (AEQ)}.
\newblock \bibinfo{journal}{\emph{Contemporary educational psychology}} \bibinfo{volume}{36}, \bibinfo{number}{1} (\bibinfo{year}{2011}), \bibinfo{pages}{36--48}.
\newblock


\bibitem[Peng et~al\mbox{.}(2020)]%
        {peng2020exploring}
\bibfield{author}{\bibinfo{person}{Zhenhui Peng}, \bibinfo{person}{Qingyu Guo}, \bibinfo{person}{Ka~Wing Tsang}, {and} \bibinfo{person}{Xiaojuan Ma}.} \bibinfo{year}{2020}\natexlab{}.
\newblock \showarticletitle{Exploring the Effects of Technological Writing Assistance for Support Providers in Online Mental Health Community}. In \bibinfo{booktitle}{\emph{Proceedings of the 2020 CHI Conference on Human Factors in Computing Systems}}. \bibinfo{pages}{1--15}.
\newblock


\bibitem[P{\'e}rez-Rosas et~al\mbox{.}(2017)]%
        {perez2017predicting}
\bibfield{author}{\bibinfo{person}{Ver{\'o}nica P{\'e}rez-Rosas}, \bibinfo{person}{Rada Mihalcea}, \bibinfo{person}{Kenneth Resnicow}, \bibinfo{person}{Satinder Singh}, \bibinfo{person}{Lawrence An}, \bibinfo{person}{Kathy~J Goggin}, {and} \bibinfo{person}{Delwyn Catley}.} \bibinfo{year}{2017}\natexlab{}.
\newblock \showarticletitle{Predicting counselor behaviors in motivational interviewing encounters}. In \bibinfo{booktitle}{\emph{Proceedings of the 15th Conference of the European Chapter of the Association for Computational Linguistics: Volume 1, Long Papers}}. \bibinfo{pages}{1128--1137}.
\newblock


\bibitem[P{\'e}rez-Rosas et~al\mbox{.}(2019)]%
        {perez2019makes}
\bibfield{author}{\bibinfo{person}{Ver{\'o}nica P{\'e}rez-Rosas}, \bibinfo{person}{Xinyi Wu}, \bibinfo{person}{Kenneth Resnicow}, {and} \bibinfo{person}{Rada Mihalcea}.} \bibinfo{year}{2019}\natexlab{}.
\newblock \showarticletitle{What makes a good counselor? learning to distinguish between high-quality and low-quality counseling conversations}. In \bibinfo{booktitle}{\emph{Proceedings of the 57th Annual Meeting of the Association for Computational Linguistics}}. \bibinfo{pages}{926--935}.
\newblock


\bibitem[Powell et~al\mbox{.}(2007)]%
        {powell2007investigating}
\bibfield{author}{\bibinfo{person}{John Powell}, \bibinfo{person}{Aileen Clarke}, {et~al\mbox{.}}} \bibinfo{year}{2007}\natexlab{}.
\newblock \showarticletitle{Investigating internet use by mental health service users: interview study}. In \bibinfo{booktitle}{\emph{Medinfo 2007: Proceedings of the 12th World Congress on Health (Medical) Informatics; Building Sustainable Health Systems}}. IOS Press, \bibinfo{pages}{1112}.
\newblock


\bibitem[Pruksachatkun et~al\mbox{.}(2019)]%
        {pruksachatkun2019moments}
\bibfield{author}{\bibinfo{person}{Yada Pruksachatkun}, \bibinfo{person}{Sachin~R Pendse}, {and} \bibinfo{person}{Amit Sharma}.} \bibinfo{year}{2019}\natexlab{}.
\newblock \showarticletitle{Moments of change: Analyzing peer-based cognitive support in online mental health forums}. In \bibinfo{booktitle}{\emph{Proceedings of the 2019 CHI Conference on Human Factors in Computing Systems}}. \bibinfo{pages}{1--13}.
\newblock


\bibitem[Quinn and Zhai(2016)]%
        {quinn2016cost}
\bibfield{author}{\bibinfo{person}{Philip Quinn} {and} \bibinfo{person}{Shumin Zhai}.} \bibinfo{year}{2016}\natexlab{}.
\newblock \showarticletitle{A cost-benefit study of text entry suggestion interaction}. In \bibinfo{booktitle}{\emph{Proceedings of the 2016 CHI conference on human factors in computing systems}}. \bibinfo{pages}{83--88}.
\newblock


\bibitem[Radford et~al\mbox{.}(2019)]%
        {GPT2}
\bibfield{author}{\bibinfo{person}{Alec Radford}, \bibinfo{person}{Jeffrey Wu}, \bibinfo{person}{Rewon Child}, \bibinfo{person}{David Luan}, \bibinfo{person}{Dario Amodei}, \bibinfo{person}{Ilya Sutskever}, {et~al\mbox{.}}} \bibinfo{year}{2019}\natexlab{}.
\newblock \showarticletitle{Language models are unsupervised multitask learners}.
\newblock \bibinfo{journal}{\emph{OpenAI blog}} \bibinfo{volume}{1}, \bibinfo{number}{8} (\bibinfo{year}{2019}), \bibinfo{pages}{9}.
\newblock


\bibitem[Raffel et~al\mbox{.}(2020)]%
        {raffel2020exploring}
\bibfield{author}{\bibinfo{person}{Colin Raffel}, \bibinfo{person}{Noam Shazeer}, \bibinfo{person}{Adam Roberts}, \bibinfo{person}{Katherine Lee}, \bibinfo{person}{Sharan Narang}, \bibinfo{person}{Michael Matena}, \bibinfo{person}{Yanqi Zhou}, \bibinfo{person}{Wei Li}, {and} \bibinfo{person}{Peter~J Liu}.} \bibinfo{year}{2020}\natexlab{}.
\newblock \showarticletitle{Exploring the limits of transfer learning with a unified text-to-text transformer}.
\newblock \bibinfo{journal}{\emph{The Journal of Machine Learning Research}} \bibinfo{volume}{21}, \bibinfo{number}{1} (\bibinfo{year}{2020}), \bibinfo{pages}{5485--5551}.
\newblock


\bibitem[Raisch and Krakowski(2021)]%
        {raisch2021artificial}
\bibfield{author}{\bibinfo{person}{Sebastian Raisch} {and} \bibinfo{person}{Sebastian Krakowski}.} \bibinfo{year}{2021}\natexlab{}.
\newblock \showarticletitle{Artificial intelligence and management: The automation--augmentation paradox}.
\newblock \bibinfo{journal}{\emph{Academy of management review}} \bibinfo{volume}{46}, \bibinfo{number}{1} (\bibinfo{year}{2021}), \bibinfo{pages}{192--210}.
\newblock


\bibitem[Rashkin et~al\mbox{.}(2019)]%
        {EmpatheticDialogues}
\bibfield{author}{\bibinfo{person}{Hannah Rashkin}, \bibinfo{person}{Eric~Michael Smith}, \bibinfo{person}{Margaret Li}, {and} \bibinfo{person}{Y-Lan Boureau}.} \bibinfo{year}{2019}\natexlab{}.
\newblock \showarticletitle{Towards Empathetic Open-domain Conversation Models: A New Benchmark and Dataset}. In \bibinfo{booktitle}{\emph{Proceedings of the 57th Annual Meeting of the Association for Computational Linguistics}}. \bibinfo{pages}{5370--5381}.
\newblock


\bibitem[Reagans et~al\mbox{.}(2005)]%
        {reagans2005individual}
\bibfield{author}{\bibinfo{person}{Ray Reagans}, \bibinfo{person}{Linda Argote}, {and} \bibinfo{person}{Daria Brooks}.} \bibinfo{year}{2005}\natexlab{}.
\newblock \showarticletitle{Individual experience and experience working together: Predicting learning rates from knowing who knows what and knowing how to work together}.
\newblock \bibinfo{journal}{\emph{Management science}} \bibinfo{volume}{51}, \bibinfo{number}{6} (\bibinfo{year}{2005}), \bibinfo{pages}{869--881}.
\newblock


\bibitem[Repper and Carter(2011)]%
        {repper2011review}
\bibfield{author}{\bibinfo{person}{Julie Repper} {and} \bibinfo{person}{Tim Carter}.} \bibinfo{year}{2011}\natexlab{}.
\newblock \showarticletitle{A review of the literature on peer support in mental health services}.
\newblock \bibinfo{journal}{\emph{Journal of mental health}} \bibinfo{volume}{20}, \bibinfo{number}{4} (\bibinfo{year}{2011}), \bibinfo{pages}{392--411}.
\newblock


\bibitem[Reynolds and Perrin(2004)]%
        {reynolds2004mismatches}
\bibfield{author}{\bibinfo{person}{Julie~S Reynolds} {and} \bibinfo{person}{Nancy~A Perrin}.} \bibinfo{year}{2004}\natexlab{}.
\newblock \showarticletitle{Mismatches in social support and psychosocial adjustment to breast cancer.}
\newblock \bibinfo{journal}{\emph{Health Psychology}} \bibinfo{volume}{23}, \bibinfo{number}{4} (\bibinfo{year}{2004}), \bibinfo{pages}{425}.
\newblock


\bibitem[Rickwood et~al\mbox{.}(2021)]%
        {rickwood2021online}
\bibfield{author}{\bibinfo{person}{Debra Rickwood}, \bibinfo{person}{Vanessa Kennedy}, \bibinfo{person}{Koki Miyazaki}, \bibinfo{person}{Nic Telford}, \bibinfo{person}{Stephen Carbone}, \bibinfo{person}{Ella Hewitt}, {and} \bibinfo{person}{Carolyn Watts}.} \bibinfo{year}{2021}\natexlab{}.
\newblock \showarticletitle{An online platform to provide work and study support for young people with mental health challenges: observational and survey study}.
\newblock \bibinfo{journal}{\emph{JMIR Mental Health}} \bibinfo{volume}{8}, \bibinfo{number}{2} (\bibinfo{year}{2021}), \bibinfo{pages}{e21872}.
\newblock


\bibitem[Robertson et~al\mbox{.}(2021)]%
        {robertson2021can}
\bibfield{author}{\bibinfo{person}{Ronald~E Robertson}, \bibinfo{person}{Alexandra Olteanu}, \bibinfo{person}{Fernando Diaz}, \bibinfo{person}{Milad Shokouhi}, {and} \bibinfo{person}{Peter Bailey}.} \bibinfo{year}{2021}\natexlab{}.
\newblock \showarticletitle{“I can’t reply with that”: Characterizing problematic email reply suggestions}. In \bibinfo{booktitle}{\emph{Proceedings of the 2021 CHI Conference on Human Factors in Computing Systems}}. \bibinfo{pages}{1--18}.
\newblock


\bibitem[Rodgers(2010)]%
        {rodgers2010review}
\bibfield{author}{\bibinfo{person}{Philip~L Rodgers}.} \bibinfo{year}{2010}\natexlab{}.
\newblock \bibinfo{booktitle}{\emph{Review of the applied suicide intervention skills training program (ASIST): rationale, evaluation results, and directions for future research}}.
\newblock \bibinfo{publisher}{LivingWorks Education Incorporated Calgary, Alberta, Canada}.
\newblock


\bibitem[Roller et~al\mbox{.}(2021)]%
        {roller2021recipes}
\bibfield{author}{\bibinfo{person}{Stephen Roller}, \bibinfo{person}{Emily Dinan}, \bibinfo{person}{Naman Goyal}, \bibinfo{person}{Da Ju}, \bibinfo{person}{Mary Williamson}, \bibinfo{person}{Yinhan Liu}, \bibinfo{person}{Jing Xu}, \bibinfo{person}{Myle Ott}, \bibinfo{person}{Eric~Michael Smith}, \bibinfo{person}{Y-Lan Boureau}, {et~al\mbox{.}}} \bibinfo{year}{2021}\natexlab{}.
\newblock \showarticletitle{Recipes for Building an Open-Domain Chatbot}. In \bibinfo{booktitle}{\emph{Proceedings of the 16th Conference of the European Chapter of the Association for Computational Linguistics: Main Volume}}. \bibinfo{pages}{300--325}.
\newblock


\bibitem[R{\o}nning and Bj{\o}rkly(2019)]%
        {ronning2019use}
\bibfield{author}{\bibinfo{person}{Solrun~Brenk R{\o}nning} {and} \bibinfo{person}{St{\aa}l Bj{\o}rkly}.} \bibinfo{year}{2019}\natexlab{}.
\newblock \showarticletitle{The use of clinical role-play and reflection in learning therapeutic communication skills in mental health education: an integrative review}.
\newblock \bibinfo{journal}{\emph{Advances in medical education and practice}} (\bibinfo{year}{2019}), \bibinfo{pages}{415--425}.
\newblock


\bibitem[Rozental(2016)]%
        {rozental2016negative}
\bibfield{author}{\bibinfo{person}{Alexander Rozental}.} \bibinfo{year}{2016}\natexlab{}.
\newblock \emph{\bibinfo{title}{Negative effects of Internet-based cognitive behavior therapy: Monitoring and reporting deterioration and adverse and unwanted events}}.
\newblock \bibinfo{thesistype}{Ph.\,D. Dissertation}. \bibinfo{school}{Department of Psychology, Stockholm University}.
\newblock


\bibitem[Rumelhart et~al\mbox{.}(1986)]%
        {rumelhart1986learning}
\bibfield{author}{\bibinfo{person}{David~E Rumelhart}, \bibinfo{person}{Geoffrey~E Hinton}, {and} \bibinfo{person}{Ronald~J Williams}.} \bibinfo{year}{1986}\natexlab{}.
\newblock \showarticletitle{Learning representations by back-propagating errors}.
\newblock \bibinfo{journal}{\emph{nature}} \bibinfo{volume}{323}, \bibinfo{number}{6088} (\bibinfo{year}{1986}), \bibinfo{pages}{533--536}.
\newblock


\bibitem[Saha et~al\mbox{.}(2022)]%
        {saha2022towards}
\bibfield{author}{\bibinfo{person}{Tulika Saha}, \bibinfo{person}{Vaibhav Gakhreja}, \bibinfo{person}{Anindya~Sundar Das}, \bibinfo{person}{Souhitya Chakraborty}, {and} \bibinfo{person}{Sriparna Saha}.} \bibinfo{year}{2022}\natexlab{}.
\newblock \showarticletitle{Towards Motivational and Empathetic Response Generation in Online Mental Health Support}. In \bibinfo{booktitle}{\emph{Proceedings of the 45th International ACM SIGIR Conference on Research and Development in Information Retrieval}}. \bibinfo{pages}{2650–2656}.
\newblock


\bibitem[Schleider et~al\mbox{.}(2020)]%
        {schleider2020acceptability}
\bibfield{author}{\bibinfo{person}{Jessica~Lee Schleider}, \bibinfo{person}{Mallory Dobias}, \bibinfo{person}{Jenna Sung}, \bibinfo{person}{Emma Mumper}, {and} \bibinfo{person}{Michael~C Mullarkey}.} \bibinfo{year}{2020}\natexlab{}.
\newblock \showarticletitle{Acceptability and utility of an open-access, online single-session intervention platform for adolescent mental health}.
\newblock \bibinfo{journal}{\emph{JMIR mental health}} \bibinfo{volume}{7}, \bibinfo{number}{6} (\bibinfo{year}{2020}), \bibinfo{pages}{e20513}.
\newblock


\bibitem[Schwalbe et~al\mbox{.}(2014)]%
        {schwalbe2014sustaining}
\bibfield{author}{\bibinfo{person}{Craig~S Schwalbe}, \bibinfo{person}{Hans~Y Oh}, {and} \bibinfo{person}{Allen Zweben}.} \bibinfo{year}{2014}\natexlab{}.
\newblock \showarticletitle{Sustaining motivational interviewing: A meta-analysis of training studies}.
\newblock \bibinfo{journal}{\emph{Addiction}} \bibinfo{volume}{109}, \bibinfo{number}{8} (\bibinfo{year}{2014}), \bibinfo{pages}{1287--1294}.
\newblock


\bibitem[Shah et~al\mbox{.}(2022)]%
        {shah2022modeling}
\bibfield{author}{\bibinfo{person}{Raj~Sanjay Shah}, \bibinfo{person}{Faye Holt}, \bibinfo{person}{Shirley~Anugrah Hayati}, \bibinfo{person}{Aastha Agarwal}, \bibinfo{person}{Yi-Chia Wang}, \bibinfo{person}{Robert~E Kraut}, {and} \bibinfo{person}{Diyi Yang}.} \bibinfo{year}{2022}\natexlab{}.
\newblock \showarticletitle{Modeling Motivational Interviewing Strategies On An Online Peer-to-Peer Counseling Platform}.
\newblock \bibinfo{journal}{\emph{Proceedings of the ACM on Human-Computer Interaction}} \bibinfo{volume}{6}, \bibinfo{number}{CSCW2} (\bibinfo{year}{2022}), \bibinfo{pages}{1--24}.
\newblock


\bibitem[Shang et~al\mbox{.}(2015)]%
        {shang2015neural}
\bibfield{author}{\bibinfo{person}{Lifeng Shang}, \bibinfo{person}{Zhengdong Lu}, {and} \bibinfo{person}{Hang Li}.} \bibinfo{year}{2015}\natexlab{}.
\newblock \showarticletitle{Neural Responding Machine for Short-Text Conversation}. In \bibinfo{booktitle}{\emph{Proceedings of the 53rd Annual Meeting of the Association for Computational Linguistics and the 7th International Joint Conference on Natural Language Processing (Volume 1: Long Papers)}}. \bibinfo{pages}{1577--1586}.
\newblock


\bibitem[Sharma et~al\mbox{.}(2021)]%
        {sharma2021towards}
\bibfield{author}{\bibinfo{person}{Ashish Sharma}, \bibinfo{person}{Inna~W Lin}, \bibinfo{person}{Adam~S Miner}, \bibinfo{person}{David~C Atkins}, {and} \bibinfo{person}{Tim Althoff}.} \bibinfo{year}{2021}\natexlab{}.
\newblock \showarticletitle{Towards facilitating empathic conversations in online mental health support: A reinforcement learning approach}. In \bibinfo{booktitle}{\emph{Proceedings of the Web Conference 2021}}. \bibinfo{pages}{194--205}.
\newblock


\bibitem[Sharma et~al\mbox{.}(2023)]%
        {sharma2023human}
\bibfield{author}{\bibinfo{person}{Ashish Sharma}, \bibinfo{person}{Inna~W Lin}, \bibinfo{person}{Adam~S Miner}, \bibinfo{person}{David~C Atkins}, {and} \bibinfo{person}{Tim Althoff}.} \bibinfo{year}{2023}\natexlab{}.
\newblock \showarticletitle{Human--AI collaboration enables more empathic conversations in text-based peer-to-peer mental health support}.
\newblock \bibinfo{journal}{\emph{Nature Machine Intelligence}} \bibinfo{volume}{5}, \bibinfo{number}{1} (\bibinfo{year}{2023}), \bibinfo{pages}{46--57}.
\newblock


\bibitem[Sharma and De~Choudhury(2018)]%
        {linguistic_accommodation}
\bibfield{author}{\bibinfo{person}{Eva Sharma} {and} \bibinfo{person}{Munmun De~Choudhury}.} \bibinfo{year}{2018}\natexlab{}.
\newblock \bibinfo{booktitle}{\emph{Mental Health Support and Its Relationship to Linguistic Accommodation in Online Communities}}.
\newblock \bibinfo{publisher}{Association for Computing Machinery}, \bibinfo{address}{New York, NY, USA}, \bibinfo{pages}{1–13}.
\newblock
\showISBNx{9781450356206}
\urldef\tempurl%
\url{https://doi.org/10.1145/3173574.3174215}
\showURL{%
\tempurl}


\bibitem[Shen et~al\mbox{.}(2020)]%
        {shen2020counseling}
\bibfield{author}{\bibinfo{person}{Siqi Shen}, \bibinfo{person}{Charles Welch}, \bibinfo{person}{Rada Mihalcea}, {and} \bibinfo{person}{Ver{\'o}nica P{\'e}rez-Rosas}.} \bibinfo{year}{2020}\natexlab{}.
\newblock \showarticletitle{Counseling-style reflection generation using generative pretrained transformers with augmented context}. In \bibinfo{booktitle}{\emph{Proceedings of the 21th Annual Meeting of the Special Interest Group on Discourse and Dialogue}}. \bibinfo{pages}{10--20}.
\newblock


\bibitem[Sinha et~al\mbox{.}(2001)]%
        {recommender_system}
\bibfield{author}{\bibinfo{person}{Swearingen Sinha}, \bibinfo{person}{Kirsten Medhurst}, {and} \bibinfo{person}{Rashmi Sinha}.} \bibinfo{year}{2001}\natexlab{}.
\newblock \showarticletitle{Beyond algorithms: An HCI perspective on recommender systems}. In \bibinfo{booktitle}{\emph{ACM SIGIR 2001 workshop on recommender systems}}, Vol.~\bibinfo{volume}{13}. \bibinfo{pages}{1--11}.
\newblock


\bibitem[Smithson et~al\mbox{.}(2011)]%
        {smithson2011problem}
\bibfield{author}{\bibinfo{person}{Janet Smithson}, \bibinfo{person}{Siobhan Sharkey}, \bibinfo{person}{Elaine Hewis}, \bibinfo{person}{Ray Jones}, \bibinfo{person}{Tobit Emmens}, \bibinfo{person}{Tamsin Ford}, {and} \bibinfo{person}{Christabel Owens}.} \bibinfo{year}{2011}\natexlab{}.
\newblock \showarticletitle{Problem presentation and responses on an online forum for young people who self-harm}.
\newblock \bibinfo{journal}{\emph{Discourse studies}} \bibinfo{volume}{13}, \bibinfo{number}{4} (\bibinfo{year}{2011}), \bibinfo{pages}{487--501}.
\newblock


\bibitem[S{\o}vold et~al\mbox{.}(2021)]%
        {sovold2021prioritizing}
\bibfield{author}{\bibinfo{person}{Lene~E S{\o}vold}, \bibinfo{person}{John~A Naslund}, \bibinfo{person}{Antonis~A Kousoulis}, \bibinfo{person}{Shekhar Saxena}, \bibinfo{person}{M~Walid Qoronfleh}, \bibinfo{person}{Christoffel Grobler}, {and} \bibinfo{person}{Lars M{\"u}nter}.} \bibinfo{year}{2021}\natexlab{}.
\newblock \showarticletitle{Prioritizing the Mental Health and Well-Being of Healthcare Workers: An Urgent Global Public Health Priority}.
\newblock \bibinfo{journal}{\emph{Frontiers in public health}}  \bibinfo{volume}{9} (\bibinfo{year}{2021}).
\newblock


\bibitem[Stade et~al\mbox{.}(2023)]%
        {stade2023artificial}
\bibfield{author}{\bibinfo{person}{Elizabeth Stade}, \bibinfo{person}{Shannon~Wiltsey Stirman}, \bibinfo{person}{Lyle~H Ungar}, \bibinfo{person}{H~Andrew Schwartz}, \bibinfo{person}{David~Bryce Yaden}, \bibinfo{person}{Jo{\~a}o Sedoc}, \bibinfo{person}{Robert DeRubeis}, \bibinfo{person}{Robb Willer}, {et~al\mbox{.}}} \bibinfo{year}{2023}\natexlab{}.
\newblock \showarticletitle{Artificial intelligence will change the future of psychotherapy: A proposal for responsible, psychologist-led development}.
\newblock  (\bibinfo{year}{2023}).
\newblock


\bibitem[Tanana et~al\mbox{.}(2016)]%
        {tanana2016comparison}
\bibfield{author}{\bibinfo{person}{Michael Tanana}, \bibinfo{person}{Kevin~A Hallgren}, \bibinfo{person}{Zac~E Imel}, \bibinfo{person}{David~C Atkins}, {and} \bibinfo{person}{Vivek Srikumar}.} \bibinfo{year}{2016}\natexlab{}.
\newblock \showarticletitle{A comparison of natural language processing methods for automated coding of motivational interviewing}.
\newblock \bibinfo{journal}{\emph{Journal of substance abuse treatment}}  \bibinfo{volume}{65} (\bibinfo{year}{2016}), \bibinfo{pages}{43--50}.
\newblock


\bibitem[Tanana et~al\mbox{.}(2019)]%
        {tanana2019development}
\bibfield{author}{\bibinfo{person}{Michael~J Tanana}, \bibinfo{person}{Christina~S Soma}, \bibinfo{person}{Vivek Srikumar}, \bibinfo{person}{David~C Atkins}, {and} \bibinfo{person}{Zac~E Imel}.} \bibinfo{year}{2019}\natexlab{}.
\newblock \showarticletitle{Development and evaluation of ClientBot: Patient-like conversational agent to train basic counseling skills}.
\newblock \bibinfo{journal}{\emph{Journal of medical Internet research}} \bibinfo{volume}{21}, \bibinfo{number}{7} (\bibinfo{year}{2019}), \bibinfo{pages}{e12529}.
\newblock


\bibitem[Th{\'e}riault et~al\mbox{.}(2009)]%
        {theriault2009feelings}
\bibfield{author}{\bibinfo{person}{Anne Th{\'e}riault}, \bibinfo{person}{Nicola Gazzola}, {and} \bibinfo{person}{Brian Richardson}.} \bibinfo{year}{2009}\natexlab{}.
\newblock \showarticletitle{Feelings of incompetence in novice therapists: Consequences, coping, and correctives}.
\newblock \bibinfo{journal}{\emph{Canadian Journal of Counselling and Psychotherapy}} \bibinfo{volume}{43}, \bibinfo{number}{2} (\bibinfo{year}{2009}).
\newblock


\bibitem[Trnka et~al\mbox{.}(2009)]%
        {trnka2009user}
\bibfield{author}{\bibinfo{person}{Keith Trnka}, \bibinfo{person}{John McCaw}, \bibinfo{person}{Debra Yarrington}, \bibinfo{person}{Kathleen~F McCoy}, {and} \bibinfo{person}{Christopher Pennington}.} \bibinfo{year}{2009}\natexlab{}.
\newblock \showarticletitle{User interaction with word prediction: The effects of prediction quality}.
\newblock \bibinfo{journal}{\emph{ACM Transactions on Accessible Computing (TACCESS)}} \bibinfo{volume}{1}, \bibinfo{number}{3} (\bibinfo{year}{2009}), \bibinfo{pages}{1--34}.
\newblock


\bibitem[Vaswani et~al\mbox{.}(2017)]%
        {vaswani2017attention}
\bibfield{author}{\bibinfo{person}{Ashish Vaswani}, \bibinfo{person}{Noam Shazeer}, \bibinfo{person}{Niki Parmar}, \bibinfo{person}{Jakob Uszkoreit}, \bibinfo{person}{Llion Jones}, \bibinfo{person}{Aidan~N Gomez}, \bibinfo{person}{{\L}ukasz Kaiser}, {and} \bibinfo{person}{Illia Polosukhin}.} \bibinfo{year}{2017}\natexlab{}.
\newblock \showarticletitle{Attention is all you need}.
\newblock \bibinfo{journal}{\emph{Advances in neural information processing systems}}  \bibinfo{volume}{30} (\bibinfo{year}{2017}).
\newblock


\bibitem[Vinyals and Le(2015)]%
        {vinyals2015neural}
\bibfield{author}{\bibinfo{person}{Oriol Vinyals} {and} \bibinfo{person}{Quoc Le}.} \bibinfo{year}{2015}\natexlab{}.
\newblock \showarticletitle{A neural conversational model}.
\newblock \bibinfo{journal}{\emph{ICML Deep Learning Workshop}} (\bibinfo{year}{2015}).
\newblock


\bibitem[Wang and Alonso(2022)]%
        {wang2022clozesearch}
\bibfield{author}{\bibinfo{person}{Mengru Wang} {and} \bibinfo{person}{Omar Alonso}.} \bibinfo{year}{2022}\natexlab{}.
\newblock \showarticletitle{ClozeSearch: A Collocation Retrieval Application to Assist in Scientific Writing}. In \bibinfo{booktitle}{\emph{Proceedings of the 31st ACM International Conference on Information \& Knowledge Management}}. \bibinfo{pages}{5009--5013}.
\newblock


\bibitem[Wang et~al\mbox{.}(2023)]%
        {wang2023metrics}
\bibfield{author}{\bibinfo{person}{Tony Wang}, \bibinfo{person}{Haard~K Shah}, \bibinfo{person}{Raj~Sanjay Shah}, \bibinfo{person}{Yi-Chia Wang}, \bibinfo{person}{Robert~E Kraut}, {and} \bibinfo{person}{Diyi Yang}.} \bibinfo{year}{2023}\natexlab{}.
\newblock \showarticletitle{Metrics for Peer Counseling: Triangulating Success Outcomes for Online Therapy Platforms}. In \bibinfo{booktitle}{\emph{Proceedings of the 2023 CHI Conference on Human Factors in Computing Systems}}. \bibinfo{pages}{1--17}.
\newblock


\bibitem[Wang et~al\mbox{.}(2015)]%
        {wang2015eliciting}
\bibfield{author}{\bibinfo{person}{Yi-Chia Wang}, \bibinfo{person}{Robert~E Kraut}, {and} \bibinfo{person}{John~M Levine}.} \bibinfo{year}{2015}\natexlab{}.
\newblock \showarticletitle{Eliciting and receiving online support: using computer-aided content analysis to examine the dynamics of online social support}.
\newblock \bibinfo{journal}{\emph{Journal of medical Internet research}} \bibinfo{volume}{17}, \bibinfo{number}{4} (\bibinfo{year}{2015}), \bibinfo{pages}{e99}.
\newblock


\bibitem[Wilson et~al\mbox{.}(2024)]%
        {wilson2024creating}
\bibfield{author}{\bibinfo{person}{Cara Wilson}, \bibinfo{person}{Larissa Pschetz}, \bibinfo{person}{Billy Dixon}, \bibinfo{person}{Sue~J Lewis}, \bibinfo{person}{Joe Revans}, {and} \bibinfo{person}{John Vines}.} \bibinfo{year}{2024}\natexlab{}.
\newblock \showarticletitle{Creating Resources for Designing with and for Care Ecologies in HCI}. In \bibinfo{booktitle}{\emph{Proceedings of the 2024 ACM Designing Interactive Systems Conference}}. \bibinfo{pages}{3161--3178}.
\newblock


\bibitem[Wolf et~al\mbox{.}(2020)]%
        {HuggingFace}
\bibfield{author}{\bibinfo{person}{Thomas Wolf}, \bibinfo{person}{Lysandre Debut}, \bibinfo{person}{Victor Sanh}, \bibinfo{person}{Julien Chaumond}, \bibinfo{person}{Clement Delangue}, \bibinfo{person}{Anthony Moi}, \bibinfo{person}{Pierric Cistac}, \bibinfo{person}{Tim Rault}, \bibinfo{person}{Rémi Louf}, \bibinfo{person}{Morgan Funtowicz}, \bibinfo{person}{Joe Davison}, \bibinfo{person}{Sam Shleifer}, \bibinfo{person}{Patrick von Platen}, \bibinfo{person}{Clara Ma}, \bibinfo{person}{Yacine Jernite}, \bibinfo{person}{Julien Plu}, \bibinfo{person}{Canwen Xu}, \bibinfo{person}{Teven~Le Scao}, \bibinfo{person}{Sylvain Gugger}, \bibinfo{person}{Mariama Drame}, \bibinfo{person}{Quentin Lhoest}, {and} \bibinfo{person}{Alexander~M. Rush}.} \bibinfo{year}{2020}\natexlab{}.
\newblock \showarticletitle{Transformers: State-of-the-art natural language processing}. In \bibinfo{booktitle}{\emph{Proceedings of the 2020 conference on empirical methods in natural language processing: system demonstrations}}. \bibinfo{pages}{38--45}.
\newblock


\bibitem[Woodnutt et~al\mbox{.}(2024)]%
        {woodnutt2024could}
\bibfield{author}{\bibinfo{person}{Samuel Woodnutt}, \bibinfo{person}{Chris Allen}, \bibinfo{person}{Jasmine Snowden}, \bibinfo{person}{Matt Flynn}, \bibinfo{person}{Simon Hall}, \bibinfo{person}{Paula Libberton}, {and} \bibinfo{person}{Francesca Purvis}.} \bibinfo{year}{2024}\natexlab{}.
\newblock \showarticletitle{Could artificial intelligence write mental health nursing care plans?}
\newblock \bibinfo{journal}{\emph{Journal of Psychiatric and Mental Health Nursing}} \bibinfo{volume}{31}, \bibinfo{number}{1} (\bibinfo{year}{2024}), \bibinfo{pages}{79--86}.
\newblock


\bibitem[Xiao et~al\mbox{.}(2014)]%
        {xiao2014modeling}
\bibfield{author}{\bibinfo{person}{Bo Xiao}, \bibinfo{person}{Daniel Bone}, \bibinfo{person}{Maarten~Van Segbroeck}, \bibinfo{person}{Zac~E Imel}, \bibinfo{person}{David~C Atkins}, \bibinfo{person}{Panayiotis~G Georgiou}, {and} \bibinfo{person}{Shrikanth~S Narayanan}.} \bibinfo{year}{2014}\natexlab{}.
\newblock \showarticletitle{Modeling therapist empathy through prosody in drug addiction counseling}. In \bibinfo{booktitle}{\emph{Fifteenth annual conference of the international speech communication association}}.
\newblock


\bibitem[Xu et~al\mbox{.}(2022)]%
        {xu2022beyond}
\bibfield{author}{\bibinfo{person}{Jing Xu}, \bibinfo{person}{Arthur Szlam}, {and} \bibinfo{person}{Jason Weston}.} \bibinfo{year}{2022}\natexlab{}.
\newblock \showarticletitle{Beyond Goldfish Memory: Long-Term Open-Domain Conversation}. In \bibinfo{booktitle}{\emph{Proceedings of the 60th Annual Meeting of the Association for Computational Linguistics (Volume 1: Long Papers)}}. \bibinfo{pages}{5180--5197}.
\newblock


\bibitem[Yang et~al\mbox{.}(2024b)]%
        {yang2024social}
\bibfield{author}{\bibinfo{person}{Diyi Yang}, \bibinfo{person}{Caleb Ziems}, \bibinfo{person}{William Held}, \bibinfo{person}{Omar Shaikh}, \bibinfo{person}{Michael~S Bernstein}, {and} \bibinfo{person}{John Mitchell}.} \bibinfo{year}{2024}\natexlab{b}.
\newblock \showarticletitle{Social skill training with large language models}.
\newblock \bibinfo{journal}{\emph{arXiv preprint arXiv:2404.04204}} (\bibinfo{year}{2024}).
\newblock


\bibitem[Yang et~al\mbox{.}(2024a)]%
        {yang2024makes}
\bibfield{author}{\bibinfo{person}{Wenjie Yang}, \bibinfo{person}{Anna Fang}, \bibinfo{person}{Raj~Sanjay Shah}, \bibinfo{person}{Yash Mathur}, \bibinfo{person}{Diyi Yang}, \bibinfo{person}{Haiyi Zhu}, {and} \bibinfo{person}{Robert~E Kraut}.} \bibinfo{year}{2024}\natexlab{a}.
\newblock \showarticletitle{What Makes Digital Support Effective? How Therapeutic Skills Affect Clinical Well-Being}.
\newblock \bibinfo{journal}{\emph{Proceedings of the ACM on Human-Computer Interaction}} \bibinfo{volume}{8}, \bibinfo{number}{CSCW1} (\bibinfo{year}{2024}), \bibinfo{pages}{1--29}.
\newblock


\bibitem[Yao et~al\mbox{.}(2022)]%
        {yao2022learning}
\bibfield{author}{\bibinfo{person}{Zheng Yao}, \bibinfo{person}{Haiyi Zhu}, {and} \bibinfo{person}{Robert~E Kraut}.} \bibinfo{year}{2022}\natexlab{}.
\newblock \showarticletitle{Learning to Become a Volunteer Counselor: Lessons from a Peer-to-Peer Mental Health Community}.
\newblock \bibinfo{journal}{\emph{Proceedings of the ACM on Human-Computer Interaction}} \bibinfo{volume}{6}, \bibinfo{number}{CSCW2} (\bibinfo{year}{2022}), \bibinfo{pages}{1--24}.
\newblock


\bibitem[Zhang et~al\mbox{.}(2019)]%
        {BERTScore}
\bibfield{author}{\bibinfo{person}{Tianyi Zhang}, \bibinfo{person}{Varsha Kishore}, \bibinfo{person}{Felix Wu}, \bibinfo{person}{Kilian~Q Weinberger}, {and} \bibinfo{person}{Yoav Artzi}.} \bibinfo{year}{2019}\natexlab{}.
\newblock \showarticletitle{BERTScore: Evaluating Text Generation with BERT}. In \bibinfo{booktitle}{\emph{International Conference on Learning Representations}}.
\newblock


\bibitem[Zhang et~al\mbox{.}(2020)]%
        {DialoGPT}
\bibfield{author}{\bibinfo{person}{Yizhe Zhang}, \bibinfo{person}{Siqi Sun}, \bibinfo{person}{Michel Galley}, \bibinfo{person}{Yen-Chun Chen}, \bibinfo{person}{Chris Brockett}, \bibinfo{person}{Xiang Gao}, \bibinfo{person}{Jianfeng Gao}, \bibinfo{person}{Jingjing Liu}, {and} \bibinfo{person}{William~B Dolan}.} \bibinfo{year}{2020}\natexlab{}.
\newblock \showarticletitle{DIALOGPT: Large-Scale Generative Pre-training for Conversational Response Generation}. In \bibinfo{booktitle}{\emph{Proceedings of the 58th Annual Meeting of the Association for Computational Linguistics: System Demonstrations}}. \bibinfo{pages}{270--278}.
\newblock


\bibitem[Zhou et~al\mbox{.}(2020)]%
        {zhou2020design}
\bibfield{author}{\bibinfo{person}{Li Zhou}, \bibinfo{person}{Jianfeng Gao}, \bibinfo{person}{Di Li}, {and} \bibinfo{person}{Heung-Yeung Shum}.} \bibinfo{year}{2020}\natexlab{}.
\newblock \showarticletitle{The design and implementation of xiaoice, an empathetic social chatbot}.
\newblock \bibinfo{journal}{\emph{Computational Linguistics}} \bibinfo{volume}{46}, \bibinfo{number}{1} (\bibinfo{year}{2020}), \bibinfo{pages}{53--93}.
\newblock


\bibitem[Ziems et~al\mbox{.}(2023)]%
        {ziems2023can}
\bibfield{author}{\bibinfo{person}{Caleb Ziems}, \bibinfo{person}{William Held}, \bibinfo{person}{Omar Shaikh}, \bibinfo{person}{Jiaao Chen}, \bibinfo{person}{Zhehao Zhang}, {and} \bibinfo{person}{Diyi Yang}.} \bibinfo{year}{2023}\natexlab{}.
\newblock \showarticletitle{Can Large Language Models Transform Computational Social Science?}
\newblock \bibinfo{journal}{\emph{arXiv preprint arXiv:2305.03514}} (\bibinfo{year}{2023}).
\newblock


\bibitem[Zohny et~al\mbox{.}(2024)]%
        {zohny2024generative}
\bibfield{author}{\bibinfo{person}{Hazem Zohny}, \bibinfo{person}{Sebastian~Porsdam Mann}, \bibinfo{person}{Brian~D Earp}, {and} \bibinfo{person}{John McMillan}.} \bibinfo{year}{2024}\natexlab{}.
\newblock \bibinfo{title}{Generative AI and medical ethics: the state of play}.
\newblock , \bibinfo{numpages}{75--76}~pages.
\newblock


\end{thebibliography}

\appendix
\newpage
\section{Appendix}

\begin{table}[!ht]
\caption{Comparison of different models on the two conditioning scenarios on 7C-HQ-small.   
% The differences in numbers are statistically non-significant (t-test, p > 0.1).   % \diyi{please mention it in the caption, whether these numbers are significant - same for other tables}
  }
  \label{tab:best_generator}
  \small
    \centering
    \begin{tabular}{llcccccc}
    \toprule
        \multicolumn{1}{l}{\begin{tabular}[l]{@{}l@{}} Conditioned on \\MI Strategies\end{tabular}}  & Model & ROUGE-1 & ROUGE-2 & ROUGE-L & BERT Score (F) & BLEU & Avg Length \\ \midrule
        No & BART & 0.118 & 0.015 & 0.111 & 0.841 & 0.061 & 7.269 \\ 
        No & GPT-2 & 0.119 & \textbf{0.018} & 0.110 & 0.839 & 0.058 & 7.496 \\ 
        No & DialoGPT & 0.119 & 0.015 & 0.111 & 0.841 & 0.060 & 7.127 \\ \midrule
        Yes & BART & 0.128 & 0.017 & \textbf{0.117} & 0.845 & 0.074 & 7.710 \\ 
        Yes & GPT-2 & 0.111 & 0.013 & 0.100 & \textbf{0.879} & \textbf{0.084} & 9.728 \\
        Yes	& DialoGPT & \textbf{0.132}	&0.012	&0.114	&0.878	&\textbf{0.084} & 8.871\\ \bottomrule
        % Yes & DialoGPT & \textbf{0.141} & \textbf{0.025} & \textbf{0.128} & 0.870 & \textbf{0.180} &\textbf{10.442} \\ \hline
    \end{tabular}
\end{table}

\begin{table*}[!ht]
  \caption{Results of the next utterance strategy predictions (trained on 7C-HQ, reported on 7C-MI).}
  \label{tab:predictor}
  \begin{tabular}{lrcccc}
    \toprule
Strategy & \# Instances & Accuracy &	Recall &	Precision &	F1 Score\\
    \midrule
Open Question (QUO)&	3,586&	0.632&	0.802	&0.599&	0.686\\
Closed Question (QUC)	&3,714&	0.612&	0.795&	0.582&	0.672\\
Persuade (PR)	&3,778&	0.692&	0.788&	0.661&0.719\\
Reflection (RF)	&3,312&	0.648&	0.804&	0.613&	0.695\\
Support (SUP)&	2,742&	0.672&	0.783&	0.640	&0.705\\
Introduction/Greeting (INT)&	430	&0.867&	0.916&	0.835&	0.874\\
Grounding (GR)&	1,926	&0.721&	0.817&	0.685 &0.745\\
Affirm (AF)&	666&	0.722&	0.820&	0.686&	0.747\\
    \midrule
Overall	&20,154&	0.664&	0.801	&0.631&0.705\\
  \bottomrule
\end{tabular}
\end{table*}

% \section{Questionnaire presented to study participants}
\begin{table}
\caption{Questionnaire presented to study participants. The participants completed this questionnaire during the user studies. In addition to the questions in this table, screenshots of individual UI components are also included in the questionnaire as a visual aid to help participants identify which component of \texttt{CARE} a question refers to.} % \slh{I commented out the questions that are not in the results.}
  \label{tab:questionnaire}
    \centering
    \resizebox{\textwidth}{!}{%
    \begin{tabular}{l l}
    \toprule
        \textbf{Questions} & \textbf{Question type} \\ \midrule
        % How confident are you about listening in the category you selected? & Likert scale from 1 to 10 with 10 being the most comfortable \\ \hline
        How often did the suggested strategies (blue hints) \textbf{suit the situation?} & \multicolumn{1}{l}{\begin{tabular}[l]{@{}l@{}}  Likert scale with six options \\ Never, Very Rarely, Rarely, Occasionally, Frequently, Very Frequenty\end{tabular}}
         \\ 
        How often did the suggested strategies (blue hints) \textbf{help?} & \multicolumn{1}{l}{\begin{tabular}[l]{@{}l@{}}  Likert scale with six options \\ Never, Very Rarely, Rarely, Occasionally, Frequently, Very Frequenty\end{tabular}}\\ \midrule
        How often did the example responses (white buttons) \textbf{look natural on 7 Cups?} &\multicolumn{1}{l}{\begin{tabular}[l]{@{}l@{}}  Likert scale with six options \\ Never, Very Rarely, Rarely, Occasionally, Frequently, Very Frequenty\end{tabular}}  \\ 
        How often did the example responses (white buttons) \textbf{fit the conversation topic?} & \multicolumn{1}{l}{\begin{tabular}[l]{@{}l@{}}  Likert scale with six options \\ Never, Very Rarely, Rarely, Occasionally, Frequently, Very Frequenty\end{tabular}} \\ 
        How often did the example responses (white buttons) \textbf{help?} & \multicolumn{1}{l}{\begin{tabular}[l]{@{}l@{}}  Likert scale with six options \\ Never, Very Rarely, Rarely, Occasionally, Frequently, Very Frequenty\end{tabular}} \\ 
        How often did the example responses (white buttons) \textbf{contain \emph{harmful} message?} &\multicolumn{1}{l}{\begin{tabular}[l]{@{}l@{}}  Likert scale with six options \\ Never, Very Rarely, Rarely, Occasionally, Frequently, Very Frequenty\end{tabular}} \\ \midrule
        As a whole, was the tool \textbf{straightforward to use?} & \multicolumn{1}{l}{\begin{tabular}[l]{@{}l@{}} Likert scale with five options \\ Strongly Disagree, Disagree, Undecided, Agree, Strongly Agree\end{tabular}}  \\
        As a whole, was the tool \textbf{helpful?} & \multicolumn{1}{l}{\begin{tabular}[l]{@{}l@{}} Likert scale with five options \\ Strongly Disagree, Disagree, Undecided, Agree, Strongly Agree\end{tabular}} \\ 
        % As a whole, was the \textbf{reliable?} & \multicolumn{1}{l}{\begin{tabular}[l]{@{}l@{}} Likert scale with five options \\ Strongly Disagree, Disagree, Undecided, Agree, Strongly Agree\end{tabular}} \\
        \midrule
        What do you like about the tool? & 
        \multicolumn{1}{l}{\begin{tabular}[r]{@{}l@{}} Check box with the following options \\Saves time \\ Saves typing \\ Simplifies decision process \\ Increases listeners' confidence\\ Reminds listeners of counseling strategies\\ Inspires better responses \\ 
        Improves the overall quality of your responses \\ 
        None of the above \\ 
        Other (also has a text box)\end{tabular}}     \\   
        
        \midrule
        What do you dislike about the tool? & \multicolumn{1}{l}{\begin{tabular}[r]{@{}l@{}} Check box with the following options \\ 
        Takes more time to read examples \\ 
        Requires more decisions \\ 
        Disrupts thinking \\ 
        Limits the variety of counseling strategies \\ 
        Limits the diversity of responses \\ 
        Reduces the overall quality of your responses \\ 
        Interferes in listener-member communication \\ 
        None of the above \\ 
        Other (also has a text box)\end{tabular}}     \\  \midrule 
        This tool helps more when  & \multicolumn{1}{l}{\begin{tabular}[r]{@{}l@{}} Check box with the following options \\ 
        the Listener is newer to Listening chats. \\ 
        The listener is less confident in the chat's category. \\ 
        The listener is less familiar with the Member's chief complaint. \\ 
        The member is less communicative. \\ 
        the Listener feels more stressed. \\ 
        None of the above \\ 
        Other (also has a text box)\end{tabular}}    \\ \midrule
        If the tool is provided to you, how often do you think you will make use of it? & \multicolumn{1}{l}{\begin{tabular}[l]{@{}l@{}}  Likert scale with six options \\ Never, Very Rarely, Rarely, Occasionally, Frequently, Very Frequenty\end{tabular}} \\ \midrule
        % If you can switch the tool on and off as you like, will you switch it on? & \multicolumn{1}{l}{\begin{tabular}[l]{@{}l@{}}  Likert scale with six options \\ Absolutely Not, Probably Not, Possibly, Probably, Very Probably, Definitely\end{tabular}}  \\ \hline
        Any other comments? & Open-ended text box \\ \bottomrule
    \end{tabular}
    }
\end{table}

\clearpage
% Positive Examples  (Chat Excerpts)
\section{Positive Example Chat Excerpts}
\label{sec:pos-exceprts}
We selected one excerpt for each common type of situation in which \texttt{CARE} was shown to the participant, and the participant checked it. 
The text in the bottom input box is the actual response the participant sent. 
Texts adopted by the participants are highlighted in \colorbox{yellow}{yellow} or \colorbox{green}{green}.

\begin{figure}[h]
    \centering
    \includegraphics[width=0.9\textwidth]{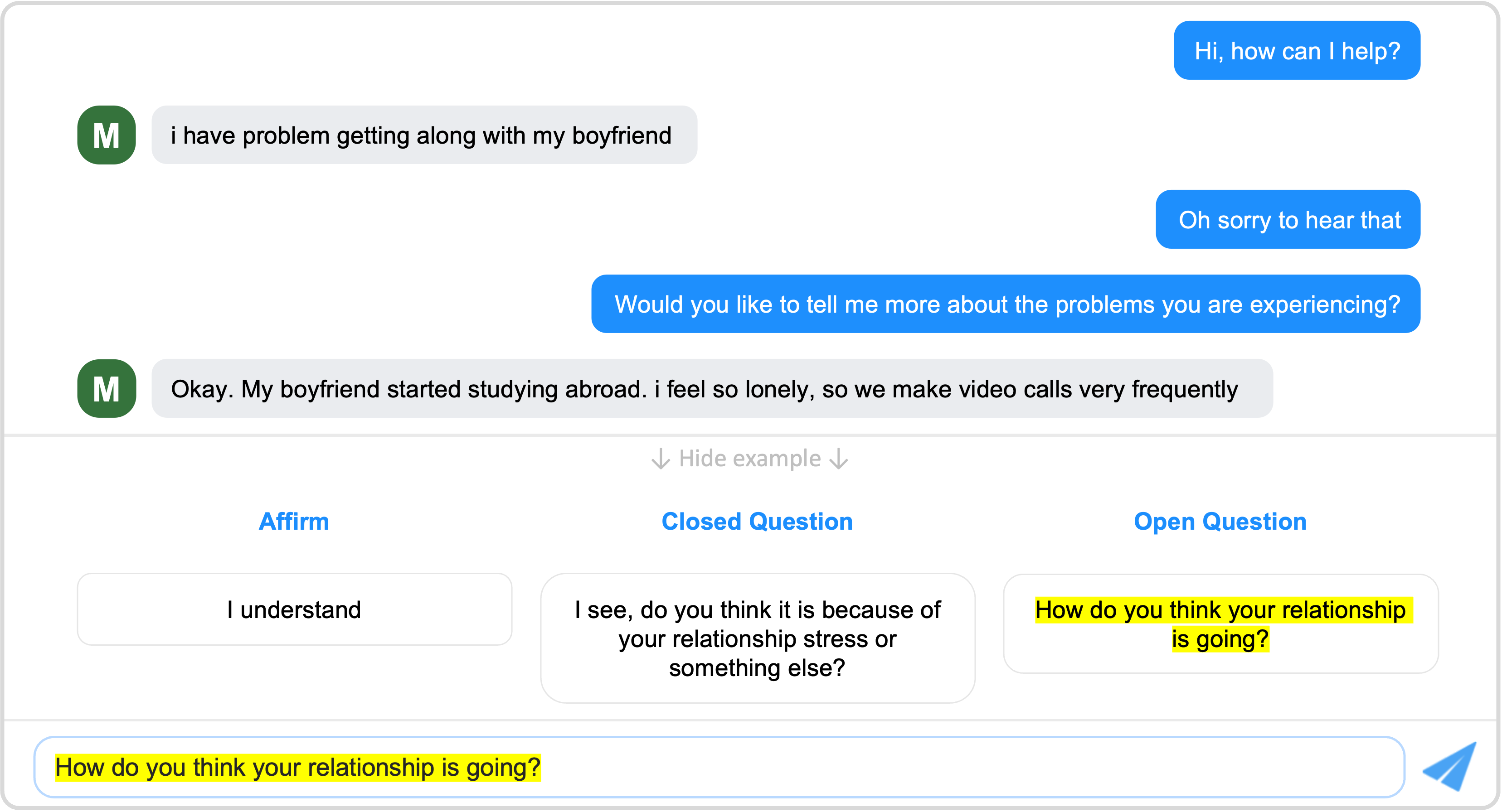}
    \caption{The participant selected and sent a response from \texttt{CARE} without any modification. 
}
    \label{fig:chat-1}
\end{figure}

\begin{figure}[h]
    \centering
    \includegraphics[width=0.9\textwidth]{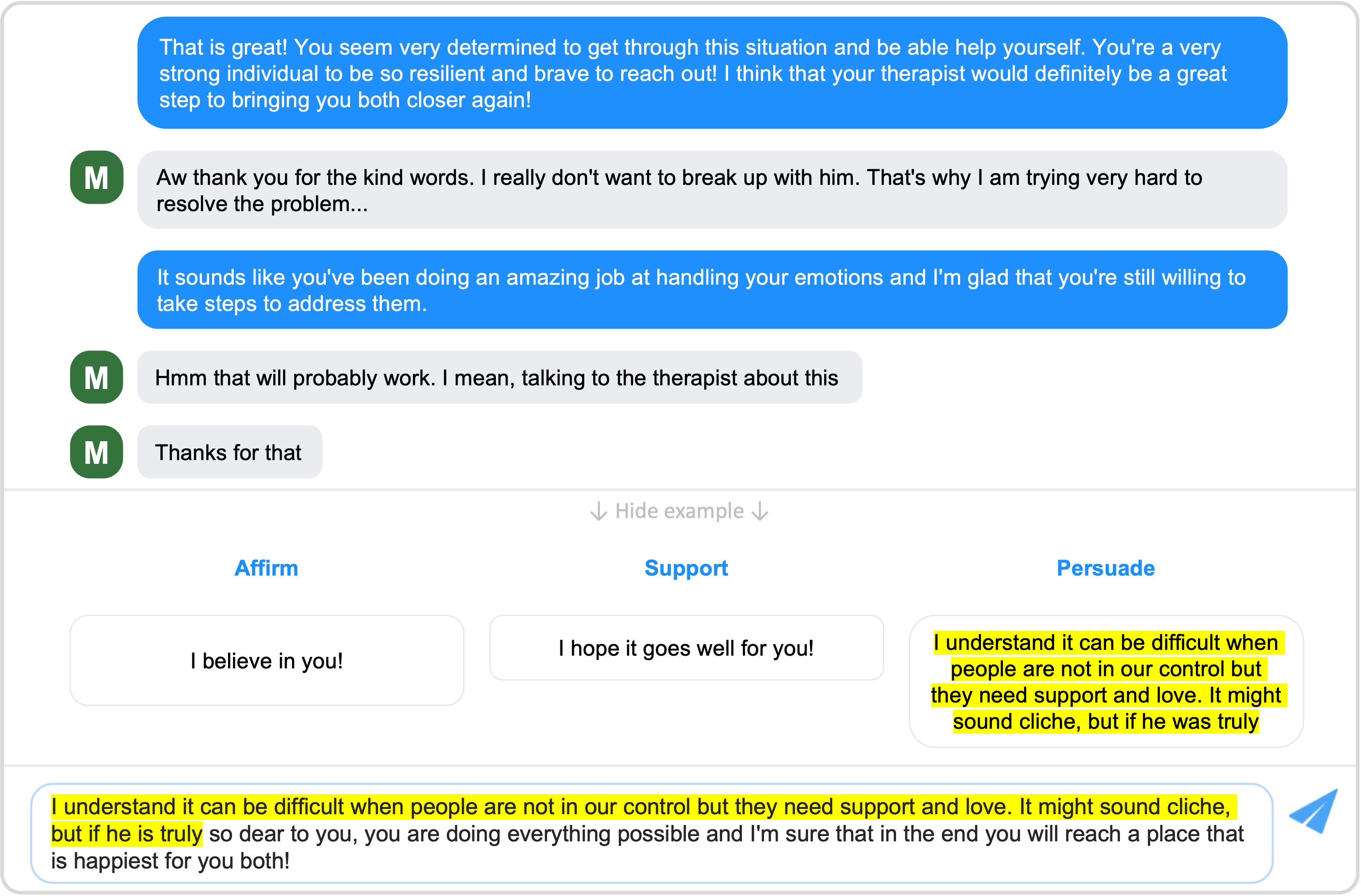}
    \caption{The participant added to a \texttt{CARE} response by completing it. That \texttt{CARE} response was truncated because of the generation length limit.}
    \label{fig:chat-2}
\end{figure}

\begin{figure}[h]
    \centering
    \includegraphics[width=0.9\textwidth]{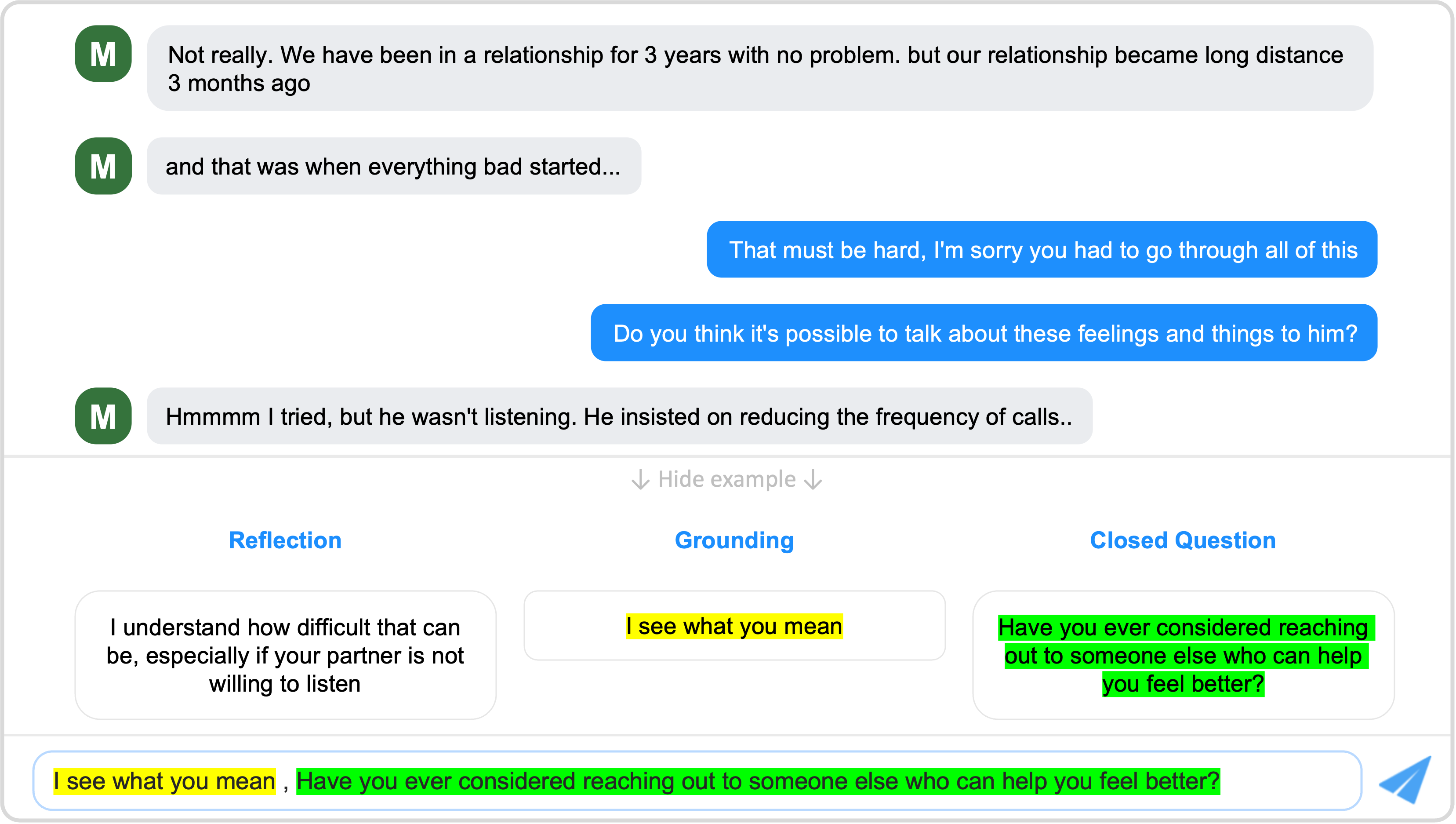}
    \caption{The participant combined two suggestions from \texttt{CARE} by concatenating them with a comma.}
    \label{fig:chat-3}
\end{figure}

\begin{figure}[h]
    \centering
    \includegraphics[width=0.9\textwidth]{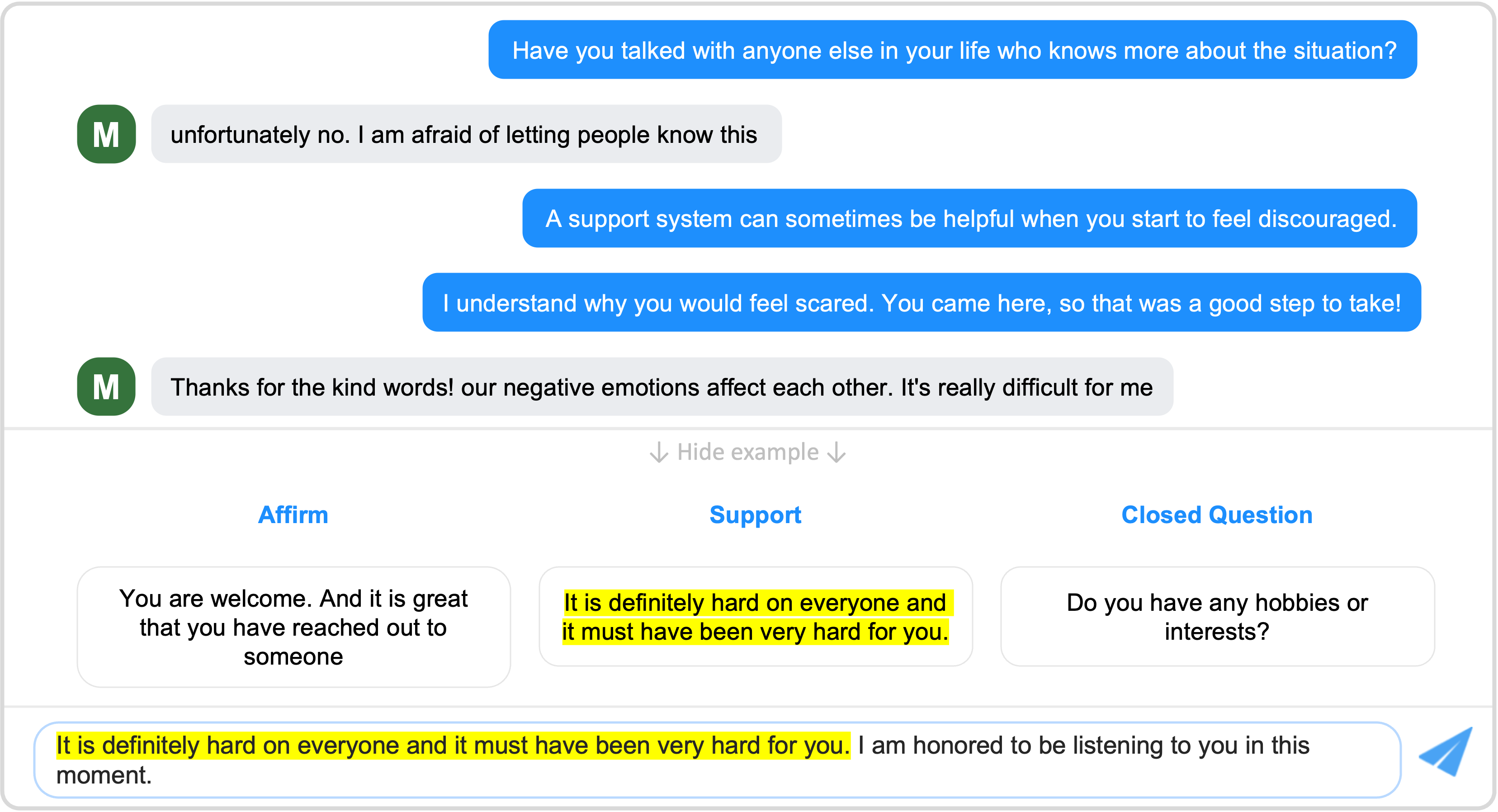}
    \caption{The participant added their feelings (about being honored) to the suggestion provided by\texttt{CARE}.}
    \label{fig:chat-4}
\end{figure}

\begin{figure}[h]
    \centering
    \includegraphics[width=0.9\textwidth]{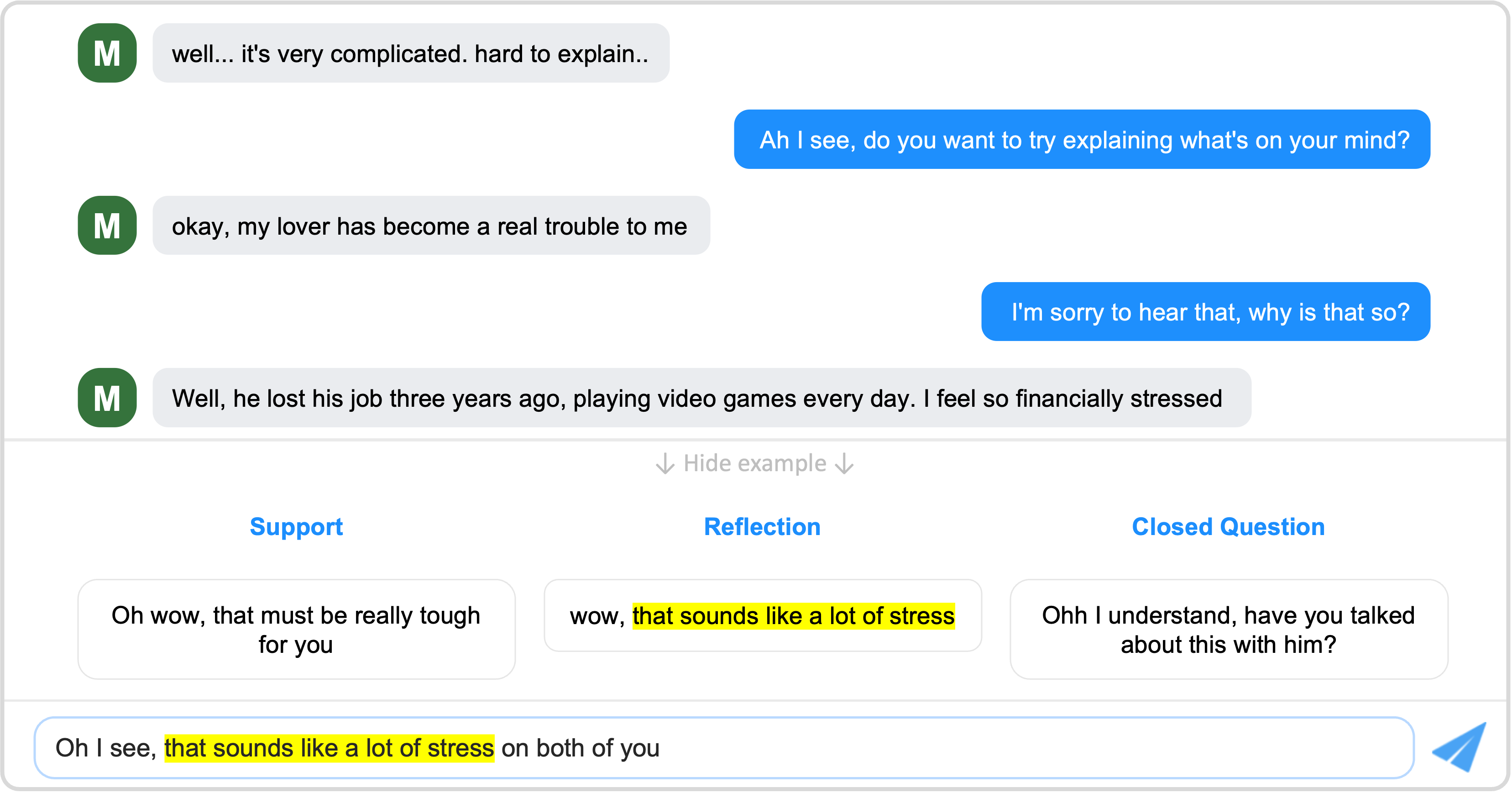}
    \caption{The participant personalized\texttt{CARE}'s suggestion to fit their style.}
    \label{fig:chat-5}
\end{figure}

\clearpage
% Negative Examples (Chat Excerpts)
\section{Negative Example Chat Excerpts}
\label{sec:neg-exceprts}
We selected excerpts for situations in which \texttt{CARE} was shown to the participant, but the participant did not select it.
\begin{figure}[h]
    \centering
    \includegraphics[width=0.9\textwidth]{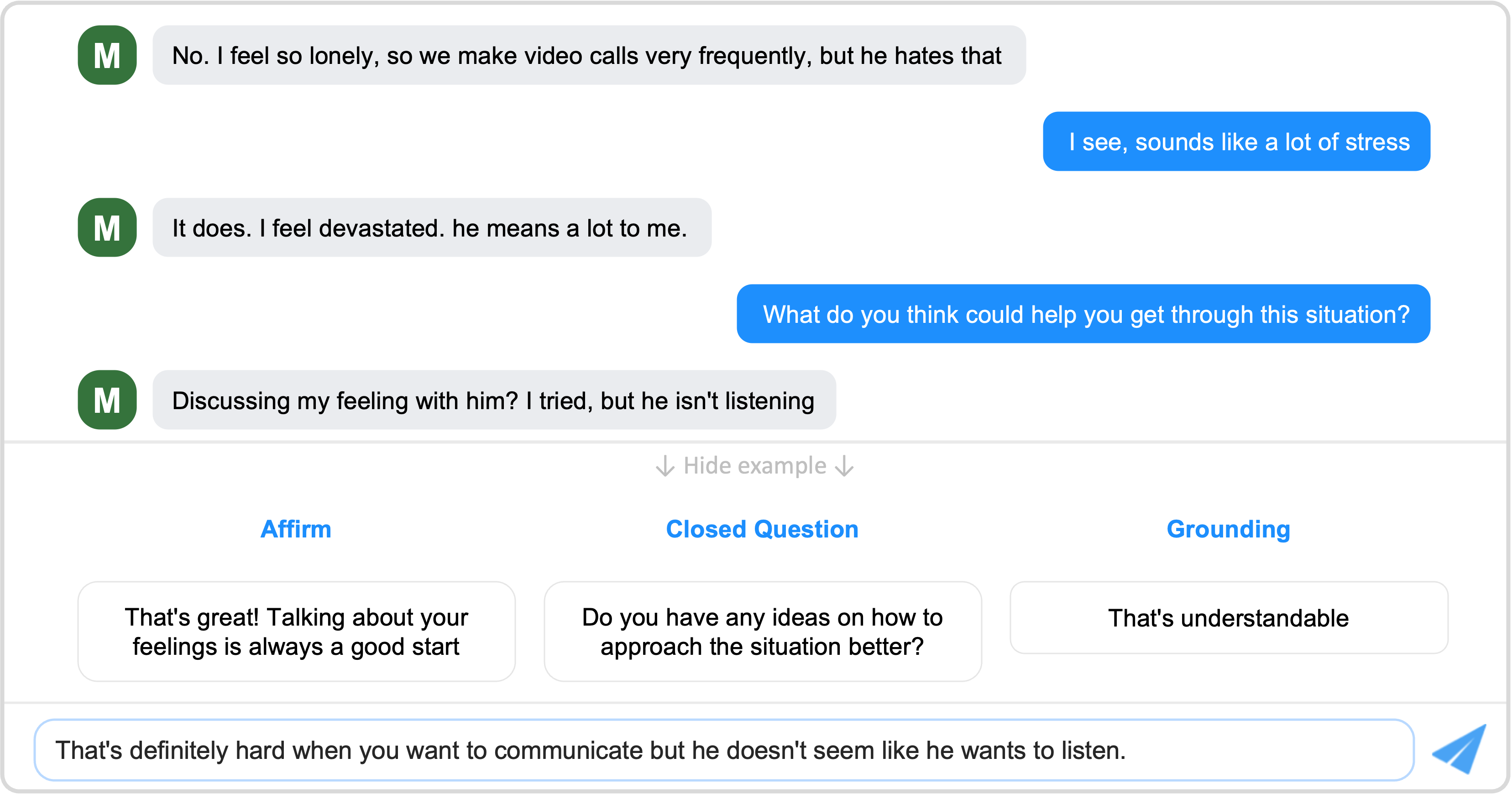}
    \caption{CARE’s suggestions were valid, but the participant made a different point from the suggested ones.}
    \label{fig:chat-6}
\end{figure}

\begin{figure}[h]
    \centering
    \includegraphics[width=0.9\textwidth]{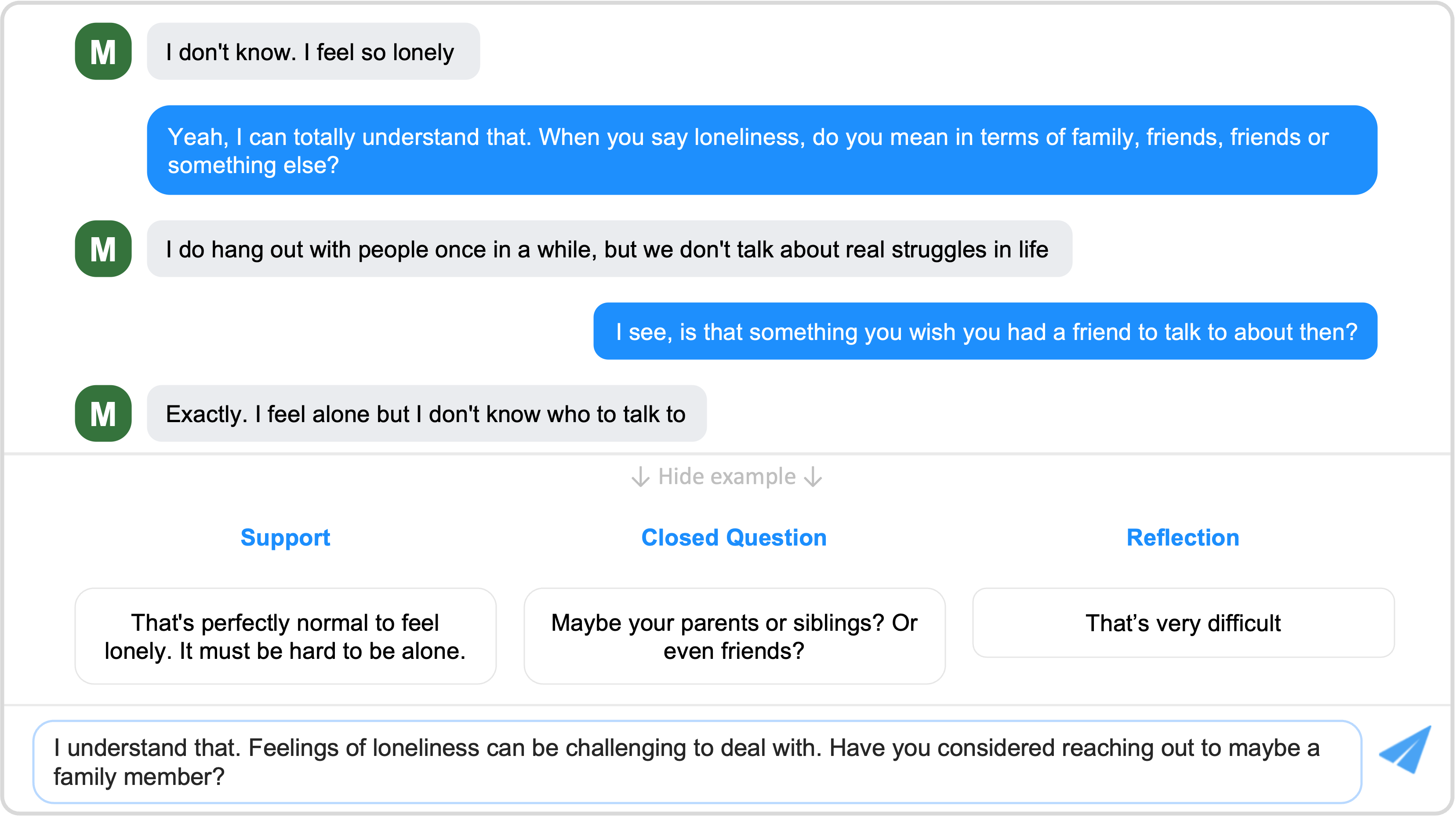}
    \caption{CARE’s suggestions were valid, but the participant combined more than one strategy in their response.}
    \label{fig:chat-7}
\end{figure}

\begin{figure}[h]
    \centering
    \includegraphics[width=0.9\textwidth]{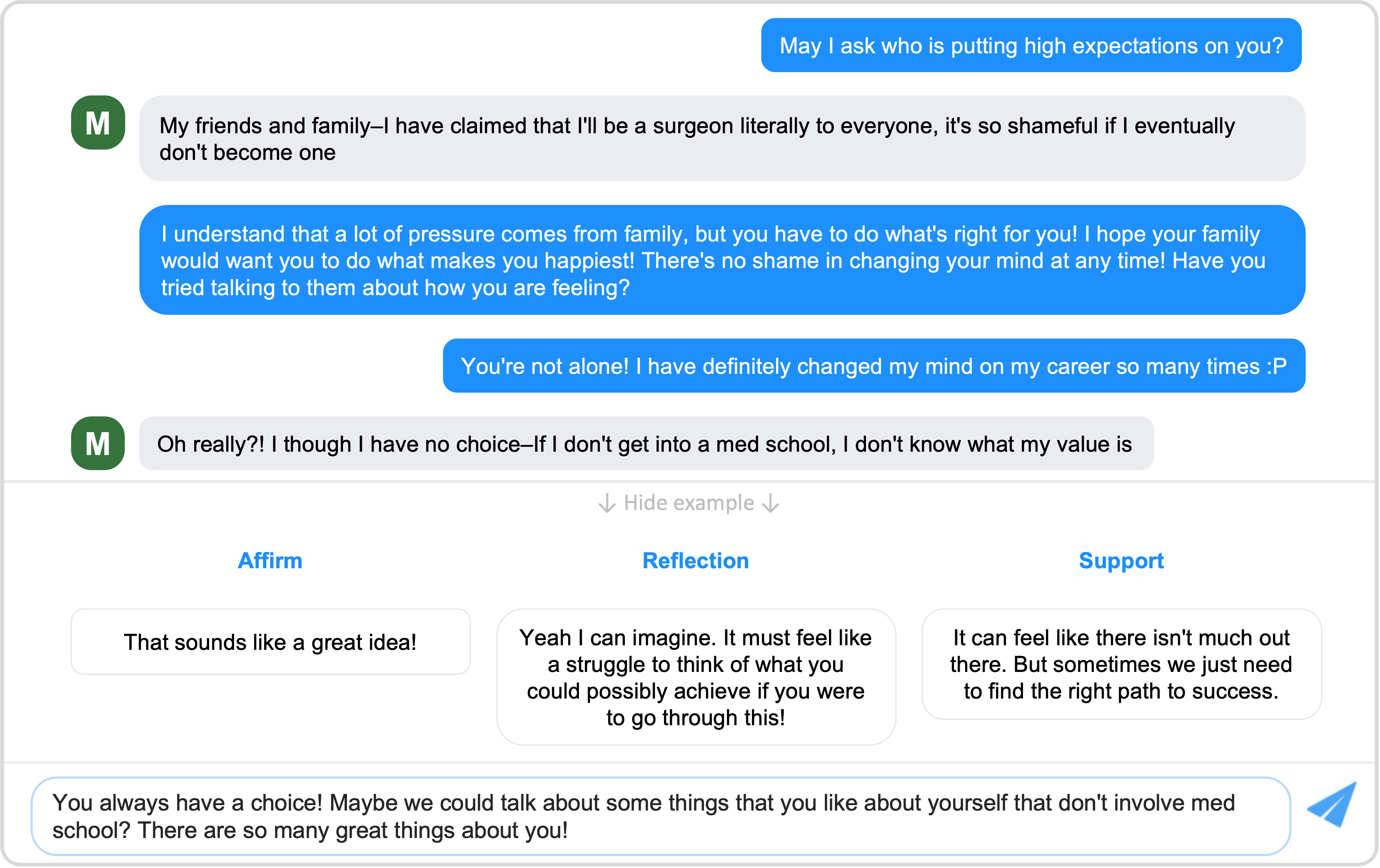}
    \caption{\texttt{CARE} replied based on the literal message without considering the underlying sentiment and implications.}
    \label{fig:chat-8}
\end{figure}

\begin{figure}[h]
    \centering
    \includegraphics[width=0.9\textwidth]{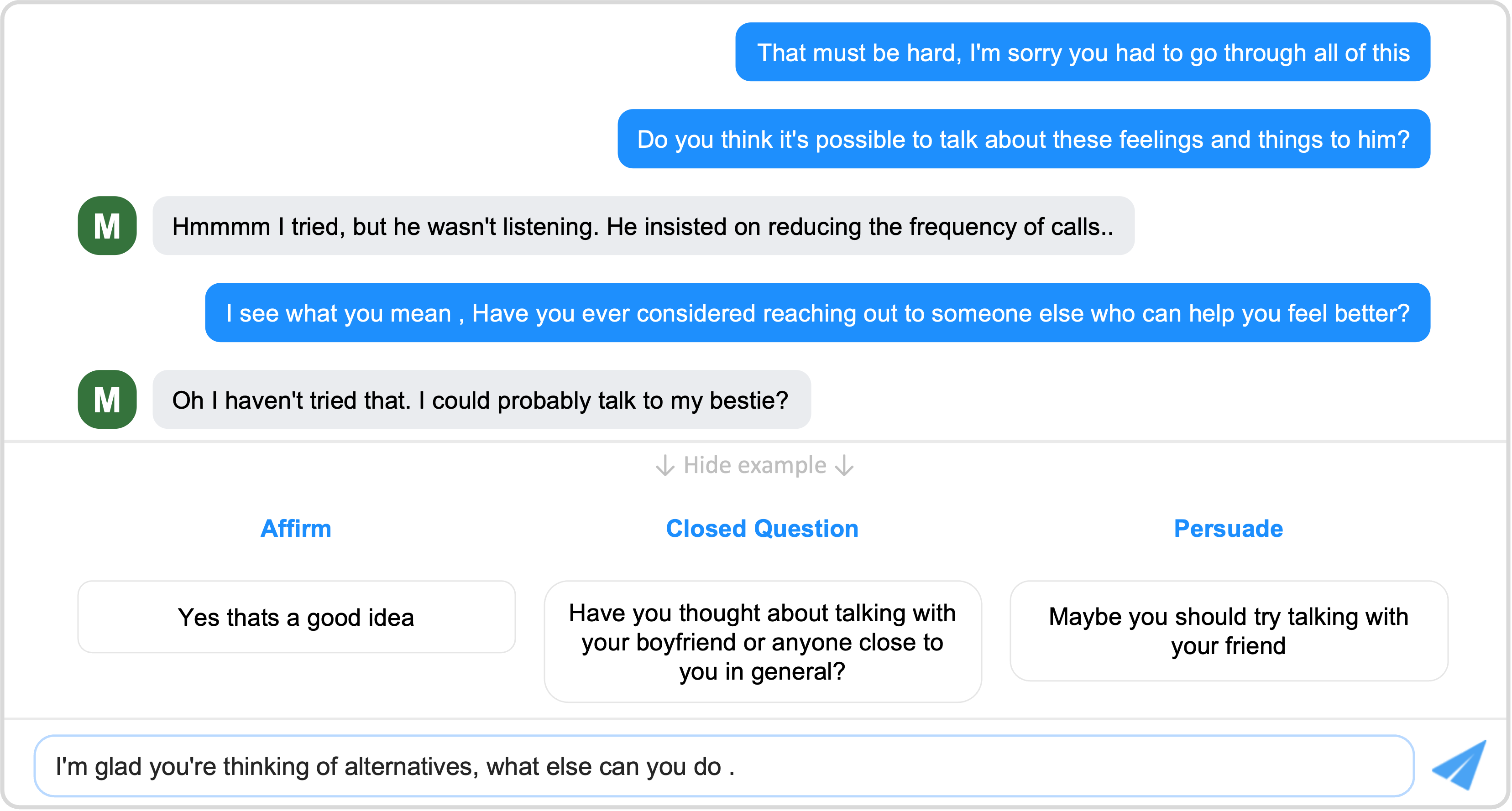}
    \caption{Neither the identity of the “he” nor the reason of the calls was given to\texttt{CARE}, making it challenging for \texttt{CARE} to respond accordingly. On the other hand, a human counselor can remember or refer to the previous messages before responding.}
    \label{fig:chat-9}
\end{figure}

\section{Details of System Development}\label{sec:sys-details}
Specifically, the system is hosted on an Amazon AWS server with an Ubuntu Operating system (32 Threads, 480GB RAM) having 8 Tesla k80 GPUs (12 GB RAM). These specifications enable \texttt{CARE} to interact seamlessly with humans. The back end is implemented with a Flask API that uses WebSockets to enable continuous bi-directional communication with the chat client. The functions provided by the back-end API fulfill the requirements of the system such as posting and pushing log-in credentials, posting chat logs for post-hoc analysis, and pushing generations from the back-end models. Server-side event handlers are invoked through API function calls made by the front-end interface. All models (DialoGPT, BERT) are hosted on the same server to enable faster communication. 
Given the confidentiality of the training data of our models, since releasing a language model is subject to potential data leakage~\cite{carlini2021extracting}, we do not publicly release our fine-tuned models.

\section{Transcript}
\label{transcript}
\begin{displayquote}
This video introduces a new system for listeners to chat with members. It's similar to the Seven Cups one, except that it will suggest generated strategies and example responses to you. We will discuss them in detail later.

Please note that the use of the suggestions and generations is voluntary. In other words, you can use them if they help, but if they don't, feel free to disregard them.

First of all, you will receive a chat ID and a user ID to log in. Then, you can start the chat with us. It looks and feels just like Seven Cups. When the tool feels confident, it will suggest strategies and example responses.

When the generations are available, you can click on the box over the message input field to see the strategies and example responses. Each suggestion comes with the corresponding example response. For example, this response uses this strategy. Click a response to fill the input box with it. You may edit the response before sending it if you wish.

Now, let's look at the strategies one by one.

[Pause]

That's it. Please feel free to ask questions if anything is unclear as we continue.
\end{displayquote}

\section{Supplemental Discussion}
\subsection{Utilization and Learning Outcomes}\label{sec:click-through}
The effectiveness of writing assistants is often assessed using \emph{click-through rate} \cite{chen2019gmail, hohenstein2018ai, kannan2016smart, robertson2021can, trnka2009user, quinn2016cost}. 
The click-through rate (CTR) helps understand the user engagement, content effectiveness on which the participant prefers strategies and acts as an overall proxy for the generation quality. 
However, it remains unclear whether this metric reflects changes in learning outcomes due to its use in writing assistants. Therefore, studies in educational contexts employ additional metrics to evaluate their impact. For instance, \citet{dewi2020implementation} and \citet{huang2020effectiveness} assess students' grades in writing exams, while \citet{nazari2021application} use established tools to analyze learners' cognitive engagement \cite{fredricks2005school}, self-efficacy for writing \cite{bruning2013examining}, and achievement emotions \cite{pekrun2011measuring}. Although evaluating the long-term outcomes of \texttt{CARE} is beyond the scope of this study, we recommend that future researchers measure the influence of \texttt{CARE} on trainees and their support seekers using peer counseling metrics (e.g., \cite{wang2023metrics}).

\subsection{Representativeness of Participants}\label{sec:representives}
The participant pool in this study consists predominantly of females (11 females, 3 males, 1 unknown). This gender distribution is consistent with prior research on 7 Cups \cite{yao2022learning, baumel2015online} and other mental health platforms \cite{schleider2020acceptability, rickwood2021online}. We hypothesize that this skewed distribution may mirror the actual population, as care work has historically been dominated by females \cite{wilson2024creating}. We recommend that future research endeavors seek to include a more diverse participant pool to provide a broader perspective (Section~\ref{sec:limit-future}).

\section{Semi-Structure Interview Scripts}\label{sec:script-interview}
\begin{displayquote}
Other than what you might have typed in the questionnaires,
What do you like about the tool? 
What do you dislike about the tool? 
What are the advantages of using this tool?
What are the disadvantages of using this tool?
In what circumstances will this tool help more?
In what circumstances did this tool help less?
Did the tool affect you in any way during the Listening chat? In what way?
Comments on components: suggestion, generation, UI, UX?

Did "whether the tool was available" affect how you feel during the chat? In what way?
Other than whether the tool was available, did you notice anything else that affected how you feel during the chat? What and how?
Does "whether such a tool is available on 7 Cups" affect how you feel about conducting Listening chats in general?
Do you want to see such a tool deployed? In what way?

"Any background in psychology, counseling, or therapy?
-> How long have you been providing therapy?"
Mentor new listeners?
Do you do counseling other than volunteering as a listener on 7 Cups?

Any other thoughts, comments, etc. on anything you'd like to share?
\end{displayquote}

\section{Scripts for Mock Chats}\label{sec:script-mock-chats}
We show the scripts for mock chats used in our user studies in Table~\ref{tab:script-a1}--\ref{tab:script-r2}.

\begin{table}[!htp]\centering
\caption{Script for Anxiety 1}\label{tab:script-a1}
% \scriptsize
\begin{tabular}{lp{0.8\textwidth}}\toprule
\multirow{4}{*}{Introduction} &hi there \\
&nice to meet you \\
&pretty bad, and that’s why I’m here \\
&i'm doing badly \\
& \\
\multirow{4}{*}{What happened} &hmm both sound fine for me. \\
&well, i'll take the exam for medical school admission next month but I don't think I can get into any at all \\
&I feel like I have sacrificed a lot, but no school on earth will admit me anyways \\
&I don't know what the purpose of living like this is \\
& \\
\multirow{4}{*}{College life} &I don't have time to enjoy my college life either \\
&I just study all day \\
&no friends, no life, not like other college students \\
&I feel like I don't deserve a break because I'm not good enough \\
& \\
\multirow{5}{*}{Expectation} &i have claimed that I'll be a surgeon literally to everyone \\
&it's so shameful if I eventually don't become one \\
&People's expectations on me are so high, making me feel so stressful... \\
&People make me feel that if i get good grades in school, I have to be a surgeon. This feeling is awful. \\
&I believe everyone is standing there, watching me fail. I hate this. \\
& \\
\multirow{4}{*}{Emotion} &I hate studying all day. I can't stand with it anymore \\
&i have no choice. If I don't get into a med school, I don't know what my value is \\
&what can i do? i literally have no idea \\
&I feel so bad \\
& \\
\multirow{3}{*}{Unconfidence} &i don't think i'm that good \\
&i am not as brilliant as those medical school students \\
&i don't think i can make it \\
\bottomrule
\end{tabular}
\end{table}

\begin{table}[!htp]\centering
\caption{Script for Anxiety 2}\label{tab:script-a2}
\begin{tabular}{lp{0.8\textwidth}}\toprule
\multirow{4}{*}{Introduction} &Hello \\
&Nice to meet you too \\
&Not very well \\
&I'm not feeling well \\
& \\
\multirow{7}{*}{What happened} &I'll try but i'm not sure \\
&my mood has been up and down in the past few days \\
&i just feel bad \\
&I don't know what to do \\
&I feel so alone \\
&I feel like I don't have real friends \\
&I do hang out with people once in a while, but we don't talk about real struggles in life. I feel so disappointed. \\
& \\
\multirow{3}{*}{Real friends} &real friends are supposed to support each other \\
&but we only chitchat and tease \\
&I don't know. I feel so lonely \\
& \\
\multirow{2}{*}{Selfish} &well, most people care only about themselves, not others \\
&We look like friends but don't really help others \\
& \\
\multirow{2}{*}{Specifics} &I can't think of a good example, and maybe that's where the problem is? \\
&I guess I will feel embarrassed if they aren't interested? \\
& \\
\multirow{4}{*}{No friends} &I feel alone but I don't know who to talk to \\
&I guess so. My classmates always tag many friends in their posts, but I don't have that many friends to tag. \\
&I feel so embarrassed \\
&Yup and I don't have that many friends to tag either \\
& \\
\multirow{2}{*}{Not coorperative} &No. I don't think they can solve my problem \\
&How can i do that? Isn't it weird to talk about that? \\
\bottomrule
\end{tabular}
\end{table}

\begin{table}[!htp]\centering
\caption{Script for Relationship Stress 1}\label{tab:script-r1}
% \small
\resizebox{\textwidth}{!}{%
\begin{tabular}{ll}\toprule
\multirow{4}{*}{Introduction} &hi there \\
&nice to meet you \\
&pretty bad, and that’s why I’m here \\
&i'm doing badly \\
\multicolumn{2}{c}{} \\
\multirow{2}{*}{What happended} &Well, I have problem getting along with my boyfriend \\
&we have a lot more quarrels than before \\
\multicolumn{2}{c}{} \\
\multirow{7}{*}{Problem} &my boyfriend started studying abroad \\
&i feel so lonely, so we make video calls very frequently \\
&he hates it, saying that I affect his academic performance. \\
&He blames his bad grades on me?! That's unfair! It's just an excuse. \\
&I feel devastated. he means a lot to me. \\
&I can't feel his love anymore. This is such a hard time for me... \\
&I think the frequency of the video calls is okay, but my boyfriend insists on reducing it \\
\multicolumn{2}{c}{} \\
\multirow{2}{*}{Long Distance} &we have been in a relationship for 3 years \\
&it has been 3 months since our relationship became long distance \\
\multicolumn{2}{c}{} \\
\multirow{4}{*}{Quarrel} &the quarrels are just over really unimportatns stuffs. so annoying \\
&I tried, but he wasn't listening \\
&We both adapting to college life \\
&Tthe quarrels over those really unimportatns stuffs are so annoying \\
\bottomrule
\end{tabular}
}
\end{table}

\begin{table}[!htp]\centering
\caption{Script for Relationship Stress 2}\label{tab:script-r2}
% \scriptsize
\resizebox{\textwidth}{!}{%
\begin{tabular}{lp{0.8\textwidth}}\toprule
\multirow{4}{*}{Introduction} &Hello \\
&Nice to meet you too \\
&Not very well \\
&I'm not feeling well \\
& \\
\multirow{2}{*}{What happened} &well... it's very complicated. hard to explain.. \\
&my lover has become a real trouble to me \\
& \\
\multirow{3}{*}{Finance} &he lost his job three years ago, playing video games every day. I feel so financially stressed \\
&he wants to be a profesisonal gamer, but he hasn't made any progress for 3 years... \\
&I really hope that he gets a real job, but he doesn't listen to me, unfortunately \\
& \\
\multirow{4}{*}{Mental health} &we both suffer from mental health problems \\
&I have depression, and i believe he has it too \\
&our negative emotions affect each other \\
&whenever i try to talk to him, he yells at me \\
& \\
\multirow{13}{*}{Uncommunicative} &you can't solve the problem for me \\
&i don't think there's any way to solve this \\
&but i don't want to break up with him \\
&unfortunately no \\
&understood. what can i do then? \\
&Will that work? I have litttle confidence \\
&It's very complicated and hard to explain, but I'll still try to describe what happened \\
&Definitely... He wants to be a profesisonal gamer, but he hasn't made any progress for 3 years \\
&Uh not really.. Whenever i try to talk to him, he yells at me \\
&I guess that was casued by his depression although he hasn't been diagno \\
&That is very true. I really hope that he gets a real job, but he doesn't listen to me, unfortunately \\
&Yeah that's exactly what I mean. \\
&Not yet. That's probably somthing I should try although I don't think there's really a solution to the program \\
\bottomrule
\end{tabular}
}
\end{table}
\section{Access to the training datasets in the study}

7Cups retains ownership of the training and evaluation data used in this study. Researchers interested in accessing these datasets are encouraged to contact 7 Cups to arrange the necessary data use agreements to access the train datasets for the study.
\received{January 2024}
\received[revised]{July 2024}
\received[accepted]{October 2024}

% For Articles V9cscw021-V9cscw212 use:

\end{document}